\documentclass{article}

\usepackage{arxiv}
\usepackage{3_preamble}

\title{Simultaneous Inference in Multiple Matrix-Variate Graphs for High-Dimensional Neural Recordings}
\renewcommand{\shorttitle}{Multiple Matrix-Variate Graphs for Neural Recordings}

\author{ 
    Zongge Liu \thanks{The authors contribute equally to this paper.} \\
    Department of Statistics \& Data Science \\
    Carnegie Mellon University\\
    Pittsburgh, PA 15213 \\
    \texttt{zonggel@andrew.cmu.edu}
    \And
    Heejong Bong \footnotemark[\value{footnote}]\\
    Department of Statistics\\
    Purdue University\\
    West Lafayette, IN 47907 \\
    \texttt{hbong@umich.edu} \\
    \And
    Zhao Ren \\
    Department of Statistics\\
    University of Pittsburgh\\
    Pittsburgh, PA 15260 \\
    \texttt{zren@pitt.edu} \\
    \AND
    Matthew A. Smith \\
    Department of Biomedical Engineering \\
    Carnegie Mellon University \\
    Pittsburgh, PA 15213 \\
    \texttt{msmith@andrew.cmu.edu} \\
    \And
    Robert E. Kass \\
    Department of Statistics \& Data Science \\
    Carnegie Mellon University \\
    Pittsburgh, PA 15213 \\
    \texttt{kass@stat.cmu.edu} \\
}

\hypersetup{
pdftitle={},
pdfsubject={},
pdfauthor={},
pdfkeywords={Gaussian graphical model; Simultaneous testing; Multiple graphs; Heterogeneous learning; Bandable matrix},
}

\begin{document}

%%%%%%%%%%%%%%%%%%%%
%%% Front matter %%%
%%%%%%%%%%%%%%%%%%%%

\maketitle

\begin{abstract}
We study simultaneous inference for multiple matrix-variate Gaussian graphical models in high-dimensional settings. Such models arise when spatiotemporal data are collected across multiple sample groups or experimental sessions, where each group is characterized by its own graphical structure but shares common sparsity patterns. A central challenge is to conduct valid inference on collections of graph edges while efficiently borrowing strength across groups under both high-dimensionality and temporal dependence. We propose a unified framework that combines joint estimation via group penalized regression with a high-dimensional Gaussian approximation bootstrap to enable global testing of edge subsets of arbitrary size. The proposed procedure accommodates temporally dependent observations and avoids naive pooling across heterogeneous groups. We establish theoretical guarantees for the validity of the simultaneous tests under mild conditions on sample size, dimensionality, and non-stationary autoregressive temporal dependence, and show that the resulting tests are nearly optimal in terms of the testable region boundary. The method relies only on convex optimization and parametric bootstrap, making it computationally tractable. Simulation studies and a neural recording example illustrate the efficacy of the proposed approach.
\end{abstract}

% keywords can be removed
\keywords{Gaussian graphical model \and Simultaneous testing \and Multiple graphs \and Heterogeneous learning \and Bandable matrix}

%%%%%%%%%%%%%%%%%
%%% Main text %%%
%%%%%%%%%%%%%%%%%

\section{Introduction} \label{sec:intro}

Graphical models provide a powerful framework for characterizing dependency structures in high-dimensional data. Their matrix-variate extensions have been widely used for analyzing high-dimensional spatiotemporal data, with prominent applications including functional connectivity networks among brain regions in neuroscience \citep{kass2023identification,bong2020dynamiccs} and financial stock price networks in econometrics \citep{chang2018confidence}. A central challenge in these settings arises from the high dimensionality of the target graph, where the prohibitive number of candidate edges leads to a severe multiple testing burden. Addressing this challenge requires not only new inferential tools but also careful and efficient use of available data. In many applications, it is necessary to combine data from multiple sample groups collected under heterogeneous conditions, such as different subjects or experimental sessions, in order to overcome the limited sample size available within each group. However, naive pooling across groups can introduce spurious dependencies due to heterogeneity, while analyzing each group separately often results in substantial loss of statistical power. In this paper, we develop a method for jointly estimating multiple matrix-variate graphical models by leveraging shared sparsity structures across sample groups, together with a unified framework for high-dimensional inference on graph edges.

The literature on inference for graphical models is extensive \citep{friedman2008sparse,allen2010,jankova2015confidence,chen2019,ye2019,liu2013,ren2015}, but relatively few frameworks address matrix-variate Gaussian graphical models (MGGMs; see \cref{sec:MGGM}). Existing work includes \citep{leng2012,YIN2012,zhou2014,chen2019,ye2019}. For example, \citet{chen2019} developed a multiple testing framework for support recovery with asymptotic guarantees on normality and false discovery rate control, while \citet{ye2019} proposed paired tests for detecting changes in matrix graph across correlated samples. Both approaches rely on multiple testing procedures %and 
with false discovery rate control. In contrast, \citet{chang2018confidence} derived a single global test statistic for assessing subgroups of edges simultaneously in vector-variate graphical models under temporal dependence.

Inference procedures for multiple matrix-variate graphs remain largely unexplored. In the vector-variate setting, several approaches have been proposed for joint estimation of multiple graphs, including optimization-based methods \citep{Danaher2014,cai2016,lee2015} and Bayesian formulations with structured priors \citep{Peterson2015}. The work most closely related to our setting is \citet{zhu2018}, who studied estimation of multiple MGGMs via nonconvex optimization with sparsity and group Lasso penalties. While their estimator exhibits some desirable properties, its convergence relies on strong assumptions and formal inference procedures are not developed.

We develop and study a multiple matrix-variate graph-based method for discovering common sparsity structures with improved sensitivity. Our approach uses group Lasso to borrow strength across sample groups, yielding more accurate partial-correlation estimates than the naive strategy of fitting models separately to each group; see \cref{sec:methd} for details. We establish improved convergence rates through theoretical analysis and simulation studies in \cref{sec:theorem,sec:numerical_studies}. In particular, we provide a self-contained analysis of the convergence rates and prediction error of the group Lasso estimator under temporal dependence in \cref{sec:group_Lasso}. Although group Lasso itself dates to 2006 \citep{Yuan2006}, its theoretical properties under temporal correlation have not previously been studied, and may be of independent interest for high-dimensional statistics with dependent data. In \cref{sec:thm_inference}, we develop theoretical guarantees for the proposed inference procedures using a Gaussian approximation bootstrap method based on \citet{Chernozhukov2012}. This represents the first attempt to formulate a global significance test for edge groups of arbitrary size in matrix-variate graphs beyond false discovery rate based methods. Finally, we apply our method to microelectrode recordings from two brain regions, V4 and PFC, and identify distinct cross-area subgroups with differential dynamics across cognitive stages (\cref{sec:experiment}). Our results reveal increased V4–PFC connectivity during visual memory retention, supporting their joint role in working memory.

{\bf Notations:} For a vector $\mathbf{x}$, let $\norm{\mathbf{x}}_p$ denote the $\ell_p$-norm. For a real matrix $X$, we use $X_{i, \cdot}$ to denote the $i$-th row of $X$ and $X_{\cdot, j}$ to denote the $j$-th column of $X$.
% , $X_{i, -j}$ denote the $i$-th row of $X$ with $j$-th column removed, $X_{-i, j}$ denote the $j$-th column of $X$ with $i$-th row removed, and $X_{-i}$ denote the original matrix with $i$-th row and $i$-th column removed. 
Let $\norm{X}_p$ denote the matrix $p$-norm, where $\norm{X}_2 $ is also referred to as the spectral or operator norm. We denote the Frobenius norm of $X$ by $\norm{X}_F$ and the entry-wise supremum norm by $\norm{X}_\infty$. For a set $A$, its cardinality is denoted by $\abs{A}$. We denote the zero vector, whose elements are all zero, by $\mathbf{0}$. For a positive integer $d$, we use $[d]$ to denote the set $\{1,2,...,d\}$.

\section{Matrix-variate Gaussian Graphical Model} \label{sec:MGGM}

% Vector-valued graphical models have been extensively employed to explore conditional dependence relationships in high-dimensional data, including neural recordings \citep{FORNITO2013, Vinci2018}. In these models, each node in an undirected graph represents a vector entry, and two nodes are connected by an edge if and only if their corresponding entries exhibit conditional dependence given all other entries. When the random vector follows a multivariate Gaussian distribution, the partial correlations, or equivalently elements of the precision matrix, encode conditional dependencies between entries \citep{meinshausen2006}. Consequently, this type of graph is also referred to as a \emph{partial-correlation graph}.

% Since spatiotemporal data involve both time ($t=1,\dots,p$) and space ($i = 1,\dots, q$), we consider a matrix-variate extension. 

Let $X \in \mathbb{R}^{p \times q}$ denote an observed sample of spatiotemporal data.
At a given time point $t \in [p]$, the row vector $X_{t,\cdot} \in \mathbb{R}^q$ represents measurements across $q$ spatial locations. A \emph{matrix-variate Gaussian graphical model} (MGGM, \citealp{DAWID1981}) is specified by two covariance matrices: the temporal (row) covariance $\Sigma^{(\mathcal{T})} \in \mathbb{R}^{p \times p}$ and the spatial (column) covariance $\Sigma^{(\mathcal{S})} \in \mathbb{R}^{q \times q}$. We say that $X$ follows this model with mean $\mu$, denoted by 
$
    X \sim \mathrm{MN}(\mu, \Sigma^{(\mathcal{T})}, \Sigma^{(\mathcal{S})}),
$
if and only if the vectorized data $\mathrm{vec}(X)$ follow a multivariate normal distribution with mean $\mathrm{vec}(\mu)$ and covariance matrix $\Sigma^{(\mathcal{S})} \otimes \Sigma^{(\mathcal{T})}$, where $\otimes$ denotes the Kronecker product. 

Because this parametrization is not identifiable up to a scaling factor, we assume without loss of generality that $\operatorname{tr}(\Sigma^{(\mathcal{T})}) = p$. Imposing the Kronecker product structure reduces the number of free covariance parameters from $pq(pq+1)/2$ to $p(p+1)/2 + q(q+1)/2$. Under this model, the covariance between entries $X_{ti}$ and $X_{sj}$ is given by 
$
    \Cov(X_{ti}, X_{sj}) = \Sigma^{(\mathcal{T})}_{ts} \, \Sigma^{(\mathcal{S})}_{ij}.
$
If we denote the corresponding precision matrices by $\Omega^{(\mathcal{T})} := (\Sigma^{(\mathcal{T})})^{-1}$ and $\Omega^{(\mathcal{S})} := (\Sigma^{(\mathcal{S})})^{-1}$, then the inverse covariance of $\mathrm{vec}(X)$ admits a simple analytic form:
$
    (\Sigma^{(\mathcal{S})} \otimes \Sigma^{(\mathcal{T})})^{-1} = \Omega^{(\mathcal{S})} \otimes \Omega^{(\mathcal{T})}.
$
In MGGM, interest primarily lies in the conditional independence graph structure among $q$ spatial locations, where each node represents a spatial location and the absence of an edge indicates that the corresponding nodes are conditionally independent given all remaining nodes. For Gaussian distributions, this graph structure is encoded by the precision matrix, or equivalently, by partial correlations. Thus, the target parameters in this work are the spatial partial correlations, which can be recovered from the spatial precision matrix: the partial correlation between spatial points $i$ and $j$ is
$
    \rho^{(\mathcal{S})}_{ij}
    = -\frac{\Omega^{(\mathcal{S})}_{ij}}{\sqrt{\Omega^{(\mathcal{S})}_{ii} \, \Omega^{(\mathcal{S})}_{jj}}},
$
which remains invariant across all time points (see Section~\ref{sec:spatial_est}). % By this way, the spatial association structure in $X$ is encoded by $\Sigma^{(\mathcal{S})}$.
% It imposes a strong assumption, implying, for instance, that the auto-correlation of measurement in every electrode is proportional to $\Sigma^{(\mathcal{T})}$, and the spatial covariance at every time point is proportional to $\Sigma^{(\mathcal{S})}$. But, as long as it is reasonable to separate temporal and spatial covariation, it has the advantage of reducing the number of parameters to $p^2 + q^2$ from $p^2q^2$ of the vector-variate formulation, and it has been applied in analyzing spatiotemporal data of biomedical imaging and financial markets \citep{zhou2014, chen2019,zhu2018}. 

\section{Simultaneous Inference Framework}\label{sec:methd}

In this section we develop an inference framework for identifying spatial partial-correlation graphs in multiple spatiotemporal observations. The full procedure is summarized in \cref{alg:procedure}.

For each sample group $l \in [m]$, we observe $n_l$ i.i.d. spatiotemporal measurements in $p \times q$ matrix-variate samples.
% or i.i.d. $p\times q$ matrix-variate samples, $X^{(1,l)},..., X^{(n_l,l)}$, 
At group $l$, we assume for $k \in [n_l]$ that the $k$-th sample $X^{(k,l)}$ follows matrix-variate Gaussian distribution $\distMN(0, \Sigma^{(\mathcal{T},l)}, \Sigma^{(\mathcal{S},l)})$, where $\Sigma^{(\mathcal{T},l)} \in \reals^{p \times p}$ and $\Sigma^{(\mathcal{S},l)} \in \reals^{q \times q}$ are the group-specific temporal and spatial covariance matrices, respectively. We denote the temporal and spatial precision matrices by $\Omega^{(\mathcal{T},l)} := (\Sigma^{(\mathcal{T},l)})^{-1}$ and $\Omega^{(\mathcal{S},l)}:=(\Sigma^{(\mathcal{S},l)})^{-1}$, respectively. % where $n_l$ can vary among different sessions. %We further define $\mathcal{D} = \{ \{X^{(k,1)}\}_{k=1}^{n_1},...,\{X^{(k,m)}\}_{k=1}^{n_m} \}$ as the collection of the full observations over these sessions. %{\color{red}Have we used this notation at all? if not in the main draft but in the proofs, we should define $\mathcal{D}$ in the appendix in front of the proof session}.

%Furthermore, let $\mat{A}_t = \mat{U}_t^{-1}$ denote the row (temporal) precision matrix, and $\mat{B}_t = \mat{V}_t^{-1}$ denote column (spatial) precision matrix. It is easy to see that $ \mat{X}_t^{(i)} \sim N(\mat{0}, (\mat{A}_t\otimes \mat{B}_t)^{-1})$, which implies that the graph structure is encoded by $\mat{A}_t\otimes \mat{B}_t$. In particular, the spatial connectivity structure is the support of the column precision matrix $\mat{B}_t$. For column precision matrix, we define the corresponding partial correlation for each entry $(i,j)$ at session $t$ as $\rho_{tij} =- \frac{{b}_{tij}}{ ({b}_{tii} {b}_{tjj})^{1/2}}$. We define $\mathcal{D} = \{ \{\mat{X}^{(i)}_1\}_{i=1}^{n_1},...,\{\mat{X}^{(i)}_d\}_{i=1}^{n_d} \}$ as the collection of the full observations over these sessions.

\begin{algorithm*}
\caption{Simultaneous Testing for Multiple Matrix-variate Gaussian Graphical Models}\label{alg:procedure}
\begin{algorithmic}[1]
\State \textbf{Input:} Multi-group data $\mathcal{D}$, edge set $\mathcal{S}$, test level $\alpha$
\State \textbf{Output:} Confidence region $\mathcal{C}_E(1-\alpha)$ % Test result $\Psi_\alpha$.
    \State \textbf{Spatial precision matrix estimation:}
    \For{ $i = 1:q$} 
    \State Estimate the regression coefficient ${\beta}^{(\mathcal{S},l)}_{\cdot,i}$ and the residual ${\epsilon}^{(\mathcal{S},k,l)}_{\cdot,i}$ using \cref{eq:spatial_est_beta}.
    % \State Estimate the regression residual $\epsilon^{(k)}_{tli}$ using \cref{eq:spatial_est_eps}.
    \EndFor
    \For{ $l=1:m$, $i=1:q$, $j=1:q$}
    \State Estimate the de-biased residual variance ${\Phi}^{(\mathcal{S},l)}_{ij}$ using \cref{eq:spatial_est_Phi}. 
    \State Estimate the spatial precision ${\Omega}^{(\mathcal{S},l)}_{ij}$ and partial correlation ${\rho}^{(\mathcal{S},l)}_{ij}$ using \cref{eq:spatial_est_omega_rho}.
    \EndFor
    \State \textbf{Temporal precision matrix estimation:}
    \For{ $l=1:m$}
    \State Estimate the temporal regression coefficient $\beta^{(\mathcal{T},l)}$, residual variance $\Phi^{(\mathcal{T},l)}$ and temporal covariance $\Sigma^{(\mathcal{T},l)}$ using \cref{eq:temporal_est_beta,eq:temporal_est_Phi,eq:temporal_est_Sigma_Omega}.
    \EndFor
    \State \textbf{Hypothesis testing based on bootstrap:}
    \State Estimate the covariance matrix $S_{EE}$ of the test statistic ${T}_E$ using the plug-in estimator $\hat{S}_{EE}$ in \cref{eq:inf_S}. 
    \State Sample $\{\hat{Z}_i\}_{i=1,\dots,B} \sim N({0}, \hat{S}_{EE})$ and calculate the confidence region in \cref{eq:inf_CI}.
    % \State Test result $\Psi_\alpha$ is given by Equation~\eqref{eqn:test1}.
    \State \textbf{return} Confidence region $\mathcal{C}_E(1-\alpha)$ % test results $\Psi_\alpha$.
\end{algorithmic}
\end{algorithm*}

\subsection{Estimation of Spatial Covariance Matrix}
\label{sec:spatial_est}

% We begin by considering a single session $l$ and time point $t$ to motivate our node-wise regression approach \citep{meinshausen2006,liu2013}. 
Under the matrix-variate Gaussian graphical model, for a given spatial location $i\in [q]$, at each time point $t \in [p]$, the conditional distribution of the random variable $X^{(k,l)}_{ti}$ given the remaining variables in $X^{(k,l)}_{t,\cdot}$ follows a normal distribution and can be modeled as a linear regression:
\begin{equation} \label{eq:spatial_reg_model}
    X^{(k,l)}_{ti} = X^{(k,l)}_{t,\cdot} \beta^{(\mathcal{S},l)}_{\cdot,i} + \epsilon^{(\mathcal{S},k,l)}_{ti},
\end{equation}
where the regression coefficients are related to the spatial precision matrix via
\(
    \beta^{(\mathcal{S},l)}_{ji} = -\frac{\Omega^{(\mathcal{S},l)}_{ji}}{\Omega^{(\mathcal{S},l)}_{ii}} \mathbb{I}(i \neq j),
\)
and $\mathbb{E}[\epsilon^{(\mathcal{S},k,l)}_{ti}] = 0$. Due to the Kronecker product structure in the model, the regression coefficients are time-independent. It is common in many applications of graphical models to assume that the graph exhibits a certain sparsity structure. Consequently, sparsity in the spatial precision matrix $\Omega^{(\mathcal{S},l)}$ implies corresponding sparsity in the regression coefficients $\beta^{(\mathcal{S},l)}$. 
%A similar relationship holds between the spatial precision matrix and 
A similar sparsity appears in the covariance of the node-wise regression residuals. Specifically, for the residuals $\epsilon^{(\mathcal{S},k,l)}_{t,\cdot} := (\epsilon^{(\mathcal{S},k,l)}_{t1}, \dots, \epsilon^{(\mathcal{S},k,l)}_{tq})^\top$, the covariance is given by
\(
\Cov[\epsilon^{(\mathcal{S},k,l)}_{t,\cdot}] = \Sigma^{(\mathcal{T},l)}_{tt} \cdot \Phi^{(\mathcal{S},l)},
\)
where % $\Phi^{(\mathcal{S},l)}$ is invariant across $k$ and $t$ with elements 
$\Phi^{(\mathcal{S},l)}_{ij} := \frac{\Omega^{(\mathcal{S},l)}_{ij}}{\Omega^{(\mathcal{S},l)}_{ii} \Omega^{(\mathcal{S},l)}_{jj}}$.
Our target, the spatial partial correlation, is expressed as
\(
\rho^{(\mathcal{S},l)}_{ij} = -\frac{\Omega^{(\mathcal{S},l)}_{ij}}{ \sqrt{\Omega^{(\mathcal{S},l)}_{ii} \Omega^{(\mathcal{S},l)}_{jj}}} = -\frac{\Phi^{(\mathcal{S},l)}_{ij}}{ \sqrt{\Phi^{(\mathcal{S},l)}_{ii} \Phi^{(\mathcal{S},l)}_{jj}}},
\)
and testing $\Phi^{(\mathcal{S},l)}_{ij} = 0$ is equivalent to testing $\rho^{(\mathcal{S},l)}_{ij} = 0$. % the partial correlation between $i$ and $j$ is zero.
%
% In \cref{eq:spatial_reg_model}, we treat each row of $X^{(k,l)}$ as a $q$-dimensional sample, giving us $p$ correlated vector-valued samples for a sparse linear regression model, where the covariance among these ``row samples'' is characterized by $\Sigma^{(\mathcal{T},l)}$. 
%
Because $\Phi^{(\mathcal{S},l)}$ is invariant across $t$ and $k$, % we estimate $\Phi^{(\mathcal{S},l)}$ by
$
\Phi^{(\mathcal{S},l)} = \Exp\left[\frac{1}{n_l p} \sum_{k=1}^{n_l}\sum_{t=1}^p \epsilon^{(\mathcal{S},k,l)}_{t,\cdot} \epsilon^{(\mathcal{S},k,l)\top}_{t,\cdot} \right],
$
noting that $\tr(\Sigma^{(\mathcal{T},l)}) = p$ under the identifiability constraint. It is worth noting that the residuals corresponding to different time points $t \in [p]$ exhibit temporal dependence.
%\begin{equation} \label{eq:spatial_Phi}
 %   \Phi^{(\mathcal{S},l)} = \Exp\Big[\frac{1}{n_l p} \sum_{k=1}^{n_l}\sum_{t=1}^p \epsilon^{(\mathcal{S},k,l)}_{t,\cdot} \epsilon^{(\mathcal{S},k,l)\top}_{t,\cdot} \Big].
%\end{equation} 
% where we effectively use $n_l p$ correlated samples in total.

% Finally, we define the average of the covariance matrices of residuals from these correlated $p$ samples as $\mat{R}_t = (r_{tij})_{q\times q}=\frac{1}{p} \sum_{l=1}^p \mat{R}^l_t$ with
% \begin{equation*}
%   r_{tij} = \frac{\mathrm{tr}(\mat{U}_t)}{p} \frac{b_{tij}}{b_{tii}b_{tjj}} = \frac{b_{tij}}{b_{tii}b_{tjj}}.  \label{eqn:rtij}  
% \end{equation*}
% To facilitate our analysis, we have assumed $\mathrm{tr}(\mat{U}_t)=p$  for each session in Equation (\ref{eqn:rtij}) to avoid any identifiability issue, which is formally stated in Assumption~\ref{as:id} of Section~\ref{sec:theory}. 

% Having discussed the model for individual sessions, 
We fit all $m$ sample groups jointly to improve the estimation accuracy of each $\rho^{(\mathcal{S},l)}_{ij}$. Under a mild assumption that the spatial precision matrices $\Omega^{(\mathcal{S},l)}_{ij}$ share the same sparsity pattern across sample groups, the corresponding regression coefficients $\beta^{(\mathcal{S},l)}_{\cdot,i}$ also exhibit a common support across the $m$ sample groups, albeit with potentially distinct values. To harness this group sparsity assumption, we treat the coefficients %$\beta^{(\mathcal{S},1)}_{ij}, \dots, \beta^{(\mathcal{S},m)}_{ij}$ 
$\beta^{(\mathcal{S},1)}_{ji}, \dots, \beta^{(\mathcal{S},m)}_{ji}$ as a group of parameters for each pair $(i,j)$ and apply group Lasso \citep{Yuan2006} % to the stacked $m$ linear models, thereby
to improve estimation by borrowing strength across sample groups:
%
% We introduce a few more notations before formally stating our procedure. Let $\mat{\beta}_{ti} = (\beta_{ti1}, \cdots, \beta_{ti(i-1)}, \beta_{ti(i+1)},\cdots, \beta_{tiq})' \in \mathbb{R}^{q-1}$. The stacked coefficient from $d$ sessions is denoted as $\mat{\beta}_{i}^0 = \left( {\mat{\beta}_{1i}}^{\T}, {\mat{\beta}_{2i}}^{\T}, \cdots, {\mat{\beta}_{di}}^{\T} \right)^{\T}\in \mathbb{R}^{(q-1)d}$. By the construction of $\mat{\beta}_i^0$, we have group sparsity structure in the sense that all but at most $s$ subvectors $\mat{\beta}^0_{i(l)}$ are none-zero where the $l$th group subvector of $\mat{\beta}^0_{i}$ is defined as $\mat{\beta}_{i(l)}^0 = ({\beta}_{1il}, ..., {\beta}_{dil})'\in \mathbb{R}^{d}$. From the definition, we observe that $\mat{\beta}_{i(l)}^0 = \mat{0}$ for all $(i,l)\in \mathcal{E}_s^c$, where $\mathcal{E}^c_s$ is the compliment set of $\mathcal{E}_s$.
%Let the stacked row samples for each session $\mat{Z}_t = \left({\mat{X}^{(1)}_t}^{\T}, {\mat{X}^{(2)}_t}^{\T}, \cdots, {\mat{X}^{(n_t)}_t}^{\T}\right)^{\T} \in \mathbb{R}^{n_t p \times q}$, and its $i$th column $\mat{Z}_{t,\cdot,i} = \left({\mat{X}_{t, \cdot, i}^{(1)}}^{\T}, {\mat{X}_{t, \cdot, i}^{(2)}}^{\T},\cdots, {\mat{X}_{t, \cdot, i}^{(n_t)}}^{\T}\right)^{\T} \in \mathbb{R}^{n_t p}$. 
%For each session/graph $t$ and each node $i$,  the residuals of the $k$th sample is denoted as 
% ${\mat{\varepsilon}}^{(k)}_{t,\cdot, i} =  \mat{X}_{t,\cdot,i}^{(k)} - \mat{X}_{t, \cdot,-i}^{(k)}{\mat{\beta}}_{ti}$.
\begin{equation} \label{eq:spatial_est_beta}
\begin{split}
    & \{\hat{\beta}^{(\mathcal{S},l)}_{\cdot,i}\}_{l=1,\dots,m}:= \\ 
    & \argmin_{\{b^{(l)}\}_{l=1,\dots,m}} \left\{
    \frac{1}{2 n_0 p} \sum_{l=1}^m\sum_{k=1}^{n_l} \norm{X^{(k,l)}_{\cdot,i} - X^{(k,l)} b^{(l)}}_2^2
    + \gamma_i \sum_{j:j \neq i}\sqrt{\sum_{l=1}^{m} \frac{\norm{\sum_k X^{(k,l)}_{\cdot,j}}_2^2}{n_l p} b^{(l)2}_j}
    \right\}, \\
    % & \text{w.r.t} ~~b^{(l)}_{i} = 0,
\end{split}
\end{equation}
with respect to $b_i^{(l)} = 0$ for all $l$, where $n_0=\min_{1\leq l\leq m} n_l$. The parameter $\gamma_i$ can be tuned using cross-validation or other model selection methods. %In our theoretical analysis, a data-driven yet conservative choice of each $\gamma_i$ can be picked. 

Once the regression coefficients are estimated, the residuals are computed as $\hat{\epsilon}^{(\mathcal{S},k,l)}_{ti} := X^{(k,l)}_{ti} - X^{(k,l)}_{t,\cdot} \hat{\beta}^{(\mathcal{S},l)}_{\cdot,i}$.
% \begin{equation*} \label{eqn:spatial_est_eps}
%   \hat{\epsilon}^{(\mathcal{S},k,l)}_{ti} = X^{(k,l)}_{ti} - X^{(k,l)}_{t,\cdot} \hat{\beta}^{(\mathcal{S},l)}_{\cdot,i}.
% \end{equation*}
Although the empirical covariance matrix of the fitted residuals provides a straightforward estimate of $\Phi_{ij}^{(\mathcal{S},l)}$, this estimate is biased due to the lasso-type penalty, leading to a larger error rate than the expected $1/\sqrt{n_l p}$ rate \citep{liu2013}. To address this bias, we %introduce 
adopt a bias-correction term in the covariance estimate:
\begin{equation} \label{eq:spatial_est_Phi}
    \hat{\Phi}^{(\mathcal{S},l)}_{ij} := \begin{cases}
        -\frac{1}{n_l p} \sum_{k=1}^{n_l} \sum_{t=1}^p \left( \hat{\epsilon}^{(\mathcal{S},k,l)}_{ti} \hat{\epsilon}^{(\mathcal{S},k,l)}_{tj} 
        + \hat{\epsilon}^{(\mathcal{S},k,l)2}_{tj} \hat{\beta}^{(\mathcal{S},l)}_{ji}
        + \hat{\epsilon}^{(\mathcal{S},k,l)2}_{ti} \hat{\beta}^{(\mathcal{S},l)}_{ij} \right), & \text{if} ~~ i \neq j, \\
        \frac{1}{n_l p} \sum_{k=1}^{n_l} \sum_{t=1}^p \hat{\epsilon}^{(\mathcal{S},k,l)}_{ti} \hat{\epsilon}^{(\mathcal{S},k,l)}_{tj}, & \text{if} ~~ i = j. 
    \end{cases}
\end{equation}
% Using this bias-corrected estimate, 
We then estimate the spatial precision matrix $\Omega^{(\mathcal{S},l)}$ and the partial correlation matrix $\rho^{(\mathcal{S},l)}$ as follows:
\begin{equation} \label{eq:spatial_est_omega_rho}
    \hat{\Omega}^{(\mathcal{S},l)}_{ij} := \frac{\hat{\Phi}^{(\mathcal{S},l)}_{ij}}{\hat{\Phi}^{(\mathcal{S},l)}_{ii}\hat{\Phi}^{(\mathcal{S},l)}_{jj}} ~~\text{and}~~
    \hat{\rho}^{(\mathcal{S},l)}_{ij} := - \frac{\hat{\Phi}^{(\mathcal{S},l)}_{ij}}{\sqrt{\hat{\Phi}^{(\mathcal{S},l)}_{ii} \hat{\Phi}^{(\mathcal{S},l)}_{jj}}}.
\end{equation}
While our partial correlation estimator is similar to the form proposed by \cite{chen2019} for a single matrix-variate Gaussian graphical model, our approach leverages information from multiple sample groups using the group Lasso estimate, leading to a faster convergence rate, as demonstrated in \cref{thm:spatial_regression}, which in turn facilitates the subsequent inference procedure.
% This improvement, which is summarized in \cref{thm:spatial_regression} and \cref{remark:sample_size_requirement}, results in enhanced testing power by a factor of $\sqrt{m}$, as demonstrated in \cref{thm:inf_bootstrap,thm:inf_power}.

%%%%%%%%%%%%%%%%%%%%%%%%%%%%%%%%%%
%%%%%%%%%%%%%%%%%%%%%%%%%%%%%%%%%%
%%%%%%%%%%%%%%%%%%%%%%%%%%%%%%%%%%

\subsection{Simultaneous Test by Parametric Bootstrap} \label{sec:inference}

\subsubsection{Single Edge Test}

By leveraging information across $m$ sample groups using group Lasso (\cref{eq:spatial_est_beta}), we not only improve the efficiency of spatial partial correlation estimates but also enhance the power of tests for detecting significant associations. We first consider testing a single edge: for a pair $(i,j)$, the null hypothesis is
\begin{equation} \label{eq:H0_single}
H_{0,ij}: \rho^{(\mathcal{S},l)}_{ij} = 0, \quad \forall l=1,\dots,m.
\end{equation}
Given the group sparsity structure, we construct a test statistic by summing the partial correlation estimates $\hat{\rho}^{(\mathcal{S},l)}_{ij}$ across all sample groups. To account for varying sample sizes across sample groups, we weight the estimates by $\sqrt{n_l p}$ and define the test statistic as
\(
    \hat{T}_{ij} := \frac{1}{\sqrt{m}} \sum_{l=1}^m \sqrt{n_l p} \, \hat{\rho}^{(\mathcal{S},l)}_{ij}.
\)
% Assuming the sign of the associations remains consistent, the mean of the test statistic $\sum_{l=1}^m \rho^{(\mathcal{S},l)}_{ij}$ equals zero if and only if the null hypothesis holds.
In \cref{thm:inf_CLT}, we show that $\hat{T}_{ij}$ asymptotically follows a normal distribution with mean
\(
T_{ij} := \frac{1}{\sqrt{m}} \sum_{l=1}^m \sqrt{n_l p} \, \rho^{(\mathcal{S},l)}_{ij}
\).
Assuming the sign of the associations remains consistent across sample groups, the mean equals zero if and only if the null hypothesis holds, so
$\hat{T}_{ij}$ should be significantly different from zero under the alternative hypothesis. Once the asymptotic variance is consistently estimated, constructing a confidence interval for $T_{ij}$ and the corresponding $p$-value for $H_{0,ij}$ is straightforward. % However, we shift our focus to the more challenging multiple edge test, treating the single edge test as a trivial special case.

\begin{remark}
% \commhb{We could put the remark on the sign information here.}
More generally, additional sign information on $\rho^{(\mathcal{S},l)}_{ij}$ may be available. With this additional knowledge, we present a test statistic based on a linear
combination of those $\hat{\rho}^{(\mathcal{S},l)}_{ij}$ for %$l = 1, \dots, m$,
$l\in [m]$, which is closely related to its $\ell_1$ norm. More specifically, with an edge-specific sign vector $\sigma_{ij}:=(\sigma^{(1)}_{ij},\dots,\sigma^{(m)}_{ij})^\top \in \{-1, 1\}^m$, we replace $\hat{T}_{ij}$ with the following sign-addressed test statistic $\hat{T}_{ij,\sigma} := \frac{1}{\sqrt{m}} \sum_{l=1}^{m} \sigma^{(l)}_{ij} \sqrt{n_l p} \hat{\rho}^{(\mathcal{S},l)}_{ij}$.
%\begin{equation} \label{eqn:inf_T_single_signed}
 %   \hat{T}_{ij,\sigma} := \frac{1}{\sqrt{m}} \sum_{l=1}^{m} \sigma^{(l)}_{ij} \sqrt{n_l p} \hat{\rho}^{(\mathcal{S},l)}_{ij}.
%\end{equation}
The normal approximation we establish in \cref{sec:theorem} also applies to the sign-addressed test statistic. % It is not the scope of our paper to provide the full technical details, but the same proof technique applies here. 
\end{remark}

\subsubsection{Simultaneous Test}

Now we shift our focus to a more challenging problem of multiple edge testing. %multiple edge test. %, treating the single edge test as a trivial special case.
In this scenario, we aim to test whether there are no edges at all in a user-specified edge set $E \subseteq [p] \times [p]$, regardless of its size, which corresponds to the following null hypothesis:
\begin{equation*} \label{eq:H0_multiple}
H_{0,E}: \rho^{(\mathcal{S},l)}_{ij} = 0, \quad \forall (i,j) \in E, \quad \forall l=1,\dots,m.
\end{equation*}
When $E$ consists of a single edge $(i,j)$, $H_{0,E}$ reduces to the single-edge null hypothesis $H_{0,ij}$ in \cref{eq:H0_single}. While multiple testing techniques, such as Bonferroni correction, can be applied to extend single-edge tests to multiple-edge tests, these methods are often overly conservative. Moreover, in neuroscience applications, the edge set $E$ typically represents connections between different brain areas, and its cardinality can grow %up to 
on the order of $q^2$. As a result, even if we establish the asymptotic normality of a single $\hat{T}_{ij}$, traditional multiple testing methods may not be valid in high-dimensional settings where $\abs{E}$ grows rapidly.

To address this challenge, we propose a simultaneous testing approach based on the supremum norm of $\hat{T}_E := ( \hat{T}_{ij} : (i,j) \in E )$, defined as:
\begin{equation} \label{eq:inf_T_multiple}
    \norm{\hat{T}_E}_\infty := \max_{(i,j) \in E} \abs{\hat{T}_{ij}}.
\end{equation}
The key idea leverages on high-dimensional central limit theorems (e.g., \citealp{Chernozhukov2012}): Although the full vector $\hat{T}_E$ may not be asymptotically jointly normal as $\abs{E}$ increases, the supremum norm $\norm{\hat{T}_E - T_E}_{\infty}$ exhibits the same limiting behavior as $\norm{Z}_\infty$, where $Z$ is a centered normal random vector with the same covariance as the asymptotic covariance of $\hat{T}_E$. Specifically, this covariance is given by the matrix of asymptotic covariances between $\hat{T}_{i_1 j_1}$ and $\hat{T}_{i_2 j_2}$ for $(i_1, j_1), (i_2, j_2) \in E$, expressed as (see \cref{app:asymp_cov} for the derivation):
\begin{equation} \label{eq:inf_S}
\begin{split}
    & S_{(i_1,j_1),(i_2,j_2)} := \\
    & \sum_{l=1}^m \frac{\norm{\Sigma^{(\mathcal{T},l)}}_F^2}{mp} 
    {\scriptstyle \left[ \begin{split} 
    & \rho^{(\mathcal{S},l)}_{i_1 i_2} \rho^{(\mathcal{S},l)}_{j_1 j_2}
    + \rho^{(\mathcal{S},l)}_{i_1 j_2} \rho^{(\mathcal{S},l)}_{i_2 j_1}
    + \frac{1}{2} \rho^{(\mathcal{S},l)}_{i_1 j_1} \rho^{(\mathcal{S},l)}_{i_2 j_2} 
    \Big( \rho^{(\mathcal{S},l)2}_{i_1 i_2} + \rho^{(\mathcal{S},l)2}_{j_1 j_2}
    + \rho^{(\mathcal{S},l)2}_{i_1 j_2} + \rho^{(\mathcal{S},l)2}_{i_2 j_1} \Big) \\
    & - \rho^{(\mathcal{S},l)}_{i_1 i_2} 
    \rho^{(\mathcal{S},l)}_{i_2 j_2} \rho^{(\mathcal{S},l)}_{i_2 j_1}
    - \rho^{(\mathcal{S},l)}_{i_1 i_2}
    \rho^{(\mathcal{S},l)}_{i_1 j_1} \rho^{(\mathcal{S},l)}_{i_1 j_2}
    - \rho^{(\mathcal{S},l)}_{j_1 j_2} 
    \rho^{(\mathcal{S},l)}_{i_2 j_2} \rho^{(\mathcal{S},l)}_{i_1 j_2}
    - \rho^{(\mathcal{S},l)}_{j_1 j_2} \rho^{(\mathcal{S},l)}_{i_2 j_1} \rho^{(\mathcal{S},l)}_{i_1 j_1}
    \end{split} \right]}.
\end{split}
\end{equation}

We approximate the distribution of $\norm{Z}_\infty$ using a parametric bootstrap based on the plug-in estimator $\hat{S}_{EE}$ of the asymptotic covariance. Generating bootstrap samples $\{\hat{Z}^{(1)}, \dots, \hat{Z}^{(B)}\}$ from $\distNorm(0, \hat{S}_{EE})$, we construct a $(1-\alpha)$-confidence region for $T_E$:
\begin{equation} \label{eq:inf_CI}
    \mathcal{C}_{E}(1-\alpha):= \{t \in \reals^{\abs{E}}: \norm{t - \hat{T}_E}_\infty \leq \hat{q}_{1-\alpha}\},
\end{equation}
% \begin{equation} \label{eq:inf_CI}
%     \mathcal{C}_{E}(1-\alpha):= \left\{ T_E: \|\hat{T}_E - T_E\|_{\infty} = \max_{(i,j) \in E} \abs*{\frac{1}{\sqrt{m}} \sum_{l=1}^m \sqrt{n_l p} (\hat{\rho}^{(\mathcal{S},l)}_{ij} - \rho^{(\mathcal{S},l)}_{ij})} \leq \hat{q}_{\norm{\hat{Z}}_\infty,1-\alpha} \right\},
% \end{equation}
where $\hat{q}_{1-\alpha}$ is the empirical $(1-\alpha)$-quantiles of % $\norm{\hat{T}+\hat{Z}}_\infty$, estimated by 
$\{\norm{\hat{Z}^{(1)}}_\infty, \dots, \norm{\hat{Z}^{(B)}}_\infty\}$. The null hypothesis $H_{0,E}$ is rejected if $\mathbf{0} \notin \mathcal{C}_{E}(1-\alpha)$. The coverage of this confidence region is studied in \cref{thm:inf_bootstrap}, with a power analysis provided in \cref{thm:inf_power}. We further discuss how borrowing information across sample groups enhances testing power in \cref{remark:sample_size_requirement}.

\subsection{Estimation of Temporal Covariance Matrix}
\label{sec:temporal_est}

The plug-in estimator $\hat{S}_{EE}$ of the asymptotic covariance (\cref{eq:inf_S}) requires the Frobenius norm of the temporal covariance matrix, $\Sigma^{(\mathcal{T},l)}$ for each sample group. We propose to estimate $\Sigma^{(\mathcal{T},l)}$ based on a modified Cholesky decomposition of its inverse, the temporal precision matrix $\Omega^{(\mathcal{T},l)}$ \citep{bickel2008,liu2017}. The Cholesky decomposition characterizes the autoregressive dependence of the signal at each time point on its past values. Specifically, suppose for each sample group $l\in [m]$ that the modified Cholesky decomposition of $\Omega^{(\mathcal{T},l)}$ is 
\(
    \Omega^{(\mathcal{T},l)} = L^{(\mathcal{T},l)} D^{(\mathcal{T},l)} L^{(\mathcal{T},l)\top},
\)
where $D^{(\mathcal{T},l)}$ is a diagonal matrix, and $L^{(\mathcal{T},l)}$ is a lower triangular matrix with diagonal entries equal to $1$. Let $\beta^{(\mathcal{T},l)} := I - L^{(\mathcal{T},l)\top}$.
%Let $\beta^{(\mathcal{T},l)} := I - L^{(\mathcal{T},l)\top}$ and $\Phi^{(\mathcal{T},l)} := \frac{\tr(\Sigma^{(\mathcal{S},l)})}{q} (D^{(\mathcal{T},l)})^{-1}$, so that the Cholesky decomposition of $\Omega^{(\mathcal{T},l)}$ can be rewritten by
%\begin{equation} \label{eq:temporal_cholesky}
%    \Omega^{(\mathcal{T},l)} = \frac{\tr(\Sigma^{(\mathcal{S},l)})}{q} (I - \beta^{(\mathcal{T},l)})^\top (\Phi^{(\mathcal{T},l)})^{-1} (I - \beta^{(\mathcal{T},l)}).
%\end{equation}
% $\beta^{(\mathcal{T},l)}$ and $\Phi^{(\mathcal{T},l)}$ associates with the auto-correlation within $X^{(k,l)}_{\cdot,i}$ across $t$ in sense that   
Then the first $(t-1)$ entries of $\beta^{(\mathcal{T},l)}$'s $t$-th row, $\beta^{(\mathcal{T},l)}_{\cdot,t}$, correspond to the regression coefficients of $X^{(k,l)}_{ti}$ on its past values, $X^{(k,l)}_{1i}, X^{(k,l)}_{2i}, \ldots, X^{(k,l)}_{(t-1)i}$ at each spatial location $i\in[q]$. That is, $X^{(k,l)}_{ti} = X^{(k,l)\top}_{\cdot,i} \beta^{(\mathcal{T},l)}_{\cdot,t} + \epsilon^{(\mathcal{T},k,l)}_{ti}$ for each time $t$. Due to the Kronecker product structure, the regression coefficients are invariant across spatial points $i\in [q]$, whereas the residuals exhibit spatial dependence characterized by $\Sigma^{(\mathcal{S},l)}$. To accommodate the matrix-variate data and the identifiability constraint $\tr(\Sigma^{(\mathcal{T},l)}) = p$, we rewrite the Cholesky decomposition of $\Omega^{(\mathcal{T},l)}$ by 
\begin{equation} \label{eq:temporal_cholesky}
    \Omega^{(\mathcal{T},l)} = \frac{\tr(\Sigma^{(\mathcal{S},l)})}{q} (I - \beta^{(\mathcal{T},l)})^\top (\Phi^{(\mathcal{T},l)})^{-1} (I - \beta^{(\mathcal{T},l)}),
\end{equation}
where $\Phi^{(\mathcal{T},l)} := \frac{\tr(\Sigma^{(\mathcal{S},l)})}{q} (D^{(\mathcal{T},l)})^{-1}$. %$i=1,\ldots, q$
% In short, this autoregression model can be written as $X^{(k,l)}_{ti} = X^{(k,l)\top}_{\cdot,i} \beta^{(\mathcal{T},l)}_{\cdot,t} + \epsilon^{(\mathcal{T},k,l)}_{ti}$ for each time $t$.

% For our matrix-variate data, we treat each column $X^{(k,l)}_{\cdot, i}$ as a $p$-dimensional sample, which leaves us $q$ correlated vector-variate samples for an autoregression model, where the covariance among these ``column samples" is characterized by $\Sigma^{(\mathcal{S},l)}$. Collecting all $n_l$ samples, with the adjustment factor in $\Phi^{(\mathcal{T},l)}$ due to the identifiability constraint, we have $\Phi^{(\mathcal{T},l)}_{tt} = \frac{1}{n_l q} \sum_{k=1}^{n_l} \sum_{i=1}^q \Var\left[ \epsilon^{(\mathcal{T},k,l)}_{ti} \right]$. Therefore,  we have $n_l q$ correlated sample in total to estimate the temporal covariance for each session or graph.

In our working example, neural recordings are not guaranteed to be temporally aligned across sample groups due to response latencies to stimuli \citep{ventura2004}. Hence we do not assume that $\Sigma^{(\mathcal{T},l)}$ is the same or even similar across sample groups $l\in[m]$. Instead, we impose a weaker bandable assumption on the modified Cholesky decomposition for each sample group. In physiological time-series signals, the dependence between time points naturally decays as the time lag increases, so a reasonable assumption is that $\beta^{(\mathcal{T},l)}_{st}$ approaches zero as $t - s \rightarrow \infty$. By applying bandable assumptions, we ignore weak dependencies between distant time points, achieving a bias-variance trade-off in the estimation of $\beta^{(\mathcal{T},l)}$.

We estimate $\beta^{(\mathcal{T},l)}$ following the procedure in \citet{liu2017}. Specifically, we treat the observed data $X^{(k,l)}$ as $q$ vector-variate ``column samples'', $X_{\cdot, 1}^{(k,l)}, \dots, X_{\cdot, q}^{(k,l)}$ rather than as $p$ vector-variate ``row samples'' as in Section \ref{sec:spatial_est} for spatial covariance matrix estimation, and fit the following linear regression model at each time point $t$:
\begin{equation} \label{eq:temporal_est_beta}
\begin{aligned}
    \hat{\beta}^{(\mathcal{T},l)}_{\cdot,t}
    := \argmin_{b \in \reals^p}
     \frac{1}{2 n_l q} \sum_{k=1}^{n_l} \norm{X^{(k,l)}_{t,\cdot} - X^{(k,l)\top} b}_2^2
    % & \text{w.r.t} ~~ b_s = 0 ~~ \text{where} ~~ s < t-h_l ~~\text{or} ~~ s \geq t,
\end{aligned}
\end{equation}
with respect to $b_s = 0$ where $s < t-h_l$ or $s \geq t$, where the bandwidth hyperparameter $h_l$ can be either user-specified or data-driven. Unlike the spatial case, the estimation of the temporal covariance matrix is performed individually for each sample group.
Having the regression coefficients estimated, we estimate $\Phi^{(\mathcal{T},l)}$ by
\begin{equation} \label{eq:temporal_est_Phi}
    \hat{\Phi}^{(\mathcal{T},l)}_{tt} := \frac{1}{n_l q} \sum_{k=1}^{n_l} \norm*{X^{(k,l)}_{t,\cdot} - X^{(k,l)\top} \beta^{(\mathcal{T},l)}_{\cdot, t}}_2^2.
\end{equation}
For technical issues, we truncate the eigenvalues of $I-\beta^{(\mathcal{T},l)}$ to ensure its invertibility, following the approach in \citet{liu2017}. Given a cut-off hyperparameter $\eta > 0$,  for a square matrix $A$, let $P_\eta(A) := U ~\max\{\min\{\Lambda,\eta\}, \eta^{-1}\} ~V^\top$, where $A$ has a singular value decomposition $A = U \Lambda V^\top$, and the $\min$ and $\max$ above are element-wise operations.
Let $\bar{\Omega}^{(\mathcal{T},l)}$ be the precursor estimator of $\Omega^{(\mathcal{T},l)}$ given by
\begin{equation} \label{eq:temporal_eig_cut}
    \bar{\Omega}^{(\mathcal{T},l)} := P_\eta(I - \hat\beta^{(\mathcal{T},l)})^\top (\hat\Phi^{(\mathcal{T},l)})^{-1} P_\eta(I - \hat\beta^{(\mathcal{T},l)}).
\end{equation}
Conforming to the identifiability constraint $\tr(\Sigma^{(\mathcal{T},l)}) = p$, we propose our estimators of $\Sigma^{(\mathcal{T},l)}$ and $\Omega^{(\mathcal{T},l)}$ as
\begin{equation} \label{eq:temporal_est_Sigma_Omega}
    \hat\Sigma^{(\mathcal{T},l)} := 
    \frac{p}{\tr(\bar\Sigma^{(\mathcal{T},l)})} \bar{\Sigma}^{(\mathcal{T},l)}
    \textand
    \hat\Omega^{(\mathcal{T},l)} := (\hat\Sigma^{(\mathcal{T},l)})^{-1} =\frac{\tr(\bar\Sigma^{(\mathcal{T},l)})}{p} \bar\Omega^{(\mathcal{T},l)},
\end{equation}
where $\bar\Sigma^{(\mathcal{T},l)} := (\bar\Omega^{(\mathcal{T},l)})^{-1}$. The estimation error bounds of the Frobenius norm of $\Sigma^{(\mathcal{T},l)}$ are provided in \cref{app:temp_cov}.

\section{Theoretical Properties} \label{sec:theorem}

We make the following assumptions on the multiple MGGM and the observed dataset $\mathcal{D}$.

\begin{assumption} \label{assmp:balanced_sample}
    $\max_{l = 1, \dots, m} \frac{n_l}{n_0} \leq \kappa_1$ for some positive constant $\kappa_1$ where $n_0 = \min_{l=1, \dots, m} n_l$.
\end{assumption}

\begin{assumption} \label{assmp:temporal_trace}
    $\tr(\Sigma^{(\mathcal{T},l)}) = p, ~\forall l = 1,\dots,m$. 
\end{assumption}

\begin{assumption} \label{assmp:eigenvalues}
    For each $l\in[m]$, %$l=1,\dots,m$, 
    let $\{\lambda^{(\mathcal{T},l)}_i\}_{i=1,\dots,p}$ be the eigenvalues of $\Sigma^{(\mathcal{T},l)}$ while $\frac{1}{\kappa_3} \leq \lambda^{(\mathcal{T},l)}_1 \leq \lambda^{(\mathcal{T},l)}_2 \leq \dots \leq \lambda^{(\mathcal{T},l)}_p \leq \kappa_3$ for some constant $\kappa_3 > 0$; define and assume $\{\lambda^{(\mathcal{S},l)}_i\}_{i=1,\dots,q}$ similarly for $\Sigma^{(\mathcal{S},l)}$.
\end{assumption}

\begin{assumption} \label{assmp:spatial_sample}
    Let $d := \max_i \abs*{\left\{j \in [q] \backslash \{i\}: \Omega^{(\mathcal{S},l)}_{ij} \neq 0 \text{ for some } l \in [m] \right\}}$ be the group-wise maximum node degree. 
    % , i.e., 
    % \begin{equation*}
    %     d := \max_i \abs*{\left\{j \in [q] \backslash \{i\}: \Omega^{(\mathcal{S},l)}_{ij} \neq 0 \text{ for some } l \in [m] \right\}}.
    % \end{equation*}
    We assume group sparsity of the partial-correlation graph in spatial association by $d \cdot \frac{\max\{m, \log(mn_0pq)\}}{(m n_0 p)^{1/2}} \rightarrow 0 $
    % \begin{equation*}
    %     d \cdot \frac{\max\{m, \log(mn_0pq)\}}{(m n_0 p)^{1/2}} \rightarrow 0 
    % \end{equation*}
    as $n_0 \rightarrow \infty$.
\end{assumption}

\begin{assumption} \label{assmp:temporal_sample}
We assume the temporal precision matrix $\Omega^{(\mathcal{T},l)}$ for each $l\in[m]$, %$l=1,\dots,m$
has Cholesky decomposition as in \cref{eq:temporal_cholesky} where $\beta^{(\mathcal{T},l)}$ satisfies $\abs{\beta^{(\mathcal{T},l)}_{st}} < \kappa_5 (t-s)^{-\alpha_l - 1}$ for any $t$ and $s$ such that $s < t$ and some $\alpha_l > 0$.
We further assume that, for $\alpha_0 = \min_{l=1,\dots,m} \alpha_l$, $\frac{\log (mn_0pq)}{(n_0q)^{1-1/(\alpha_0+1)}} \rightarrow 0$
as $n_0 \rightarrow \infty$.
\end{assumption}

\cref{assmp:balanced_sample} assumes that the sample sizes across sample groups are balanced, with $n_0$ representing this common level. \cref{assmp:temporal_trace} ensures identifiability. \cref{assmp:eigenvalues} is a standard assumption on eigenvalues commonly used in covariance estimation and graphical models \citep{cai2016survey}. \cref{assmp:spatial_sample} assumes that the spatial (column) precision matrices are sparse, and it imposes a constraint on the spatial dimension relative to the number of samples, the temporal dimension, and the number of graphs. In the special case of a single vector-variate graphical model with $m=1, p=1$, this assumption coincides with the necessary condition for root-consistent estimation of individual edges \citep{ren2015}. Moreover, a larger temporal dimension $p$ can further relax the required sample size. The first part of \cref{assmp:temporal_sample} is reasonable for neural time series, as these data
are often modeled as a low-order autoregressive process, a common assumption in the literature \citep{bickel2008, liu2017}. The second part, similar to \cref{assmp:spatial_sample}, restricts the temporal dimension. Likewise, increasing the spatial dimension $q$ can further relax the sample size requirement.

In the following, $C(\dots)$ indicates a constant that depends on the other constants within the parentheses, with values that may change across lines. For universal constants without any dependency, we omit the parentheses and denote them simply by $C$.

\subsection{Non-asymptotic error bound for the group Lasso estimate} \label{sec:group_Lasso}

We first provide a theoretical justification for our group Lasso procedure proposed in \cref{sec:spatial_est}. Although with correlated rows, our results below demonstrate that the optimal convergence rates for the estimation error, defined as $\Delta^{(\mathcal{S},l)}_{\cdot,i} := \hat{\beta}^{(\mathcal{S},l)}_{\cdot,i} - {\beta}^{(\mathcal{S},l)}_{\cdot,i}$, and prediction can be still obtained compared to the case with i.i.d. samples. The proof is provided in \cref{app:pf_spatial_regression}.

\begin{theorem} \label{thm:spatial_regression}
Suppose that $\gamma_i$ satisfies $\frac{1}{C(\kappa_1,\kappa_3)}\sqrt{\frac{m+\log (mn_0pq)}{n_0 p}} \leq \gamma_i \leq C(\kappa_1,\kappa_3) \sqrt{\frac{m+\log (mn_0pq)}{n_0 p}}$ for some sufficiently large $C(\kappa_1,\kappa_3)$. 
Then, under \crefrange{assmp:balanced_sample}{assmp:temporal_sample},
\begin{equation*}
    \Pr\left[\begin{aligned}
    & \max_i \sum_{j:j \neq i} \norm*{\underline\Delta^{(\mathcal{S},\cdot)}_{ji}}_2^2 
    \leq C(\kappa_1, \kappa_3) ~d ~\frac{m + \log(mn_0pq)}{n_0 p}, \\
    &\max_i \sum_{j:j \neq i} \norm*{\underline\Delta^{(\mathcal{S},\cdot)}_{ji}}_2 
    \leq C(\kappa_1,\kappa_3) ~d ~\sqrt{\frac{m + \log(mn_0pq)}{n_0 p}}, \\
    & \max_i \frac{1}{2 n_0 p} \sum_{l=1}^m \sum_{k=1}^{n_l} \norm*{X^{(k,l)}\Delta^{(\mathcal{S},l)}_{\cdot,i}}_2^2
    \leq C(\kappa_1,\kappa_3) ~d ~\frac{m + \log(mn_0pq)}{n_0 p}
    \end{aligned}\right] \geq 1 - C(mn_0pq)^{-1/2},
\end{equation*}
for a sufficiently large $n_0$, where $\underline{\Delta}^{(\mathcal{S},l)}_j := \frac{\norm{X^{(\mathcal{S},l)}_{\cdot,j}}_2}{\sqrt{n_l p}} \Delta^{(\mathcal{S}, l)}_j$.
% \commhb{$C_1$, $C_2$ for different parameters?}
\end{theorem}

\subsection{Theoretical justification for the simultaneous edge testing} \label{sec:thm_inference}

We now present the theoretical results for the simultaneous test in \cref{sec:inference}. In the proof of \cref{thm:inf_CLT}, we show that the error in the partial correlation estimate $\hat{\rho}^{(\mathcal{S},l)}_{ij}$ from \cref{eq:spatial_est_omega_rho} is driven by a leading term
\begin{equation} \label{eq:spatial_Theta}
    \Theta^{(\mathcal{S},l)}_{ij} := \frac{\tilde{\phi}^{(\mathcal{S},l)}_{ij}}{\sqrt{\Phi^{(\mathcal{S},l)}_{ii} \Phi^{(\mathcal{S},l)}_{jj}}} 
    - \frac{\Phi^{(\mathcal{S},l)}_{ij} \tilde{\phi}^{(\mathcal{S},l)}_{jj}}{2 \Phi^{(\mathcal{S},l)}_{jj} \sqrt{\Phi^{(\mathcal{S},l)}_{ii} \Phi^{(\mathcal{S},l)}_{jj}}}
    - \frac{\Phi^{(\mathcal{S},l)}_{ij} \tilde{\phi}^{(\mathcal{S},l)}_{ii}}{2 \Phi^{(\mathcal{S},l)}_{ii} \sqrt{\Phi^{(\mathcal{S},l)}_{ii} \Phi^{(\mathcal{S},l)}_{jj}}},
\end{equation}
where $\tilde{\phi}_{ij}^{(\mathcal{S},l)} := \tilde{\Phi}^{(\mathcal{S},l)}_{ij} - \Phi^{(\mathcal{S},l)}_{ij} = \frac{1}{n_l p} \sum_{k=1}^{n_l} \epsilon^{(\mathcal{S},k,l)\top}_{\cdot,i} \epsilon^{(\mathcal{S},k,l)}_{\cdot,j} - \Phi^{(\mathcal{S},l)}_{ij}$.
Since %the leading error term $\Theta_{ij}^{(\mathcal{S},l)}$ is a linear functional of 
$\tilde{\phi}^{(\mathcal{S},l)}_{ij}$ is an average over $t\in[p]$ and $k\in[n_l]$, the central limit theorem ensures that the partial correlation estimate $\hat{\rho}^{(\mathcal{S},l)}_{ij}$ and the single-edge test statistic $\hat{T}_{ij}$ converge in distribution to Gaussian limits for each edge $(i,j)$.

The Gaussian approximation error for the multiple-edge test statistic $\norm{\hat{T}_E}_\infty$ is governed by Berry--Esseen bounds over hyper-rectangles. The seminal work of \cite{Chernozhukov2012} established these bounds in high-dimensional settings, where the dimension can exceed the sample size. Building on this, we use the near-optimal convergence rate for the Berry--Esseen bound from \citet{chernozhukov2020nearly} to derive a sharp Gaussian approximation error bound for $\norm{\hat{T}_E}_\infty$. The proof is provided in \cref{app:pf_inf_CLT}.

\begin{proposition} \label{thm:inf_CLT}
    Let $Z \sim \distNorm(0, S_{EE})$ where the elements of $S_{EE}$ are given as in \cref{eq:inf_S}, and $\hat{T}_E$ is estimated based on $\gamma_i$'s given as in \cref{thm:spatial_regression}. Then under \crefrange{assmp:balanced_sample}{assmp:temporal_sample},
    \begin{equation*}
    \begin{aligned}
        & \sup_{x > 0} \abs*{\Pr[\norm{\hat{T}_E }_\infty > x] - \Pr[\norm{T_E + Z}_\infty > x]} \\ 
        & \leq \frac{C(\kappa_1, \kappa_3)}{\sqrt{mn_0p}} \max\left\{
            (\log \abs{E})^{2} \log(m n_0 p), 
            ~ (\log \abs{E})^{5/2}, 
            ~ d\sqrt{\log\abs{E}} (m+\log(mn_0pq))
        \right\},
    \end{aligned}
    \end{equation*}
    for a sufficiently large $n_0$.
\end{proposition}

% \begin{remark}
% Proposition~\ref{thm:inf_CLT} indicates that the Kolmogorov distance between the distributions of $\norm{\hat{T}_E}_{\infty}$ and $\norm{Z}_{\infty}$ converges to zero with rate $O\left((n_0 p d)^{-c}\right)$ for $c>0$ as shown in the proof. However, while increasing $d$ shrinks the distance, without proper adjustment of $n_0p$, naively changing $d$ will break \cref{as:sparse} and fail sample size requirement as stated in Remark~\ref{remark:sample_size_requirement}. %In practice, to achieve better results via increasing number of graphs, keeping $d$ proportional to $n_0 p$ is suggested.
% \end{remark}

The following theorem mirrors the previous proposition, with the key difference being that we replace the population covariance $S_{EE}$ with its plug-in estimator $\hat{S}_{EE}$. In essence, as long as $S_{EE}$ is well-estimated under the $\norm{\cdot}_\infty$-norm, the Gaussian approximation results remain valid. Using the convergence rate of $\hat{S}_{EE}$ in the $\norm{\cdot}_\infty$-norm, we derive the following bootstrap theorem, based on the relationship between convergence rate and bootstrap error as given in Lemma 2.1 of \citet{chernozhukov2020nearly}.

\begin{theorem} \label{thm:inf_bootstrap}
    Let $\hat{Z} \sim \distNorm(0, \hat{S}_{EE})$ where the plug-in estimator $\hat{S}_{EE}$ and $\hat{T}_E$ are estimated based on $\gamma_i$'s given in \cref{thm:spatial_regression}. Then, under \crefrange{assmp:balanced_sample}{assmp:temporal_sample},
    \begin{equation*}
    \begin{aligned}
        & \sup_{x > 0} \abs*{\Pr[\norm{\hat{T}_E}_\infty > x] - \Pr[\norm{T_E + \hat{Z}}_\infty > x|\mathcal{D}]} \\
        & \leq C(\kappa_1, \kappa_3, \kappa_5) \max\left\{
        \begin{aligned}
            & \frac{(\log \abs{E})^{2} \log(m n_0 p)}{\sqrt{mn_0p}}, 
            ~ \frac{(\log \abs{E})^{5/2}}{\sqrt{mn_0p}},
            ~ d\sqrt{\log\abs{E}}\frac{m+\log(mn_0pq)}{\sqrt{m n_0 p}}, \\
            & \log\abs{E} \log(n_0pq)
            \sqrt{\frac{\log(mn_0pq)}{(n_0q)^{1-\frac{1}{2(\alpha_0+1)}}}
            + \frac{m+\log(mn_0pq)}{mn_0p}}
        \end{aligned}
        \right\}
    \end{aligned}
    \end{equation*}
    with probability at least $1 - C(mn_0pq)^{-1/2}$ for a sufficiently large $n_0$.
\end{theorem}

The above theorem establishes the theoretical foundation for the simultaneous multiple edge testing procedure in \cref{alg:procedure}. 
Next, we formally state the validity of our testing procedure as well as a power analysis.
% under a general testing $H_{0,\mathcal{S}}: \mat{\rho}^0_{ij}=\mat{c}^0_{ij}, \text{ }\forall (i,j) \in \mathcal{S}$.
The proof is given in \cref{app:pf_inf_power}

\begin{theorem}\label{thm:inf_power}
    Suppose that $n_0$ increases at a faster rate than
    \begin{equation*}
        \frac{1}{mp} \max\left\{\begin{aligned}
            & (\log\abs{E})^4 (\log(mn_0p))^2,
            ~ (\log\abs{E})^5, 
            ~ d^2\log\abs{E}(m+\log(mn_0pq))^2,\\
            & (\log\abs{E}\log(n_0p))^2 (m + \log(mn_0pq))
        \end{aligned}\right\}
    \end{equation*}
    and $\frac{1}{q} ((\log\abs{E} \log(n_0q))^2 \log(mn_0pq))^{1+1/(2\alpha_0+1)}$.
    Under the null $H_{0,E}$ and \crefrange{assmp:balanced_sample}{assmp:temporal_sample}, the confidence region $\mathcal{C}_{E}(1-\alpha)$ estimated based on $\gamma_i$'s given in \cref{thm:spatial_regression} satisfies
    \(
        \mathbb{P}[\mathbf{0} \notin \mathcal{C}_{E}(1-\alpha)] \overset{p}{\rightarrow} \alpha.
    \)
    On the other hand, as an alternative, if 
    \begin{equation*}
        % \max_{(i,j)\in E} \abs*{\sum_{l=1}^m \sqrt{\frac{n_l p}{m}} \rho^{(\mathcal{S},l)}_{ij}}
        \norm{T_E}_\infty
        \geq C(\kappa_1,\kappa_3,\kappa_5) \sqrt{(\log q + \log(1/\alpha)) \max_{(i,j)\in E} S_{(i,j),(i,j)}},
    \end{equation*} 
    then under the same assumptions and $\gamma_i$'s, we have $\mathbb{P}[\mathbf{0} \notin \mathcal{C}_E(1-\alpha)] \overset{p}{\rightarrow} 1$.
\end{theorem}

\begin{remark} \label{remark:sample_size_requirement}
\Cref{thm:inf_power} implies that the test power converges to $1$ as $n_0 \to \infty$, provided that 
$\sum_{l=1}^m \rho^{(\mathcal{S},l)}_{ij}$ exceeds the order of $\sqrt{m \log q / (n_0 p)}$ for some $(i,j) \in E$. 
When the partial correlations $\rho^{(\mathcal{S},l)}_{ij}$ have the same sign and comparable magnitude across sample groups, this condition simplifies to 
$\max_{(i,j)\in E} |\rho^{(\mathcal{S},l)}_{ij}|$ being larger than $\sqrt{\log q / (m n_0 p)}$. 
In contrast, without aggregating multiple graphs, the corresponding detection boundary is $\sqrt{\log q / (n_0 p)}$. 
Hence, by pooling information across $m$ sample groups, the detectable signal strength is reduced by a factor of $\sqrt{m}$, thereby improving detection power. 

Moreover, when the temporal covariance matrices satisfy $\Sigma^{(\mathcal{T},l)} = I$ for all $l \in [m]$, the model simplifies to the multiple vector-variate Gaussian graphical model studied in \citet{ren2019}, with an effective sample size of $n_0 p$. 
According to their Theorem~2.3, the optimal separating rate for detecting a single edge, measured in terms of the sum of partial correlations, is $\sqrt{m / (n_0 p)}$. 
This comparison indicates that our test achieves nearly optimal performance in terms of the testable region boundary. We further conjecture that the extra $\sqrt{\log q}$ factor is intrinsic when the edge set size $|E|$ grows at a polynomial rate in $q$.
\end{remark}

% The validity of our c-level test in Equation (\ref{eqn:test2}) is summarized below before we move to the theoretical properties for Section \ref{sec:temporal_est}.
% \begin{corollary}\label{col:test_power_c_level}
% Under the null $H'_{0, \mathcal{S}}$, the test is an $\alpha$ level test, i.e., $\mathbb{P}_{H'_0}(\Psi_\alpha) \leq \alpha$. On the other hand, in the alternative case,
% if $\max_{(i,j)\in \mathcal{S}} \frac{\sum_{t=1}^d \sqrt{n_tp } (
% |{\rho}_{tij}| - c)}{\sqrt{d}} \geq C \sqrt{\log q} \max_{1\leq j\leq r} (w^{P}_{jj})^{1/2}$ and C is a large enough constant, we have $\mathbb{P}_{H'_1}(\Psi_\alpha) \rightarrow 1$.

% \end{corollary}

\section{Numerical Studies} \label{sec:numerical_studies}

In \cref{sec:sim}, we present simulation studies to validate our theoretical results of \cref{sec:theorem} and evaluate the performance of our inference procedure for multiple matrix-variate graphical models in comparison to several baseline approaches. In \cref{sec:experiment}, we apply our method to neural recordings from two Utah arrays implanted in monkey brains during a spatial working memory task. The implementation of our proposed method is available in the \texttt{mmge} R package, along with vignettes for reproducing our results, accessible at \url{github.com/HeejongBong/mmge}.

\subsection{Simulation Studies} \label{sec:sim}

\begin{figure}[t]
    \centering
    \includegraphics[width=0.7\textwidth]{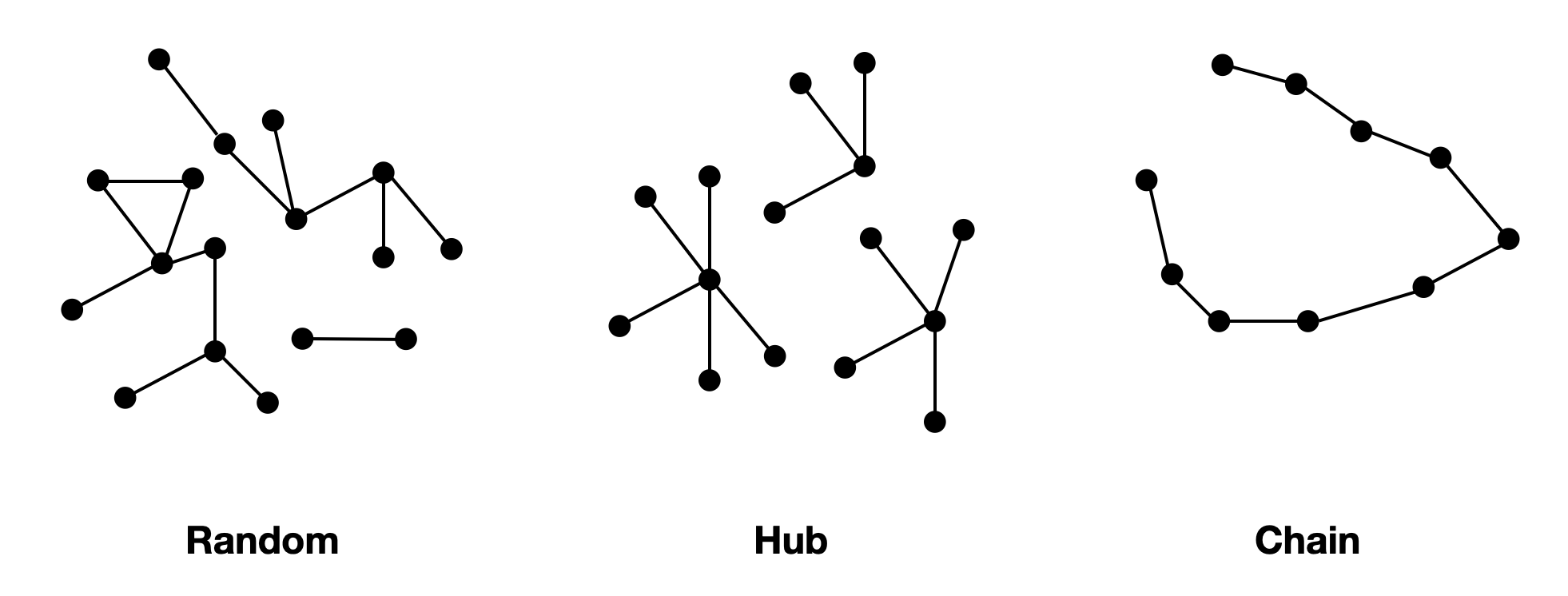}
    \caption{Simulated spatial graphs. \label{fig:graph_demo}}
\end{figure}

We evaluated the performance of our method under three spatial graph structures, illustrated in Figure~\ref{fig:graph_demo}: (1) a random graph, where edges between nodes were generated with probability $\sqrt{\frac{3}{q}}$ for each pair $(i, j)$; (2) a hub graph, where the nodes were divided into $\ceil{\frac{q}{20}}$ hub groups; and (3) a chain graph. 
Given each spatial graph structure, we generated the spatial precision matrix by sampling the non-zero entries from $\text{Unif}(0, \frac{0.3}{2^{l-1}})$, independently across sample groups $l \in[m]$. For the temporal precision matrix, we generated $\Sigma^{(\mathcal{T},l)}$ using \cref{eq:temporal_cholesky}. Following \cref{assmp:temporal_sample}, we set $\beta^{(\mathcal{T},l)}_{st} = \kappa_5 (t-s)^{-\alpha_l - 1}$ for $1 \leq s < t \leq p$, with $\kappa_5 = 0.2$ and $\alpha_l = 1$. For $\Phi^{(\mathcal{T},l)}$, we used the $p \times p$ identity matrix for all $l \in[m]$.

In \cref{sec:sim_est}, we compare our method's performance with existing methods in terms of edge detection and precision matrix estimation. In \cref{sec:sim_simultaneous}, we assess the coverage of the proposed bootstrap confidence region for $T_E$.

\subsubsection{Edge-wise Estimation Comparison}\label{sec:sim_est}

% Since simultaneous testing in multiple matrix graphs is largely missing in literature, we demonstrate the effectiveness of our method (M0) via comparison with multiple matrix-variate graph estimation method by \cite{zhu2018} (M1) and several multi-graph estimation methods for ordinary Gaussian Graphical Model (M2-M4) , as is shown in Section~\ref{sec:sim_est}. 

We compared our method (M0) with the following Gaussian graphical model estimation procedures:
\begin{itemize}
    \item (M1): matrix-variate Gaussian multi-graph estimation method by \cite{zhu2018}
    \item (M2): regression-based Gaussian graph estimation method by \cite{ren2019}
    \item (M3 \& M4): optimization based Gaussian graph estimation methods by \cite{cai2016} and \cite{lee2015}, respectively
\end{itemize}

Methods (M0) and (M1) are based on the matrix-variate Gaussian graphical model, while the others are designed for the vector-variate model. Since the vector-variate model does not account for temporal correlation, methods (M2), (M3), and (M4) are at a disadvantage when applied to spatiotemporal data. 
% (1) the most recent matrix-variate Gaussian multi-graph estimation method proposed by \cite{zhu2018} (M1), (2) multi-graph estimation methods for ordinary Gaussian Graphical Model, such as regression based method by \cite{ren2019} (M2), and optimization based methods by \cite{cai2016} (M3) and \cite{lee2015} (M4). For methods in the second category, 
To ensure a fair comparison, we used the \texttt{whiten} function from the \texttt{whitening} R package to whiten the simulated data before applying the vector-variate methods.

The edge detection performance of the methods was evaluated using receiver operating characteristic (ROC) curves. An ROC curve plots the true positive rate (TPR) against the false positive rate (FPR) as the detection threshold is varied by adjusting specific hyperparameters. For Methods (M0) and (M2), which detect non-zero spatial edges based on rejecting edge-wise null hypotheses (\cref{eq:H0_single}), we generated ROC curves by varying the $p$-value threshold while keeping other hyperparameters fixed. For the sparsity hyperparameter of Method (M0), we fixed $\gamma_i = 1\text{e}-4$ for all $i$. For the other methods, ROC curves were plotted by varying the sparsity hyperparameter, with other hyperparameters held constant.

% \begin{equation*}
% \begin{aligned}
%     TPR & = \frac{1}{m} \sum_{l=1}^m  
%     \frac{\sum_{1 \leq i < j \leq q} \mathbbm{1} (\Sigma^{(\mathcal{S},l)}_{ij}\neq 0, \hat{\Sigma}^{(\mathcal{S},l)}_{ij}\neq 0) }{\sum_{1\leq i < j \leq q} \mathbbm{1} (\Sigma^{(\mathcal{S},l)}_{ij}\neq 0)},  \\
%     FPR & = \frac{1}{d}\sum_{t=1}^d  \frac{\sum_{1 \leq i < j \leq q} \mathbbm{1} (\Sigma^{(\mathcal{S},l)}_{ij}= 0, \hat{\Sigma}^{(\mathcal{S},l)}_{ij}\neq 0) }{\sum_{1 \leq i < j \leq q} \mathbbm{1} (\Sigma^{(\mathcal{S},l)}_{ij}= 0)} . 
% \end{aligned}
% \end{equation*}
% For our method, under different regression tuning parameters, by varying test level $\alpha$, we get different ROC curves. 
%We discovered that our method is much better than baseline regardless of the tuning parameter, and one example is shown in Figure~\ref{fig:roc_li_multiple}.

\begin{figure}
    \centering
    \includegraphics[width=0.95\textwidth]{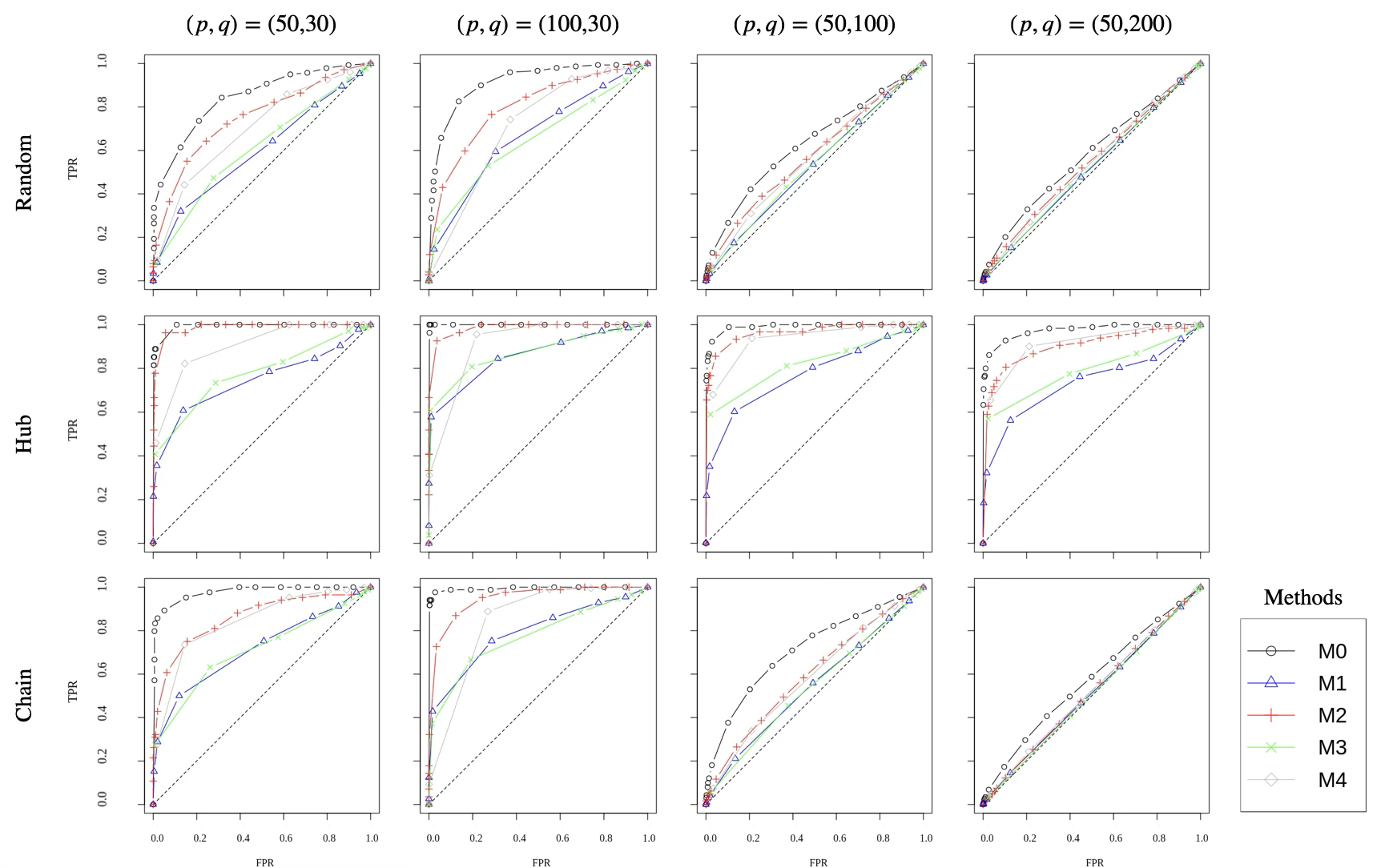}
    \caption{Simulation results are shown for different graph configurations and temporal dimensions with $n=5$ and $m=5$. Rows represent different graph types, while columns correspond to varying spatial and temporal dimensions. Curves represent our method (M0) and the other baseline methods. %Our method consistently outperforms the baselines, especially for large $p$.
    }\label{fig:simulation}
\end{figure}

The results with $m = 5$, $n=5$, and four different pairs of $(p, q)$ are shown in \cref{fig:simulation}. % The results with other values of $\gamma$'s are summarized in \cref{fig:roc_li_multiple}.
The ROC curves indicate that our method consistently outperformed the other methods, 
%Moreover, comparing methods designing for ordinary Gaussian graph, our method is much better when temporal dimension $p$ is large, thanks to the efficient use of spatial observations and temporal precision estimation based on Cholesky decomposition in \cref{sec:temporal_est}. 
% \cref{fig:sim_highd} presents the results in the settings where $m=5$, $n=5$, $p=50$, and $q \in \{100, 200\}$. 
% Our method consistently outperformed the other methods, 
thanks to the efficient use of spatial observations across sample groups. % and precise temporal precision matrix estimation based on Cholesky decomposition. 
In an additional simulation study with varying $\gamma_i$ values, our method demonstrated moderate sensitivity to the choice of the group Lasso hyperparameter, while it consistently outperformed the baseline methods across the range of tuning parameters considered, particularly in settings with high temporal and spatial dimensions.
%See \cref{fig:roc_sim_lambda} for the ROC curve results at different values of $\gamma$.

\subsubsection{Simultaneous Test}
\label{sec:sim_simultaneous}

In this section, we evaluate the coverage of the proposed bootstrap confidence interval (\cref{eq:inf_CI}) using simulated data. 
The simulations were conducted with $m$ sample groups of $n$ i.i.d. matrix-variate data, with temporal dimension $p = 50$ and spatial dimension $q = 30$. We generated 3000 bootstrap samples to construct confidence intervals for $\norm{T_E}_\infty$ over two edge sets: $E_\text{off} = \{(i,j): i \neq j\}$ and $E_\text{zero} = \{(i,j): \Omega^{(\mathcal{S},l)}_{ij} = 0, \ \forall l = 1, \dots, m\}$.

We repeated this procedure over 1000 datasets generated from the multiple matrix-variate Gaussian graphical models for each of the three spatial partial-correlation graphs (random, hub, and chain). For each graph, we computed the empirical coverage of the bootstrap confidence intervals. \cref{tab:simul_d5} presents the mean and standard deviation of the empirical coverage across the 1000 repetitions for each spatial graph and nominal coverage level.
%
% \begin{itemize}
%     \item For each type of graph (random, hub, chain), we generate one set of corresponding temporal precision matrix $\{ \mat{A}_t\}_{t=1}^d$ and spatial precision matrices $\{ \mat{B}_t\}_{t=1}^d$.
%     \item Given the precision matrices, for $i=1,\dots, 1000$, we generate one data realization, which we can apply our estimation procedure and calculate $\lVert \Delta \mat{P}_{{S_{off}}} \rVert_{\infty}$ and $\lVert \Delta \mat{P}_{{S_{zero}}}\rVert_{\infty}$. Since their true distributions are unknown, the empirical distributions based on 1000 realizations are denoted by $F_{off}(\cdot)$ and $F_{zero}(\cdot)$.
%     \item For each data realization $i$, we can apply our simultaneous testing procedure and estimate the $\mat{W}^P$ corresponding to $S_{off}$ and $S_{zero}$. By sampling $\hat{\mat{\zeta}} \sim N(0,\hat{\mat{W}}^P)$, we approximate $\lVert \Delta \mat{P}_{{S_{off}}} \rVert_{\infty}$ and $\lVert \Delta \mat{P}_{{S_{zero}}}\rVert_{\infty}$ based on the distribution of 3000 bootstrap samples at quantile $\alpha=0.925$, $\alpha=0.950$, $\alpha=0.975$, denoted by $\hat{q}_{{off}, \alpha, i}$ and $\hat{q}_{{zero}, \alpha, i}$.
%     \item Finally, we can calculate the mean and standard deviation for $\{F_{off}(\hat{q}_{{off}, \alpha, i})\}_{i=1}^{1000}$ and $\{F_{zero}(\hat{q}_{{zero}, \alpha, i})\}_{i=1}^{1000}$.
% \end{itemize}
% The results are shown in Table~\ref{tab:simul_d3} and Table~\ref{tab:simul_d5} for $d=3$ and $d=5$, respectively. 
We observe that the empirical coverages are close to the nominal values and converge further as the sample size $n$ increases. This result demonstrates the coverage of the bootstrap confidence region shown in \cref{thm:inf_bootstrap}. %Comparing \cref{tab:simul_d3} and \cref{tab:simul_d5}, we observe that our method performs better with a smaller number of sessions, aligning with our theoretical results.

\begin{table}[t]
\centering
\caption{Empirical coverage of the confidence region when $m=5, p=50,  q=30$.}\label{tab:simul_d5}
\scriptsize
\begin{tabular}{|p{1cm}|p{1.5cm}||p{1.5cm}|p{1.5cm}|p{1.5cm}|p{1.5cm}|p{1.5cm}|p{1.5cm}|}
\hline
\multirow{2}{*}{$n$} & \multirow{2}{*}{\begin{tabular}{l} Nominal \\ Coverage \end{tabular}} &\multicolumn{2}{c}{Random} & \multicolumn{2}{|c|}{Hub}& \multicolumn{2}{c|}{Chain}\\
\cline{3-8}
{} & {} & $E_\text{off}$ & $E_\text{zero}$ & $E_\text{off}$ & $E_\text{zero}$ & $E_\text{off}$ & $E_\text{zero}$ \\
\hline
\multirow{3}{*}{5}  & 0.925   & 0.906(0.009) & 0.900(0.009)   & 0.901(0.006) & 0.904(0.006)  & 0.886(0.008) & 0.888(0.007)\\
\cline{2-8}
{}                  & 0.95    & 0.935(0.006) & 0.938(0.006)   & 0.932(0.006) & 0.934(0.007)  & 0.922(0.007) & 0.923(0.008)\\
\cline{2-8}
{}                  & 0.975   & 0.971(0.004) & 0.962(0.005)   & 0.962(0.004) & 0.963(0.004)  & 0.959(0.005) & 0.959(0.005)\\
\hline
\multirow{3}{*}{10} & 0.925   & 0.908(0.006) & 0.913(0.005)   & 0.931(0.005) & 0.928(0.006)  & 0.931(0.006) & 0.928(0.007)\\
\cline{2-8}
{}                  & 0.95    & 0.934(0.004) & 0.937(0.004)   & 0.951(0.003) & 0.950(0.003)  & 0.953(0.004) & 0.953(0.005)\\
\cline{2-8}
{}                  & 0.975   & 0.961(0.004) & 0.961(0.003)   & 0.971(0.003) & 0.971(0.003)  & 0.974(0.003) & 0.975(0.003)\\
\hline
\multirow{3}{*}{20} & 0.925   & 0.917(0.007) & 0.920(0.004)   & 0.934(0.004) & 0.930(0.006)  & 0.931(0.006) & 0.928(0.007)\\
\cline{2-8}
{}                  & 0.95    & 0.947(0.006) & 0.948(0.005)   & 0.959(0.004) & 0.955(0.004)  & 0.953(0.004) & 0.953(0.005)\\
\cline{2-8}
{}                  & 0.975   & 0.978(0.002) & 0.975(0.004)   & 0.985(0.002) & 0.984(0.002)  & 0.974(0.003) & 0.975(0.002)\\
\hline
\end{tabular}
\end{table}

\subsection{Experimental Data Analysis}\label{sec:experiment}

\begin{figure}
    \centering
    \includegraphics[width=0.6\textwidth]{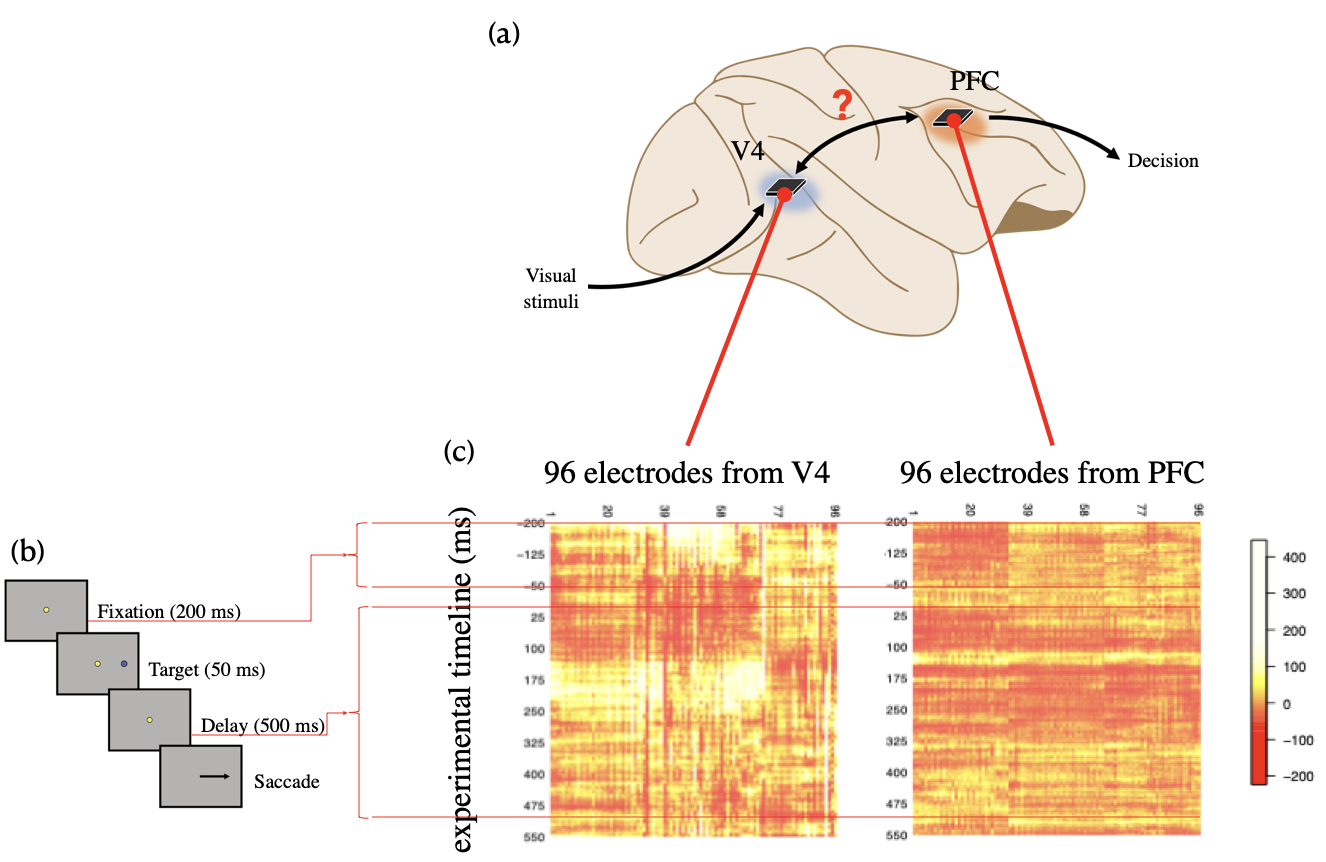}
     \caption{(a) The positions of the analyzed cortical areas, V4 and PFC in a primate brain. % Area PFC has been associated with top-down control of attention while area V4 is a mid-level visual area with robust visual responses.
     % (b) Utah array with $10 \times 10$ recording electrodes with $400\mu m$ interval \citep{utah_array_2019}. 
     The neuroelectrical activity in each area was recorded by a 96-electrode Utah array. 
     (b) The timeline of one experimental trial. % \citep{khanna2020dynamic}. % We are interested in the fixation, target presentation and delay stages of $750$ ms in total.
     (c) LFP recordings for one experimental trial. Each x-axis indicates $96$ electrodes in each brain area, and the y-axis is time in ms. Time $t=0$ was aligned at the start of the delay period.}\label{fig:experiment}
\end{figure}

We analyzed local field potential recordings (LFPs) from two Utah arrays implanted in the prefrontal cortex (PFC) and visual area V4 of an experimetnal subject (a Macaque monkey, \cref{fig:experiment}(a)). Each Utah array had $96$ electrodes, and the neuroelectrical activity in the two areas were simultaneously recorded during a memory-guided saccade task. One experimental trial of the task consisted of the following stages (see \cref{fig:experiment}(b)):
\begin{enumerate}
    \item The subject fixated at the center of the screen for $200$ ms.
    \item A circular target appeared at a randomly chosen location out of the forty possible spots on the screen (8 directions and 5 amplitudes) and turned off after $50$ ms.
    \item The subject had to remember the target location during a delay period of $500$ ms while maintaining fixation.
    \item After the delay period, the fixation point turned off, and the subject had to make a saccade to the remembered target location.
\end{enumerate}
See \citet{khanna2020dynamic} for the details of the experiment and data collection.
Previous studies in similar tasks have shown that area V4 encodes higher order visual features such as color and shape and is modulated by visual attention \citep{orban2008higher,fries2001modulation}, while the prefrontal cortex (PFC) plays a central role in cognitive control and working memory \citep{miller2001integrative}. Despite their anatomical separation and distinct functional roles, coordinated activity between V4 and PFC has been reported during visual memory retention \citep{sarnthein1998synchronization,liebe2012theta}. Building on these findings, our objective is to examine how the spatial correlation structure within and between these regions evolves across four experimental stages: fixation ($200$ ms), target presentation ($50$ ms), early delay (the first $250$ ms of the delay period), and late delay (the last $250$ ms of the delay period).

The dataset was collected across $m = 5$ experimental sessions indexed by $l$, containing $n_l = 2000, 2995, 3000, 3000,$ and $3000$ successful trials, respectively. For each trial, we observed matrix-variate data with temporal dimension $p = 750$, corresponding to $750$ ms recorded at a sampling rate of $1$ kHz, and spatial dimension $q = 192$, corresponding to $2 \times 96$ electrodes. An example trial is shown in \cref{fig:experiment}(c), where the heatmap color indicates the intensity of the local field potential signal. Although the strength of partial correlations may vary across experimental sessions due to session-specific variability, we expect the underlying sparsity pattern of the connectivity graph to be largely shared across sessions within each experimental stage. To exploit this structure, we applied the proposed method in \cref{sec:methd} separately to each stage, jointly fitting multiple matrix-variate graphs across sessions by treating trials within each session as a single sample group. This joint modeling strategy allows us to borrow information across sessions, thereby improving statistical power for detecting common connectivity patterns while accommodating heterogeneity in correlation magnitudes.

\subsubsection{Correlated neural connectivity vs. Physical distance}

\begin{figure}
    \centering
    \includegraphics[width=0.45\textwidth]{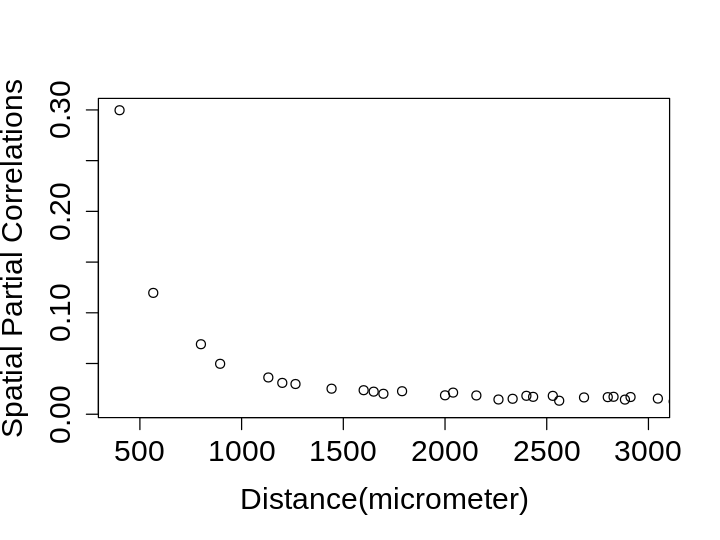}
    \caption{The averaged $\hat\rho^{(\mathcal{S})}_{ij}$ within a group of spatially equidistant within-area edges during the late delay stage in V4. % Notice that the spatial partial correlation declined as the spatial distance increased. This phenomenon is consistently identified over all experimental stages and both brain areas.
    }\label{fig:stats_dist}
\end{figure}

As an initial sanity check, we examined the relationship between neural connectivity and physical distance. Previous studies have shown that LFP signals from neighboring electrodes in monkey brains tend to exhibit strong correlations and that the correlation decays sharply as the distance between electrodes increases \citep{leopold2003spatial}.
To quantify this effect, we applied our method to the raw LFP signals from each brain region at each experimental stage. For each electrode pair $(i, j)$, we aggregated the spatial partial correlation estimates across sessions by calculating 
\(
\hat{\rho}_{ij}^{(\mathcal{S})} := \frac{1}{\sum_{l=1}^m \sqrt{n_l}} \sum_{l=1}^m \sqrt{n_l} ~\hat{\rho}^{(\mathcal{S},l)}_{ij},
\)
where the weight $\sqrt{n_l}$ accounts for the varying number of trials across sessions. We then grouped electrode pairs by spatial distance and calculated the average of $\hat{\rho}_{ij}^{(\mathcal{S})}$ within each distance group.
\cref{fig:stats_dist} shows the average spatial partial correlations within each equidistant group in area V4 during the late delay stage, plotted against physical distance. The results demonstrate a clear monotonic decrease in average $\hat{\rho}^{(\mathcal{S})}_{ij}$ with increasing distance, consistent with prior findings. We observed similar trends in other stages and in the PFC.

\subsubsection{Within-area Inference}

Next, we examined how within-region connectivity changes across the experimental stages. Using the same output from our method applied to each brain region at each stage, we quantified the within-region average connectivity of each electrode $i$ by computing $\sum_{j}\abs{\hat{\rho}_{ij}^{(\mathcal{S})}}$, where $\sum_j$ denotes summation over all other electrodes within the same region as $i$.

\begin{figure}
    \centering
    \includegraphics[width=0.85\textwidth]{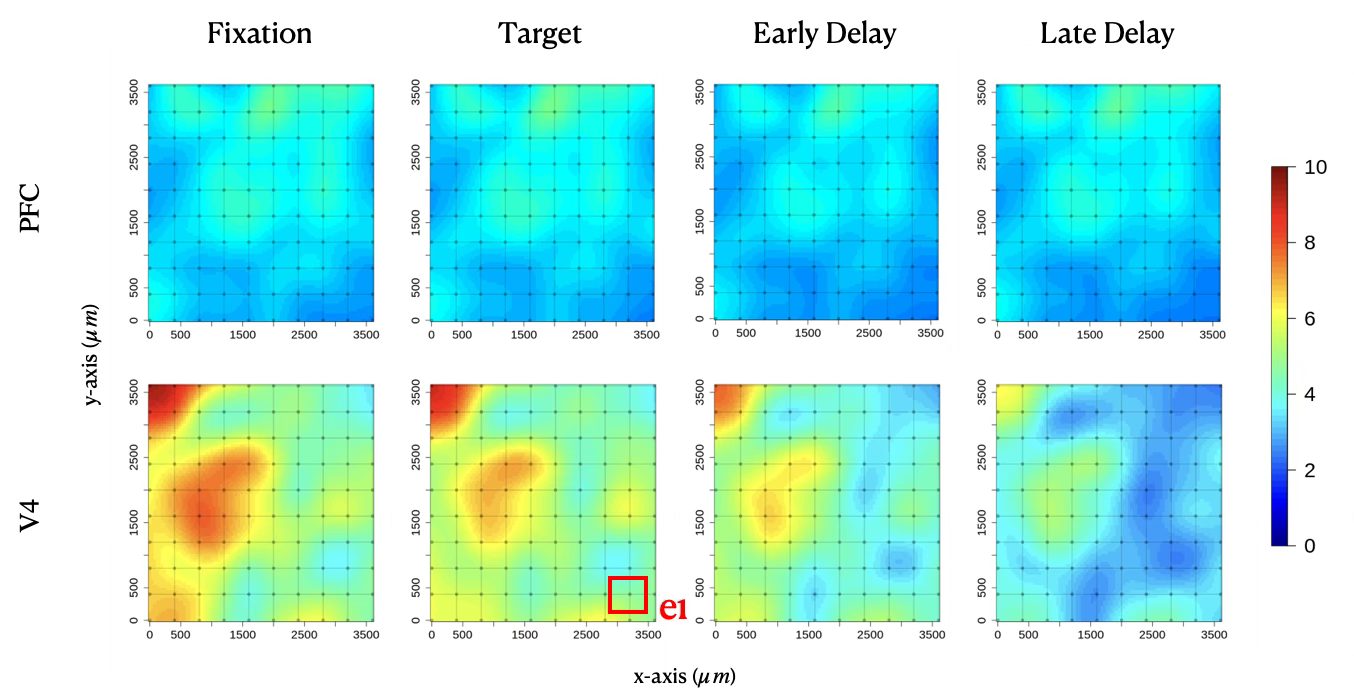}
    \caption{Spatially smoothed within-area connectivity strength distribution over the spatial electrode array for PFC and V4 over the experimental stages. The connectivity within V4 decayed over the stages, while the level of within-connectivity in PFC was stable.}\label{fig:within_region}
\end{figure}

\cref{fig:within_region} presents the smoothed distributions of within-area connectivity across the four experimental stages in both brain regions. We observed that V4 exhibited higher within-area connectivity than PFC. In V4, connectivity was strongest during the fixation and cue stages, then declined during the delay stages. In contrast, the level of connectivity in PFC remained relatively stable across experimental stages.

\subsubsection{Cross-area Inference}
\begin{figure}
    \centering
    \includegraphics[width=1.0\textwidth]{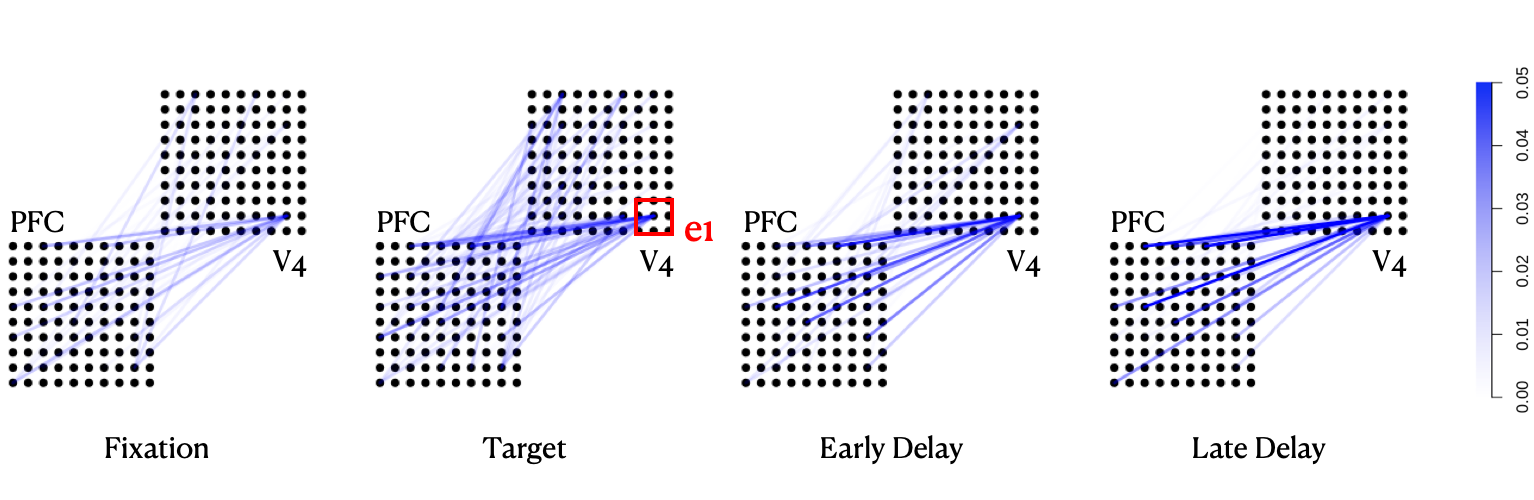}
    \caption{Estimated cross-area connectivity between PFC and V4, shown as edges color-coded by partial correlations $\hat{\rho}_{ij}^{(\mathcal{S})}$ between electrode $i$ in V4 and electrode $j$ in PFC over the four experimental stages. The x- and y-axes are spatial coordinates of the electrodes on each array. 
    %Lower left pane shows electrodes in PFC, while the upper right shows electrodes in V4. 
    The plot shows a spatial distinction in the evolution of the cross-area connection. % The connection in the edges with electrode $e_1$ (shown red in V4) strengthened over time, where the connection in the other edges was strongest during the target presentation.
    }
    \label{fig:between_region}
\end{figure}

% The declining connectivity for within area drives us to investigate cross-area connectivity. 
We applied our method to the recordings from both areas and identified significant cross-area connectivity. To reduce computation time and collinearity, we subsampled the electrodes in each area by taking every other node along the spatial dimension, reducing the spatial dimension $q$ from $192$ to $50$ in total. This reduced the number of cross-area edges to $625$. Also, because beta oscillations are often associated with communication across the two brain areas \citep{klein2020torus,miller2018wm2},
we first band-pass filtered the LFP recordings %using complex Morlet wavelets 
at the beta band $15 - 30$ Hz and downsampled the filtered data to $200$ Hz. 
% The electrode subsampling can be justified by the fact that adjacent nodes are usually strongly connected, thus skipping one neighbour will not negatively impact our results, but ensures numerical stability. Besides, it saves computational resources and make visualization of cross-area edges cleaner. 
\cref{fig:between_region} shows the cross-area edges, color-coded by the estimated spatial partial correlations $\hat\rho^{(S)}_{ij}$ between electrode $i$ in V4 and electrode $j$ in PFC. We find that cross-area connectivity between V4 and PFC is markedly heterogeneous across electrode pairs and varies substantially across experimental stages. In particular, edges involving electrode $e_1$ (shown in red in \cref{fig:between_region}) are consistently strong across stages and further strengthen during the late delay period, when working memory demands are highest. This pattern suggests that connectivity through electrode $e_1$ may reflect attentional signals directed by PFC during visual memory retention. In contrast, other cross-area edges are strongest during the target presentation stage, which may reflect transient top-down signaling from V4 to PFC associated with stimulus processing. These observations are broadly consistent with prior findings that V4–PFC interactions are engaged during attention tasks \citep{squire2013prefrontal,snyder2021stable}, and that long-range corticocortical connections are sparse and exhibit substantial variability in strength depending on anatomical distance and functional properties \citep{ts1986relationships,el2013visual,smith2008}.
% \cref{fig:between_region} shows the cross-area edges, color-coded by the estimated spatial partial correlations $\hat\rho^{(S)}_{ij}$ between electrode $i$ in V4 and electrode $j$ in PFC.
% We found that the cross-area connectivity between V4 and PFC was much stronger between some electrode pairs than others, and changed substantially over time. In particular, the edges with electrode $e_1$ (shown red in \cref{fig:between_region}) strengthened over time, whereas the other edges were strongest during the target presentation. This is broadly consistent with the finding that there are interactions between V4 and PFC during attention tasks \citep{squire2013prefrontal,snyder2021stable} and that corticocortical connections at a distance are quite sparse and highly variable in strength depending on distance and response properties \citep{ts1986relationships,el2013visual,smith2008}.

Based on this observation, we further analyzed the cross-area connectivity in two distinct edge sets: (i) edges incident to $e_1$ and (ii) edges excluding $e_1$. Cross-area connectivity was quantified by the maximal partial correlations within each edge set across the four experimental stages. For statistical inference, we constructed 95\%-confidence intervals for the maximal partial correlation in each edge set and stage, as defined in \cref{eq:inf_CI}, using 100,000 Gaussian bootstrap samples, i.e., $\|\hat{T}_E + \hat{Z}\|_\infty$ where $\hat{Z} \sim \distNorm(0, \hat{S}_{EE})$.

\begin{figure}
    \centering
    \includegraphics[width=0.9\textwidth]{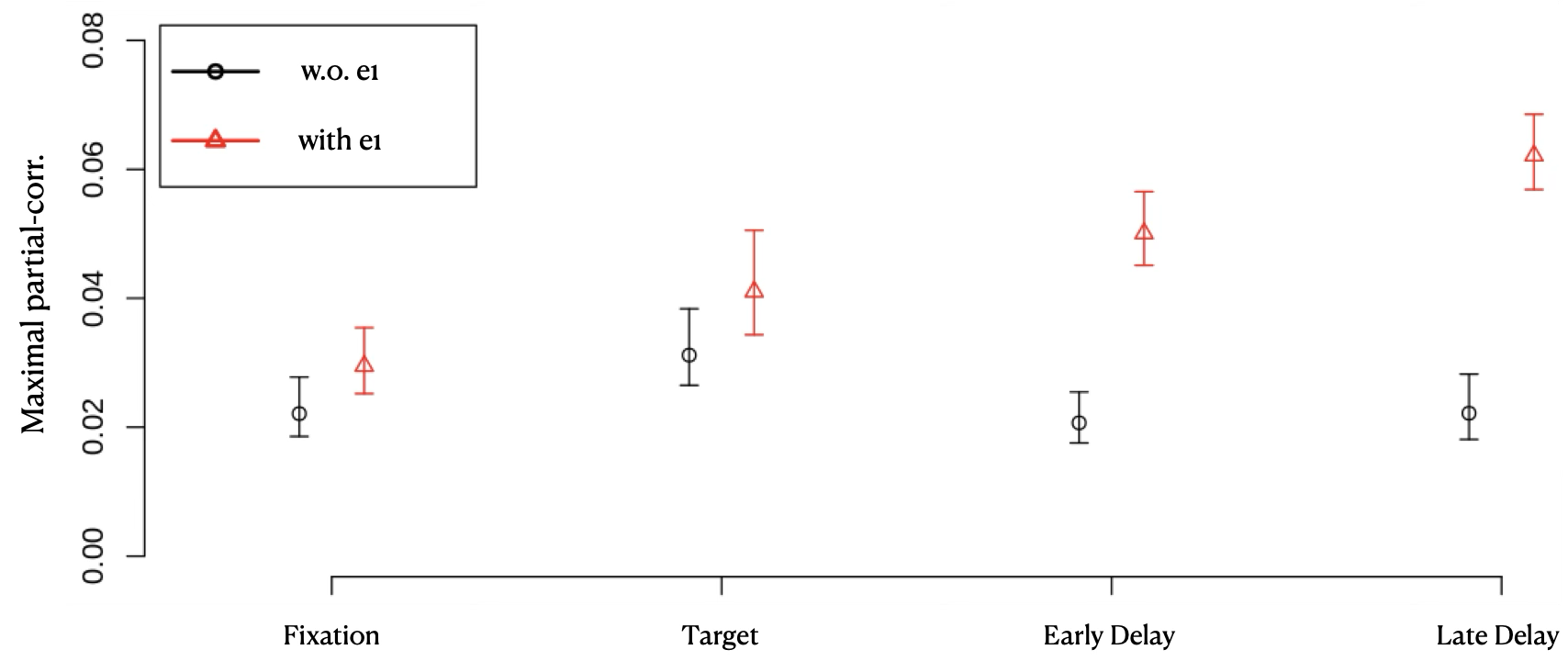}
    \caption{Inference on the maximal partial correlations over the two edge sets: (triangle) the cross-area edges with electrode $e_1$, marked in \cref{fig:between_region} and (circle) the other cross-area edges. The points are the median values of the bootstrap samples obtained by \cref{alg:procedure} with $100,000$ bootstrap samples, and the error bars are the 95 \% bootstrap confidence intervals. % The mean spatial partial correlation over the edges with electrode $e_1$ was shown to be significantly monotonic increasing by the linear trend test ($p < 1\text{e-}6$).
    }\label{fig:rho_E_infty}
\end{figure}

\cref{fig:rho_E_infty} presents these confidence intervals, with points indicating the median values of the bootstrap samples $\norm{\hat{T}_E + \hat{Z}}_\infty$ for the two edge sets across experimental stages. For edges excluding $e_1$, we reaffirm that cross-area connectivity was strongest during target presentation. %, potentially reflecting V4–PFC communication about the target stimulus. 
In contrast, edges incident to $e_1$ exhibited increased connectivity during the delay stage, when the animal processed the visual signal and prepared for the subsequent decision.

We tested the monotonic trend in cross-area connectivity for edges incident to $e_1$ using a linear trend test. In this test, we regressed connectivity levels at the four stages against the respective time points on the experimental timeline. A positive slope indicated increasing connectivity, while a negative slope indicated decreasing connectivity. The uncertainty in the slope estimate was assessed by repeating the linear trend test over the Gaussian bootstrap samples $\|\hat{T}_E + \hat{Z}\|_\infty$. The $95\%$-confidence interval of the slope was $(5.03 \mathrm{e-}5, 8.17 \mathrm{e-}5)$ $\text{ms}^{-1}$, indicating that connectivity in edges incident to $e_1$ increased significantly over time. 

A similar trend was observed when using the average partial correlation, rather than the maximal partial correlations, over the edge set (\cref{fig:rho_E_avg}). This contrasts with the decline in within-area connectivity in V4 during the delay stages, as shown in \cref{fig:within_region}. When we applied the same trend test to averaged partial correlations within V4, we found that within-area connectivity in V4 decreased significantly (\cref{fig:rho_E_v4}).

% Our inference results support the previous studies that neural variability in the spiking of neurons declines during the stimulus onset \citep{Churchland2010}, and visual stimuli cause a substantial decrease in the correlation of cortical neurons \citep{smith2008}. We also discovered robust sustained within-area connectivity in PFC during the delay stage, compared to V4, which was also reported by \cite{leavitt2017}. 
% {\color{red} \bf Rob: you may revise the scientific implications here.} 

In summary, (a) there is strong within-area functional connectivity (based on partial correlations) for V4 that diminishes across the four stages, while there is much weaker within-area connectivity for PFC that remains stable across the stages (\cref{fig:within_region}); (b) the cross-area connectivity is relatively sparse in V4 compared with PFC, except during target illumination; the signal from electrode $e_1$ which, across all four stages, has the largest partial correlation (among those in V4) with those from PFC, during the final stage (late delay) has very much stronger connectivity with PFC than the signals from other electrodes in V4; also its connectivity strengthens across the stages, while connectivity of the other electrodes with PFC stays roughly the same (\cref{fig:between_region,fig:rho_E_infty}); and (c) the signal from electrode $e_1$ does not have especially strong connectivity within V4.

\section{Conclusion} \label{sec:conc}

We proposed a method for jointly estimating multiple matrix-variate graphical models that leverages shared sparsity structures across sample groups, together with a unified framework for high-dimensional inference on graph edges. Estimation is carried out using group Lasso to exploit the shared sparsity structures, while inference is designed for high-dimensional settings via state-of-the-art central limit theorems for the supremum of weighted sums. The proposed approach accommodates settings in which the spatial and temporal dimensions, as well as the number of graphs, may all diverge and potentially exceed the available sample size.

Both our model and our assumptions are motivated by practical concerns in neural data analysis. In real data, we observed that the within-area connectivity and cross-area connectivity change as the animal progresses through different experimental stages. From a scientific perspective, the data analysis suggests the intriguing possibility that there is some degree of localization in the memory task-relevant neural signals connecting V4 with PFC. However, justification for a strong scientific conclusion would require additional experimental work. The findings are most interesting as illustrations of the way changing connectivity structure, across multiple stages of a task, can be determined from high-dimensional electrode recordings.

Our method represents the first attempt to address the simultaneous testing problem in multiple matrix-variate Gaussian graphs. An interesting future direction would be to extend this approach to other commonly used non-Gaussian graph types, such as Poisson networks. Additionally, our current implementation uses group Lasso for the regression model, which requires one tuning parameter. In the future, a tuning-free or scale-free approach, such as the self-tuned Dantzig selector or scaled Lasso, would be desirable to address issues of heterogeneity and correlation in regression with data from multiple matrix-variate Gaussian graphical models. These extensions, while beyond the scope of this work, offer promising directions for future research.

%%%%%%%%%%%%%%%%%%%%
%%% Bibliography %%%
%%%%%%%%%%%%%%%%%%%%

\bibliographystyle{apalike}
\bibliography{2_ref.bib}

%%%%%%%%%%%%%%%%%%
%%% Appendices %%%
%%%%%%%%%%%%%%%%%%

\newpage\appendix
\counterwithin{figure}{section}
\counterwithin{table}{section}

\section{Supplementary results for simulation study}

\Cref{tab:simul_d3} presents the average empirical coverages under the same simulation setting as in \cref{sec:sim_simultaneous} with the number of sessions fixed at $m=3$. 

\begin{table}[!htbp]
\centering
\caption{Average of empirical coverages and their standard deviations for $m=3, p=50,  q=30$.}\label{tab:simul_d3}
\scriptsize
\begin{tabular}{|p{1cm}|p{1cm}||p{1.5cm}|p{1.5cm}|p{1.5cm}|p{1.5cm}|p{1.5cm}|p{1.5cm}|}
\hline
\multirow{2}{*}{$n$} & \multirow{2}{*}{Quantile} &\multicolumn{2}{c}{Random} & \multicolumn{2}{|c|}{Hub}& \multicolumn{2}{c|}{Band}\\
\cline{3-8}
{} & {} & $E_\text{off}$ & $E_\text{zero}$ & $E_\text{off}$ & $E_\text{zero}$ & $E_\text{off}$ & $E_\text{zero}$\\
\hline
\multirow{3}{*}{5}  & 0.925   & 0.897(0.012) & 0.898(0.009)   & 0.908(0.010) & 0.908(0.009)   & 0.903(0.009) & 0.907(0.010)\\
\cline{2-8}
{}                  & 0.95    & 0.935(0.006) & 0.932(0.010)   & 0.939(0.007) & 0.939(0.007)   & 0.934(0.008) & 0.939(0.007)\\
\cline{2-8}
{}                  & 0.975   & 0.962(0.006) & 0.969(0.002)   & 0.970(0.004) & 0.971(0.004)   & 0.971(0.004) & 0.971(0.003)\\
\hline
\multirow{3}{*}{10} & 0.925   & 0.926(0.005) & 0.924(0.006)   & 0.923(0.005) & 0.923(0.006)   & 0.929(0.005) & 0.927(0.006)\\
\cline{2-8}
{}                  & 0.95    & 0.944(0.003) & 0.949(0.004)   & 0.945(0.004) & 0.945(0.005)   & 0.953(0.005) & 0.950(0.005)\\
\cline{2-8}
{}                  & 0.975   & 0.967(0.004) & 0.970(0.002)   & 0.978(0.003) & 0.977(0.005)   & 0.980(0.003) & 0.977(0.003)\\
\hline
\multirow{3}{*}{20} & 0.925   & 0.926(0.005) & 0.921(0.004)   & 0.926(0.005) & 0.926(0.006)   & 0.928(0.004) & 0.925(0.005)\\
\cline{2-8}
{}                  & 0.95    & 0.954(0.005) & 0.944(0.003)   & 0.948(0.004) & 0.948(0.004)   & 0.951(0.005) & 0.943(0.005)\\
\cline{2-8}
{}                  & 0.975   & 0.978(0.001) & 0.975(0.004)   & 0.977(0.005) & 0.975(0.004)   & 0.978(0.003) & 0.975(0.003)\\
\hline
\end{tabular}
\end{table}

\newpage 

\section{Supplementary results for experimental data analysis}

%\begin{figure}[H]
%    \centering
%    \includegraphics[width=0.7\textwidth]{fig/roc_sim_lambda.png}
%    \caption{ROC curve under 3 graph types with $n=5$, $q=30$ and $m=5$ at tuning parameters $\lambda \in (1, 0.01, 1e-4, 1e-6)$. Our method is consistently better regardless of tuning parameters.}\label{fig:roc_sim_lambda}
%\end{figure}
%  \begin{figure}[H]
%     \centering
%     \includegraphics[height=0.8\textwidth]{fig/session_corr.png}
%     \caption{Sample estimate of $\mat{B}_t$ for each session plotting against each other. For each panel, we plot $\text{vec}(\hat{\mat{B}}^{samp}_t)$ vs. $\text{vec}(\hat{\mat{B}}^{samp}_{t+1})$, for $1\leq t \leq 4$. Each blue dot corresponds to an edge value in precision matrix. The observation is similar to Figure~\ref{fig:session_partial_cor} and the signs of connectivity should be consistent across sessions. }\label{fig:session_cor}
% \end{figure}

\begin{figure}
    \centering
    \includegraphics[width=1.0\textwidth]{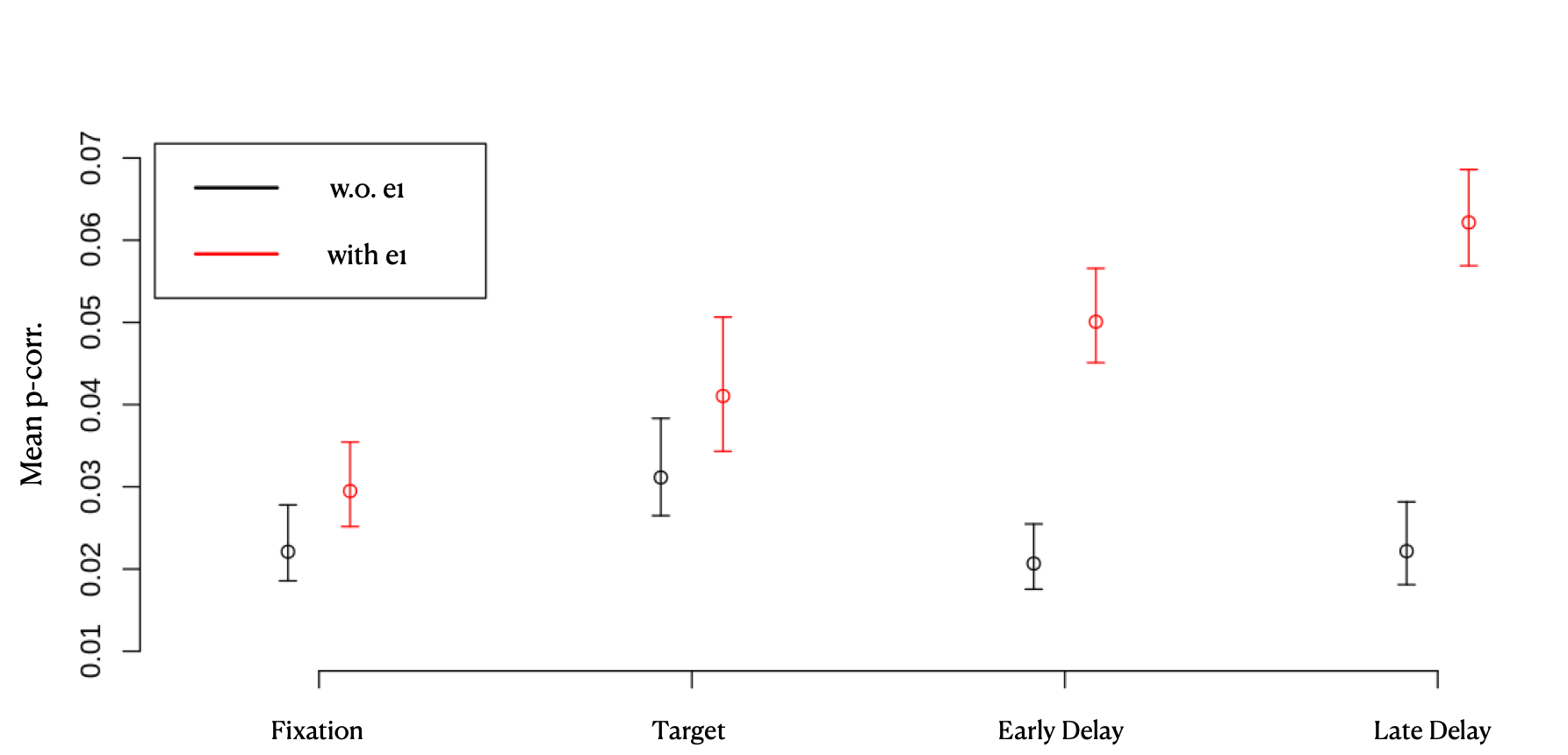}
    \caption{Inference on the mean partial correlations over the two edge sets: (red) the cross-areas edges with electrode $e_1$, marked in \cref{fig:between_region} and (black) the other cross-areas edges. The points are the median values of the bootstrap samples obtained by \cref{alg:procedure} with $100,000$ bootstrap samples, and the error bars are the 95 \% bootstrap confidence intervals.}\label{fig:rho_E_avg}
\end{figure}

\Cref{fig:rho_E_avg} presents the confidence intervals for the mean partial correlations over the two edge sets in the four experimental stages. The mean spatial partial correlation over the edges with electrode $e_1$ was shown to be significantly monotonic increasing by the linear trend test ($p < 1\text{e-}6$).

\begin{figure}
    \centering
    \includegraphics[width=1.0\textwidth]{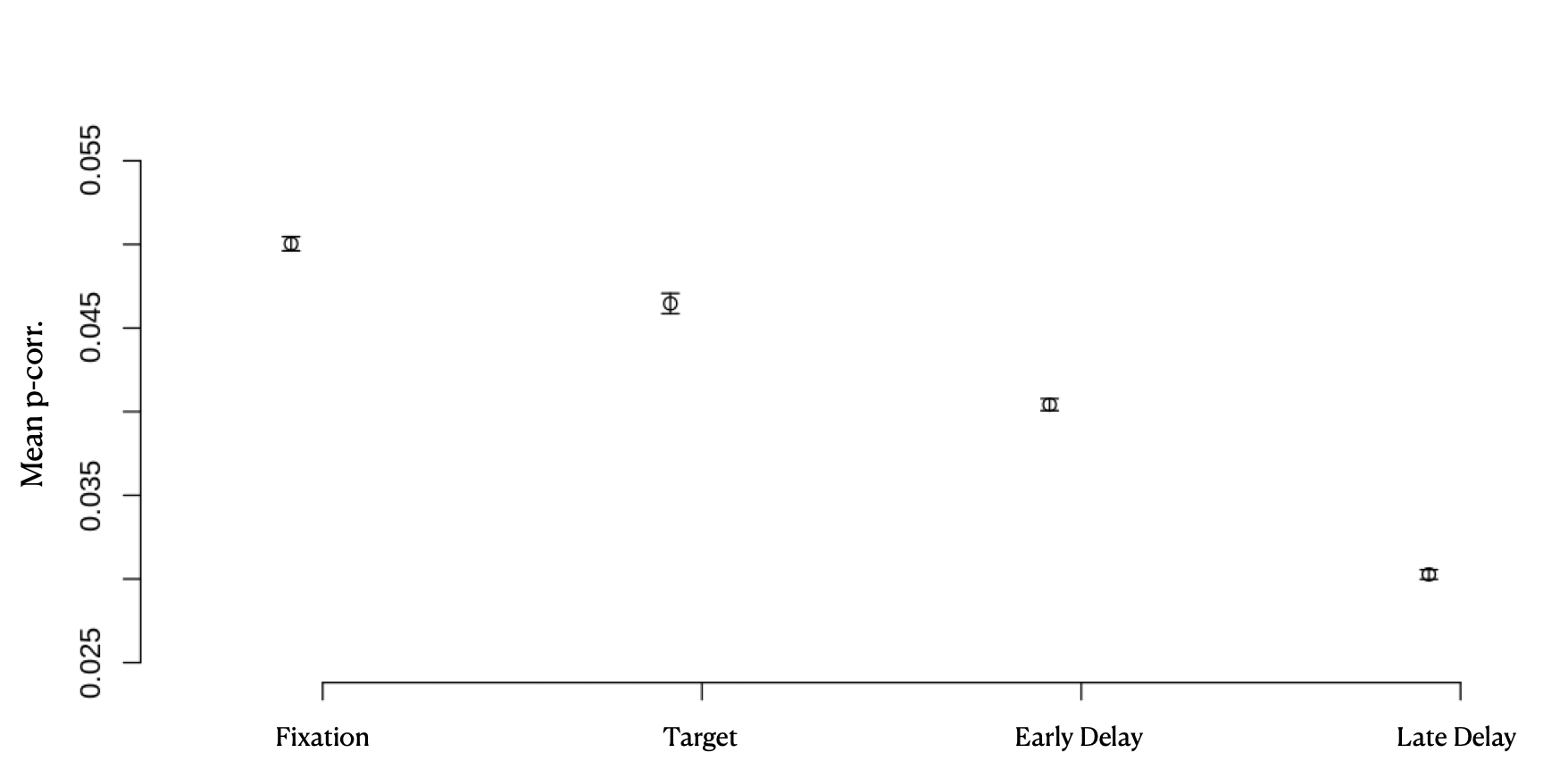}
    \caption{Inference on the mean partial correlations within V4. The points are the median values of the bootstrap samples obtained by \cref{alg:procedure} with $100,000$ bootstrap samples, and the error bars are the 95 \% bootstrap confidence intervals.}\label{fig:rho_E_v4}
\end{figure}

\Cref{fig:rho_E_v4} presents the confidence intervals for the mean partial correlations within region V4 in the four experimental stages. The mean spatial partial correlation within V4 was shown to be significantly monotonic decreasing by the linear trend test ($p < 1\text{e-}6$).

The cross-area connectivity trend was further evidenced by the $c$-level testing. Our simultaneous testing approach of multiple edges can also be extended to $c$-level tests \citep{Qiu2018}. By testing if $\norm{\rho_{E_2}}_\infty$ exceeds threshold level $c$, i.e.,
\begin{equation*}
    H_{0,c}: \norm{\rho_{E_2}}_\infty \leq c
\end{equation*}
we compared the cross-area connectivity strengths in the four stages. \cref{tab:c_level} shows the test results at four levels $c = 0, 0.02, 0.03, 0.04$. At $c = 0$, the cross-area connectivity in $E_1$ was significant with $\alpha = 0.05$ at all experimental stages. As the level increased, the tests at the early stages began unrejected. At $c = 0.04$, the test was significant only at the late delay stage. These results are consistent with our finding in \cref{fig:rho_E_infty} and the subsequent linear trend test.

\begin{table}[!htbp]
\centering
\caption{The $c$-level test results for $\norm{\hat{\rho}_{E_2}}_\infty$ at $c$-levels $\{0, 0.02, 0.03, 0.04\}$. Entries with * represent the significance at $\alpha = 0.05$.}\label{tab:c_level}
\small
\begin{tabular}{|p{2cm}||p{2cm}|p{2cm}|p{2cm}|p{2cm}|}
\hline
$c$-level &Fixation & Cue & Early Delay & Late Delay \\
\hline 
0    & * & * & * & * \\
0.02 &   & * & * & * \\
0.03 &   &   & * & * \\
0.04 &   &   &   & * \\
\hline
\end{tabular}
\end{table}

\section{Proofs}

Throughout the following theoretical arguments, we use $X^{(\mathcal{S},l)}$ (or $X^{(\mathcal{T}, l)}$) to notate stacked spatial (or temporal) observation over temporal (or spatial) components and samples within session $l$:
\begin{equation*}
\begin{split}
    X^{(\mathcal{S},l)} &= \left[X^{(1,l)\top}, \dots, X^{(n_l,l)\top}\right]^\top \in \reals^{n_l p \times q} \\
    X^{(\mathcal{T},l)} &= \left[X^{(1,l)}, \dots, X^{(n_l,l)}\right]^\top \in \reals^{n_l q \times p}.
\end{split}
\end{equation*}
These notations help us with connecting matrix-variate Gaussian graphical models to node-wise regression models:
\begin{equation*}
    X^{(\mathcal{S},l)}_{ti} =  X^{(\mathcal{S},l)}_{t,\cdot} \beta^{(\mathcal{S},l)}_{\cdot,i} + \epsilon^{(\mathcal{S},l)}_{ti}
\end{equation*}
for $t = 1, \dots, n_l p$ where $X^{(\mathcal{S},l)}_{t}$ are now dependent and not identically distributed due to the temporal association, unlike the usual regression regime. Also, for any vector $b^{(\cdot)} := \{b^{(l)} \in \reals^p\}_{l=1,\dots,m}$, we use $\underline{b}^{(\cdot)}$ to denote the collection of standardized elements, i.e.,
\begin{equation*}
    \underline{b}^{(\cdot)}_j := \left\{\frac{\norm{X^{(\mathcal{S},l)}_{\cdot,j}}_2}{\sqrt{n_l p}} b^{(l)}_j \right\}_{l=1,\dots,m}
\end{equation*}
so that \cref{eq:spatial_est_beta} is expressed in a more canonical form:
\begin{equation} \label{eq:spatial_est_beta_S}
\begin{split}
    & \hat{\beta}^{(\mathcal{S},\cdot)}_{\cdot,i} := \argmin_{\{b^{(l)}\}_{l=1,\dots,m}} \left\{
    \frac{1}{2 n_0 p} \sum_{l=1}^m \norm{X^{(\mathcal{S},l)}_{\cdot,i} - X^{(\mathcal{S},l)} b^{(l)}}_2^2
    + \gamma_i \sum_{j:j \neq i} \norm*{ \underline{b}^{(\cdot)}_j }_2
    \right\}, \\
    % & \text{w.r.t} ~~b^{(l)}_{i} = 0.
\end{split}
\end{equation}
with respect to $b_i^{(l)} = 0$ for all $l$. In the following proofs, for a real matrix $M$, we use $\norm{M}_\mathrm{op}$ and $\norm{M}_\mathrm{2}$ interchangeably to denote the matrix 2-norm, also known as the operator norm.

\subsection{Preliminary Lemmas}

Here we introduce several lemmas that will help us prove propositions and theorems later. To apply the classical techniques and results in penalized regression analyses, we need to address the dependency among $X^{(\mathcal{S},l)}_{t}$'s. To do so, we use the fact that, for $k=1, \dots, n_l$,
\begin{equation} \label{eq:iid_XX}
    \sum_{t=1}^{p} X^{(k,l)}_t X^{(k,l)^\top}_t =
    \sum_{t=1}^{p} \lambda^{(\mathcal{T},l)}_{t} W^{(k,l)}_{t} W^{(k,l)\top}_t
\end{equation}
where $W^{(k,l)}_{t} \distiid N(0, \Sigma^{(\mathcal{S},l)})$ for all $t = 1, \dots, p$.
We further let $\xi^{(k,l)}_{ti} = W^{(k,l)}_{ti} - W^{(k,l)}_{t,\cdot}\beta^{(\mathcal{S},l)}_{\cdot,i}$ so that
\begin{equation} \label{eq:iid_EX_EE}
\begin{aligned}
    \sum_{t=1}^{p} \epsilon^{(\mathcal{S},k,l)}_t X^{(k,l)^\top}_t 
    & =
    \sum_{t=1}^{p} \lambda^{(\mathcal{T},l)}_{t} \xi^{(k,l)}_{t} W^{(k,l)\top}_t, \\
    \sum_{t=1}^{p} \epsilon^{(\mathcal{S},k,l)}_t \epsilon^{(\mathcal{S},k,l)^\top}_t 
    & =
    \sum_{t=1}^{p} \lambda^{(\mathcal{T},l)}_{t} \xi^{(k,l)}_{t} \xi^{(k,l)\top}_t,
\end{aligned}
\end{equation}
and $\xi^{(k,l)}_t \distiid N(0,\Phi^{(\mathcal{S},l)})$ across $t=1,\dots, p$ and $k=1,\dots,n_l$.

The subsequent lemmas follow the application to the above equations of sub-exponential concentration inequalities, which could be found in mathematical statistics literature such as \citet{vershynin2018}. The proofs for the lemmas are given in \crefrange{app:pf_l_inf_XX}{app:pf_l_inf_EE}.

\begin{lemma} \label{lem:l_inf_XX}
\begin{equation*}
    \Pr\left[ 
        % \max_{l = 1, \dots, m} 
        \norm*{\frac{1}{n_l p} X^{(\mathcal{S},l)\top} X^{(\mathcal{S},l)} - \Sigma^{(\mathcal{S},l)}}_\infty 
        \geq C(\kappa_3) \sqrt{\frac{\nu}{n_l p}} 
    \right] 
    \leq q e^{-\nu} 
\end{equation*}
for $0 \leq \frac{\nu}{n_l p} \leq C(\kappa_3)$.
% and
% \begin{equation}
%     \Pr\left[ 
%         \max_{l = 1, \dots, m} \norm*{\frac{1}{n_l p} X^{(\mathcal{S},l)\top} X^{(\mathcal{S},l)} - \Sigma^{(\mathcal{S},l)}}_\mathrm{op} 
%         \geq C(\kappa_3) \sqrt{\frac{q+\log(mn_0pq)}{n_l p}} 
%     \right] 
%     \leq (mn_0pq)^{-1/2}. 
% \end{equation}
\end{lemma}

%%%%%%%%%%%%%%%%%%%%%%%%%%%%%%%%%%%%%

\begin{lemma} \label{lem:l_2_EX}
\begin{equation*}
\begin{aligned}
    \Pr\left[ 
        % \max_{i,j: j \neq i}
        \frac{1}{n_0 p}\left[\sum_{l=1}^m \left(\frac{\sqrt{n_l p}}{\norm{X^{(\mathcal{S},l)}_{\cdot,j}}_2} \epsilon^{(\mathcal{S},l)\top}_{\cdot,i} X^{(\mathcal{S},l)}_{\cdot,j}\right)^2\right]^{1/2}
        \geq C(\kappa_1, \kappa_3) \sqrt{\frac{m + \nu}{n_0 p}}
    \right] \leq e^{-\nu}.
\end{aligned}
\end{equation*}
and
\begin{equation*}
\begin{aligned}
    \Pr\left[ 
        % \max_{i=1,\dots,q} 
        \frac{1}{n_0 p}\left[\sum_{l=1}^m \left(\frac{\sqrt{n_l p}}{\norm{X^{(\mathcal{S},l)}_{\cdot,i}}_2} \epsilon^{(\mathcal{S},l)\top}_{\cdot,i} X^{(\mathcal{S},l)}\beta^{(\mathcal{S},l)}_{\cdot,i} \right)^2 \right]^{1/2}
        \geq C(\kappa_1, \kappa_3) \sqrt{\frac{m + \nu}{n_0 p}}
    \right] \leq e^{-\nu}. 
\end{aligned}
\end{equation*}

% Let $\lambda_i= c\frac{\xi+1}{\xi-1} \sqrt{\frac{d+\log q}{n_0 p}} $ for constant $\xi>1$ and a large enough constant $c$ depending on $\delta,\ c_e$ and $M_0$ only. Then with  probability $1-q^{-\delta}$ that event $E_i$ holds where
% $$E_{i} = \{\max_{l\neq i}\frac{\left[\sum_{t=1}^d (\bar{\mat{Z}}'_{t, \cdot, l} \mat{E}_{t i})^2\right]^{1/2} }{n_0 p}\leq \frac{\xi-1}{\xi+1}\lambda_i \}. $$
% Moreover, define $\widetilde{\mat{Z}}_{t, \cdot, i} \mat{D}_{ti}^{-1/2} = \bar{\widetilde{\mat{Z}}}_{t, \cdot, i} $; with the same probability that event $\widetilde{E}_i$ holds where
% $$\widetilde{E}_{i} = \{\max_{ 1\leq i\leq q}\frac{\left[\sum_{t=1}^d (\bar{\widetilde{\mat{Z}}}'_{t, \cdot, i} \mat{E}_{t i})^2\right]^{1/2} }{n_0 p}\leq \frac{\xi-1}{\xi+1}\lambda_i \}. $$
\end{lemma}

%%%%%%%%%%%%%%%%%%%%%%%%%%%%%

\begin{lemma} \label{lem:l_inf_EX}
\begin{equation*}
    \Pr\left[ 
        % \max_{l=1,\dots,m} 
        \max_{i,j: j \neq i}\abs*{\frac{1}{n_l p} \epsilon^{(\mathcal{S},l)\top}_{\cdot,i} X^{(\mathcal{S},l)}_{\cdot,j}}
        \geq C(\kappa_3) \sqrt{\frac{\nu}{n_l p}}
    \right]  
    \leq q e^{-\nu}, 
\end{equation*}
and
\begin{equation*}
    \Pr\left[ 
        % \max_{l=1,\dots,m} 
        \max_{i=1,\dots,q} \abs*{\frac{1}{n_l p} \epsilon^{(\mathcal{S},l)\top}_{\cdot,i} X^{(\mathcal{S},l)}\beta^{(\mathcal{S},l)}_{\cdot,i}}
        \geq C(\kappa_3) \sqrt{\frac{\nu}{n_l p}}
    \right]  
    \leq q e^{-\nu}
\end{equation*}
for $0 \leq \frac{\nu}{n_0 p} \leq C(\kappa_3)$.

% $$\mathbb{P}(\max_{1\leq i \leq q}\max_{1\leq h \leq q, h\neq i}|\frac{1}{n_t p}\sum_{k=1}^{n_t}\sum_{l=1}^p   {\varepsilon}^{(k)}_{tlj}{X}_{tlh}^{(k)}|\geq C\sqrt{\frac{\log q}{n_t p}}) \leq q^{-\delta},$$
% and 
% $$\mathbb{P}(\max_{1\leq i \leq q} |\frac{1}{n_t p}\sum_{k=1}^{n_t} \sum_{l=1}^p  {\varepsilon}^{(k)}_{tli}\mat{X}_{tl,-i}^{(k)}\mat{\beta}_{ti}|\geq C\sqrt{\frac{\log q}{n_t p}}) \leq q^{-\delta}. $$

\end{lemma}

%%%%%%%%%%%%%%%%%%%%%%%%%%%%%%%

\begin{lemma} \label{lem:l_inf_EE}
\begin{equation*}
    \Pr\left[ 
        % \max_{l=1,\dots,m} 
        \norm*{\frac{1}{n_l p} \epsilon^{(\mathcal{S},l)\top} \epsilon^{(\mathcal{S},l)} - \Phi^{(\mathcal{S},l)}}_{\infty}
        \geq C(\kappa_3) \sqrt{\frac{\nu}{n_l p}}
    \right]  
    \leq q e^{-\nu} 
\end{equation*}
for $0 \leq \frac{\nu}{n_0 p} \leq C(\kappa_3)$.

% Define  $\tilde{r}_{tij} = \frac{1}{n_t p}\sum_{k=1}^{n_t} \sum_{l=1}^p {\varepsilon}^{(k)}_{tli}{\varepsilon}^{(k)}_{tlj}$, then for any $M>0$ , there exists a constant C which depends on $\delta$, $c_e$, $M_0$ only, such that
% $$\mathbb{P} \left(\max_{1\leq i\leq j\leq q} | \tilde{r}_{tij} -  \frac{b_{tij}}{b_{tii}b_{tjj}}|\geq C\sqrt{\frac{\log q}{n_tp}} \right) \leq q^{-\delta}, $$
% and
% $$\mathbb{P} \left(\max_{1\leq i\leq j\leq q} |\sum_{t=1}^d (\tilde{r}_{tij} -  \frac{b_{tij}}{b_{tii}b_{tjj}})|\geq C\sqrt{\frac{d\log q}{n_0p}} \right) \leq q^{-\delta}. $$
\end{lemma}

%%%%%%%%%%%%%%%%%%%%%%%%%%%%%%%%%%%%%%%%%%%%%%%%%%%%%%%
%%%%%%%%%%%%%%%%%%%%%%%%%%%%%%%%%%%%%%%%%%%%%%%%%%%%%%%
%%%%%%%%%%%%%%%%%%%%%%%%%%%%%%%%%%%%%%%%%%%%%%%%%%%%%%%
%%%%%%%%%%%%%%%%%%%%%%%%%%%%%%%%%%%%%%%%%%%%%%%%%%%%%%%

\subsection{Non-asymptotic error bound for the temporal covariance matrix estimate}
\label{app:temp_cov}
In this section, we present the estimation error bounds for the session-specific temporal covariance and precision matrices obtained in \cref{sec:temporal_est}. While results for i.i.d. samples are available in \cite{liu2017}, no such results exist for the correlated samples considered in our model. Therefore, we provide a self-contained analysis, which may be of independent interest. The proof is detailed in \cref{app:pf_temporal_est_Sigma}.

\begin{proposition} \label{thm:temporal_regression}
    Suppose that $h_l = \floor{(n_l q)^{1/(1+\alpha_0)}}$. Then, following the procedure defined in \cref{sec:temporal_est}, 
    \begin{equation*}
       \mathbb{P} \left[ \max_l \max_{1\leq t \leq p} \norm{\hat{\beta}^{(\mathcal{T},l)}_{\cdot,t}-\beta^{(\mathcal{T},l)}_{\cdot,t}}_{2}
       \geq C(\kappa_1, \kappa_3, \kappa_5)
       \sqrt{\frac{\log (mn_0pq)}{(n_0q)^{1-1/(2\alpha_0+2)}}} \right] 
       \leq C (mn_0pq)^{-1/2}, 
    \end{equation*}
    \begin{equation*}
       \mathbb{P} \left[ \max_l \max_{1\leq t \leq p}
       \abs*{\hat{\Phi}^{(\mathcal{T},l)}_{tt} - \frac{\tr(\Sigma^{(\mathcal{S},l)})}{q} \Phi^{(\mathcal{T},l)}_{tt}}
       \geq C(\kappa_1, \kappa_3, \kappa_5)
       \sqrt{\frac{\log (mn_0pq)}{(n_0q)^{1-1/(2\alpha_0+2)}}}\right] 
       \leq C(mn_0pq)^{-1/2}.
    \end{equation*}
\end{proposition}

\begin{theorem} \label{thm:temporal_est_Sigma}
    Suppose that $h_l = \floor{(n_l q)^{1/(1+\alpha_0)}}$ and $\eta = C(\kappa_3)$ satisfies $\eta \leq \lambda_1(I - \beta^{(\mathcal{T},l)})$ for $l = 1, \dots, m$, where $\lambda_1(I - \beta^{(\mathcal{T},l)})$ is the smallest eigenvalue of $I - \beta^{(\mathcal{T},l)}$. Then,
    \begin{equation*}
    \begin{aligned}
        \Pr\left[\begin{aligned}
            & \max_l \frac{1}{p} \norm*{
            \bar\Sigma^{(\mathcal{T},l)}
            - \frac{\tr(\Sigma^{(\mathcal{S},l)})}{q} \Sigma^{(\mathcal{T},l)}}_F^2 
            \geq C(\kappa_1,\kappa_3,\kappa_5) \frac{\log(mn_0pq)}{(n_0q)^{1-1/(2\alpha_0+2)}}, \\
            & \max_l \frac{1}{p} \norm*{
            \bar\Omega^{(\mathcal{T},l)}
            - \frac{q}{\tr(\Sigma^{(\mathcal{S},l)})} \Omega^{(\mathcal{T},l)}}_F^2 
            \geq C(\kappa_1,\kappa_3,\kappa_5) \frac{\log(mn_0pq)}{(n_0q)^{1-1/(2\alpha_0+2)}}
        \end{aligned}\right]
        \leq C(mn_0pq)^{-1/2}.
    \end{aligned}
    \end{equation*}
\end{theorem}

Consequently, the Frobenius norms of $\Sigma^{(\mathcal{T},l)}$ can be consistently estimated, which is sufficient for our main result \cref{thm:inf_bootstrap}. The proof is given in \cref{app:pf_temporal_est_F}.

\begin{corollary} \label{thm:temporal_est_F}
    Suppose that $h_l = \floor{(n_l q)^{1/(1+\alpha_0)}}$ and $\eta = C(\kappa_3)$ satisfies $\eta \leq \lambda_1(I - \beta^{(\mathcal{T},l)})$ for $l = 1, \dots, m$. Then,
    \begin{equation*}
        \Pr\left[
            \max_l \abs*{
            \frac{\norm{\hat\Sigma^{(\mathcal{T},l)}}_F^2 
            - \norm{\Sigma^{(\mathcal{T},l)}}_F^2}{p}} 
            \geq C(\kappa_1,\kappa_3,\kappa_5) \frac{\log(mn_0pq)}{(n_0q)^{1-1/(2\alpha_0+2)}}
        \right] \leq C(mn_0pq)^{-1/2}
    \end{equation*}
    for sufficiently large $n_0$.
\end{corollary}

%%%%%%%%%%%%%%%%%%%%%%%%%%%%%%%%%%%%%%%%%%%%%%%%%%%%%%%
%%%%%%%%%%%%%%%%%%%%%%%%%%%%%%%%%%%%%%%%%%%%%%%%%%%%%%%
%%%%%%%%%%%%%%%%%%%%%%%%%%%%%%%%%%%%%%%%%%%%%%%%%%%%%%%
%%%%%%%%%%%%%%%%%%%%%%%%%%%%%%%%%%%%%%%%%%%%%%%%%%%%%%%

\subsection{Proof of Theorem~\ref{thm:spatial_regression}} \label{app:pf_spatial_regression}

Based on the optimality of the group Lasso estimate in \cref{eq:spatial_est_beta_S}, we have the following basic inequality:
\begin{equation} \label{eq:basic_ineq}
\begin{aligned}
    & \frac{1}{2 n_0 p} \sum_{l=1}^m \left( 
    \norm{X^{(\mathcal{S},l)}_{\cdot,i} - X^{(\mathcal{S},l)} \hat{\beta}^{(\mathcal{S},l)}_{\cdot,i}}_2^2
    - \norm{X^{(\mathcal{S},l)}_{\cdot,i} - X^{(\mathcal{S},l)} \beta^{(\mathcal{S},l)}_{\cdot,i}}_2^2 \right) \\
    & \leq \gamma_i \sum_{j:j \neq i} \left( 
    \norm*{ \underline{\beta}^{(\mathcal{S},\cdot)}_{ji}}_2
    - \norm*{ \hat{\underline\beta}^{(\mathcal{S},\cdot)}_{ji}}_2 \right)
\end{aligned}
\end{equation}
for every $i = 1, \dots, q$. For the left hand side, by decomposing $\hat{\beta}^{(\mathcal{S},l)}_{\cdot,i} = \Delta^{(\mathcal{S},l)}_{\cdot,i} + {\beta}^{(\mathcal{S},l)}_{\cdot,i}$,
\begin{equation} \label{eq:basic_ineq_left}
\begin{aligned}
    & \frac{1}{2 n_0 p} \sum_{l=1}^m \left( 
    \norm{X^{(\mathcal{S},l)}_{\cdot,i} - X^{(\mathcal{S},l)} \hat{\beta}^{(\mathcal{S},l)}_{\cdot,i}}_2^2
    - \norm{X^{(\mathcal{S},l)}_{\cdot,i} - X^{(\mathcal{S},l)} \beta^{(\mathcal{S},l)}_{\cdot,i}}_2^2 \right) \\
    & = \frac{1}{2 n_0 p} \sum_{l=1}^m \left( 
    \norm{X^{(\mathcal{S},l)} \Delta^{(\mathcal{S},l)}_{\cdot,i}}_2^2 
    - 2 (X^{(\mathcal{S},l)}\Delta^{(\mathcal{S},l)}_{\cdot,i})^\top \epsilon^{(\mathcal{S},l)}_{\cdot,i} \right).
\end{aligned}
\end{equation}
For the first term,
\begin{equation*}
\begin{aligned}
    & \frac{1}{2 n_0 p} \sum_{l=1}^m 
    \norm{X^{(\mathcal{S},l)} \Delta^{(\mathcal{S},l)}_{\cdot,i}}_2^2 \\
    & = \frac{1}{2 n_0 p} \sum_{l=1}^m \Delta^{(\mathcal{S},l)\top}_{\cdot,i} X^{(\mathcal{S},l)\top} X^{(\mathcal{S},l)} \Delta^{(\mathcal{S},l)}_{\cdot,i} \\
    & = \sum_{l=1}^m \frac{2 n_l p}{2 n_0 p} \left[ \Delta^{(\mathcal{S},l)\top}_{\cdot,i} \Sigma^{(\mathcal{S},l)} \Delta^{(\mathcal{S},l)}_{\cdot,i}
    + \Delta^{(\mathcal{S},l)\top}_{\cdot,i} \left\{ \frac{X^{(\mathcal{S},l)\top} X^{(\mathcal{S},l)}}{2 n_l p} - \Sigma^{(\mathcal{S},l)} \right\} \Delta^{(\mathcal{S},l)}_{\cdot,i} \right] \\
    & \geq \frac{1}{\kappa_3} \sum_{l=1}^m \norm{\Delta^{(\mathcal{S},l)}_{\cdot,i}}_2^2 
    - \frac{\kappa_1}{\kappa_3} \sum_{l=1}^m \Delta^{(\mathcal{S},l)\top}_{\cdot,i} \left\{ \frac{X^{(\mathcal{S},l)\top} X^{(\mathcal{S},l)}}{2 n_l p} - \Sigma^{(\mathcal{S},l)} \right\} \Delta^{(\mathcal{S},l)}_{\cdot,i}
\end{aligned}    
\end{equation*}
where the last inequality resorts to \cref{assmp:balanced_sample,assmp:eigenvalues}. To provide a further lower-bound, we use the fact based on \cref{lem:l_inf_XX} that
\begin{equation*}
    \frac{1}{\kappa_3} \sum_{l=1}^m \norm{\Delta^{(\mathcal{S},l)}_{\cdot,i}}_2^2 
    \geq C(\kappa_3) \sum_{j:j \neq i} \norm{\underline\Delta^{(\mathcal{S},\cdot)}_{ji}}_2^2
\end{equation*}
and 
\begin{equation} \label{eq:basic_ineq_left_first}
\begin{aligned}
    & \sum_{l=1}^m \Delta^{(\mathcal{S},l)\top}_{\cdot,i} \left\{ \frac{X^{(\mathcal{S},l)\top} X^{(\mathcal{S},l)}}{2 n_l p} - \Sigma^{(\mathcal{S},l)} \right\} \Delta^{(\mathcal{S},l)}_{\cdot,i} \\
    & \leq \sum_{l=1}^m \sum_{j_1,j_2} \Delta^{(\mathcal{S},l)}_{j_1 i} \left\{ \frac{X^{(\mathcal{S},l)\top}_{\cdot,j_1} X^{(\mathcal{S},l)}_{\cdot,j_2}}{2 n_l p} - \Sigma^{(\mathcal{S},l)}_{j_1 j_2} \right\} \Delta^{(\mathcal{S},l)}_{j_2 i} \\
    & \leq \sum_{j_1,j_2} \norm*{\Delta^{(\mathcal{S},\cdot)}_{j_1i}}_2 
    \max_l \abs*{ \frac{X^{(\mathcal{S},l)\top}_{\cdot,j_1} X^{(\mathcal{S},l)}_{\cdot,j_2}}{2 n_l p} - \Sigma^{(\mathcal{S},l)}_{j_1 j_2} }  \norm*{\Delta^{(\mathcal{S},\cdot)}_{j_2i}}_2 \\
    & \leq \max_l \abs*{ \frac{X^{(\mathcal{S},l)\top}_{\cdot,j_1} X^{(\mathcal{S},l)}_{\cdot,j_2}}{2 n_l p} - \Sigma^{(\mathcal{S},l)}_{j_1 j_2} } 
    \left( \sum_{j} \norm*{\Delta^{(\mathcal{S},\cdot)}_{ji}}_2 \right)^2 \\
    & \leq C(\kappa_3) \sqrt{\frac{\log(mn_0pq)}{n_l p}} \left( \sum_{j:j \neq i} \norm*{\underline{\Delta}^{(\mathcal{S},\cdot)}_{ji}}_2\right)^2
\end{aligned}    
\end{equation}
uniformly over $i=1,\dots,q$ with probability at least $1 - (mn_0pq)^{-1/2}$.
For the second term,
\begin{equation} \label{eq:basic_ineq_left_second}
\begin{aligned}
    & \frac{1}{n_0 p} \abs*{\sum_{l=1}^m (X^{(\mathcal{S},l)}\Delta^{(\mathcal{S},l)}_{\cdot,i})^\top  \epsilon^{(\mathcal{S},l)}_{\cdot,i}} 
    = \frac{1}{n_0 p} \sum_{j:j \neq i} \abs*{\sum_{l=1}^m (\epsilon^{(\mathcal{S},l)\top}_{\cdot,i} X^{(\mathcal{S},l)}_{\cdot,j}) \Delta^{(\mathcal{S},l)}_{ji} } \\
    & \leq \frac{1}{n_0 p} \sum_{j:j \neq i} \norm*{\left\{\frac{\sqrt{n_l p}}{\norm{X^{(\mathcal{S},l)}_{\cdot,j}}_2} \epsilon^{(\mathcal{S},l)\top}_{\cdot,i} X^{(\mathcal{S},l)}_{\cdot,j}\right\}_{l=1,\dots,m}}_2
    \norm*{ \underline{\Delta}_{ji}^{(\mathcal{S},\cdot)}}_2 
    \\
    & \leq \frac{1}{n_0 p} \max_{j:j \neq i} \norm*{\left\{\frac{\sqrt{n_l p}}{\norm{X^{(\mathcal{S},l)}_{\cdot,j}}_2} \epsilon^{(\mathcal{S},l)\top}_{\cdot,i} X^{(\mathcal{S},l)}_{\cdot,j}\right\}_{l=1,\dots,m}}_2
    ~ \sum_{j:j \neq i} \norm*{ \underline{\Delta}_{ji}^{(\mathcal{S},\cdot)}}_2
     \\
    & \leq C(\kappa_1, \kappa_3) ~ \sqrt{\frac{m + \log(mn_0pq)}{n_0 p}} 
    ~ \sum_{j:j \neq i} \norm*{ \underline{\Delta}_{ji}^{(\mathcal{S},\cdot)}}_2
\end{aligned}
\end{equation}
uniformly over $i=1,\dots,q$ with probability at least $1 - (mn_0pq)^{-1/2}$ where the first inequality is due to the Cauchy-Schwarz inequality, and the last inequality is due \cref{lem:l_2_EX}. Suppose that $\gamma_i \asymp \sqrt{\frac{m + \log(mn_0pq)}{n_0 p}}$ satisfies
\begin{equation*}
\begin{aligned}
    & \frac{1}{n_0 p} \abs*{\sum_{l=1}^m (X^{(\mathcal{S},l)}\Delta^{(\mathcal{S},l)}_{\cdot,i})^\top  \epsilon^{(\mathcal{S},l)}_{\cdot,i}} 
    \leq \frac{\gamma_i}{2} \sum_{j:j \neq i} \norm*{ \underline{\Delta}_{ji}^{(\mathcal{S},\cdot)}}_2.
\end{aligned}
\end{equation*}
For the right-hand side of \cref{eq:basic_ineq},
\begin{equation} \label{eq:basic_ineq_right}
\begin{aligned}
    & \gamma_i \sum_{j:j \neq i} \left( 
    \norm*{  \underline{\beta}^{(\mathcal{S},\cdot)}_{ji}}_2
    - \norm*{ \hat{\underline\beta}^{(\mathcal{S},\cdot)}_{ji}}_2 \right)
    \leq \gamma_i \bigg( \sum_{j: \beta^{(\mathcal{S},l)}_{ji} \neq 0}  
    \norm*{ \underline{\Delta}^{(\mathcal{S},\cdot)}_{ji}}_2
    - \sum_{j: \beta^{(\mathcal{S},l)}_{ji} = 0} \norm*{  \underline\Delta^{(\mathcal{S},\cdot)}_{ji}}_2 \bigg).
\end{aligned}
\end{equation}
On the one hand, because $\frac{1}{2 n_0 p} \sum_{l=1}^m \norm{X^{(\mathcal{S},l)} \Delta^{(\mathcal{S},l)}_{\cdot,i}}_2^2 \geq 0 $ a.s., 
% \cref{eq:basic_ineq,eq:basic_ineq_left,eq:basic_ineq_left_second,eq:basic_ineq_right} earn
\begin{equation} \label{eq:basic_ineq_one}
\begin{aligned}
    % & \frac{1}{\kappa_3} \sum_{l=1}^m \norm{\Delta^{(\mathcal{S},l)}_{\cdot,i}}_2^2
    % - C(\kappa_3, \nu) \sqrt{\frac{\log(mq)}{n_l p}} \left( \sum_{j:j \neq i} \norm*{\left\{\frac{\norm{X^{(\mathcal{S},l)}_{\cdot,j}}_2}{\sqrt{n_l p}} \Delta^{(\mathcal{S},l)}_{ji}\right\}_{l=1,\dots,m}}_2 \right)^2 \\
    & - \frac{\gamma_i}{2} 
    ~ \sum_{j:j \neq i} \norm*{\underline\Delta_{ji}^{(\mathcal{S},\cdot)}}_2 
    \overset{\eqref{eq:basic_ineq_left_second}}{\leq} - \frac{1}{n_0 p} \abs*{\sum_{l=1}^m (X^{(\mathcal{S},l)}\Delta^{(\mathcal{S},l)}_{\cdot,i})^\top  \epsilon^{(\mathcal{S},l)}_{\cdot,i}} \\
    & \leq \frac{1}{2 n_0 p} \sum_{l=1}^m \left( 
    \norm{X^{(\mathcal{S},l)} \Delta^{(\mathcal{S},l)}_{\cdot,i}}_2^2 
    - 2 (X^{(\mathcal{S},l)}\Delta^{(\mathcal{S},l)}_{\cdot,i})^\top \epsilon^{(\mathcal{S},l)}_{\cdot,i} \right) \\
    & \overset{\eqref{eq:basic_ineq_left}}{=} \frac{1}{2 n_0 p} \sum_{l=1}^m \left( 
    \norm{X^{(\mathcal{S},l)}_{\cdot,i} - X^{(\mathcal{S},l)} \hat{\beta}^{(\mathcal{S},l)}_{\cdot,i}}_2^2
    - \norm{X^{(\mathcal{S},l)}_{\cdot,i} - X^{(\mathcal{S},l)} \beta^{(\mathcal{S},l)}_{\cdot,i}}_2^2 \right) \\
    & \overset{\eqref{eq:basic_ineq}}{\leq} \gamma_i \sum_{j:j \neq i} \left( 
    \norm*{  \underline{\beta}^{(\mathcal{S},\cdot)}_{ji}}_2
    - \norm*{ \hat{\underline\beta}^{(\mathcal{S},\cdot)}_{ji}}_2 \right)
    \overset{\eqref{eq:basic_ineq_right}}{\leq} \gamma_i \bigg( \sum_{j: \beta^{(\mathcal{S},l)}_{ji} \neq 0}  
    \norm*{ \underline{\Delta}^{(\mathcal{S},\cdot)}_{ji}}_2
    - \sum_{j: \beta^{(\mathcal{S},l)}_{ji} = 0} \norm*{\underline\Delta^{(\mathcal{S},\cdot)}_{ji}}_2 \bigg)
\end{aligned}
\end{equation}
uniformly over $i=1,\dots,q$ with probability at least $1 - 2(mn_0pq)^{-1/2}$. 
% Due to \cref{assmp:spatial_sample}, $\sqrt{\frac{m + \log(mn_0pq)}{n_0 p}} \rightarrow 0$ as $n_0 \rightarrow \infty$, and t
Therefore
\begin{equation} \label{eq:delta_zero_vs_nonzero}
\begin{aligned}
    \sum_{j: \beta^{(\mathcal{S},l)}_{ji} = 0} \norm*{\underline\Delta^{(\mathcal{S},\cdot)}_{ji}}_2
    \leq 2 \sum_{j: \beta^{(\mathcal{S},l)}_{ji} \neq 0} \norm*{\underline\Delta^{(\mathcal{S},\cdot)}_{ji}}_2
\end{aligned}
\end{equation}
uniformly over $i=1,\dots,q$ with the same probability. %as long as $\min_i \gamma_i \geq C(\kappa_1,\kappa_3) \sqrt{\frac{m + \log(mn_0pq)}{n_0 p}} $ for some large enough $C(\kappa_1,\kappa_3)$. 
On the other hand, \crefrange{eq:basic_ineq_left}{eq:basic_ineq_right} earn
\begin{equation} \label{eq:basic_ineq_other}
\begin{aligned}
    & \gamma_i \bigg( \sum_{j: \beta^{(\mathcal{S},l)}_{ji} \neq 0}  
    \norm*{ \underline{\Delta}^{(\mathcal{S},\cdot)}_{ji}}_2
    - \sum_{j: \beta^{(\mathcal{S},l)}_{ji} = 0}  \norm*{\underline\Delta^{(\mathcal{S},\cdot)}_{ji}}_2 \bigg) \\
    & \overset{\eqref{eq:basic_ineq_right}}{\geq} \gamma_i \sum_{j:j \neq i} \left( 
    \norm*{  \underline{\beta}^{(\mathcal{S},\cdot)}_{ji}}_2
    - \norm*{ \hat{\underline\beta}^{(\mathcal{S},\cdot)}_{ji}}_2 \right) \\
    & \overset{\eqref{eq:basic_ineq}}{\geq} \frac{1}{2 n_0 p} \sum_{l=1}^m \left( 
    \norm{X^{(\mathcal{S},l)}_{\cdot,i} - X^{(\mathcal{S},l)} \hat{\beta}^{(\mathcal{S},l)}_{\cdot,i}}_2^2
    - \norm{X^{(\mathcal{S},l)}_{\cdot,i} - X^{(\mathcal{S},l)} \beta^{(\mathcal{S},l)}_{\cdot,i}}_2^2 \right) \\
    & \overset{\eqref{eq:basic_ineq_left}}{=} \frac{1}{2 n_0 p} \sum_{l=1}^m \left( 
    \norm{X^{(\mathcal{S},l)} \Delta^{(\mathcal{S},l)}_{\cdot,i}}_2^2 
    - 2 (X^{(\mathcal{S},l)}\Delta^{(\mathcal{S},l)}_{\cdot,i})^\top \epsilon^{(\mathcal{S},l)}_{\cdot,i} \right) \\
    & \overset{\eqref{eq:basic_ineq_left_first},\eqref{eq:basic_ineq_left_second}}{\geq} C(\kappa_3) \sum_{j:j \neq i} \norm{\underline\Delta^{(\mathcal{S},\cdot)}_{ji}}_2^2
    - C(\kappa_1,\kappa_3) \sqrt{\frac{\log(mn_0pq)}{n_l p}} \bigg( \sum_{j:j \neq i} \norm*{ \underline\Delta^{(\mathcal{S},\cdot)}_{ji}}_2 \bigg)^2 \\
    & \quad \quad - \gamma_i \norm*{\underline\Delta_{ji}^{(\mathcal{S},\cdot)}}_2 \\
\end{aligned}
\end{equation}
uniformly over $i=1,\dots,d$ with probability at least $1 - 2(mn_0pq)^{-1/2}$. Because
\begin{equation*}
\begin{aligned}
    \bigg( \sum_{j:j \neq i} \norm*{ \underline\Delta^{(\mathcal{S},\cdot)}_{ji}}_2 \bigg)^2
    & \overset{\eqref{eq:delta_zero_vs_nonzero}}{\leq} 2
    \bigg( \sum_{j: \beta^{(\mathcal{S},l)}_{ji} \neq 0} \norm*{\underline\Delta^{(\mathcal{S},\cdot)}_{ji}}_2 \bigg)^2 
    \leq 2 ~d \sum_{j: \beta^{(\mathcal{S},l)}_{ji} \neq 0} \norm*{\underline\Delta^{(\mathcal{S},\cdot)}_{ji}}_2^2  
\end{aligned}
\end{equation*}
with probability at least $1 - 2(mn_0pq)^{-1/2}$,
%, and $d \sqrt{\frac{\log(mn_0pq)}{n_o p}} \rightarrow 0$ as $n_0 \rightarrow \infty$ due to \cref{assmp:spatial_sample}, then 
\cref{eq:basic_ineq_other} implies that
\begin{equation*}
\begin{aligned}
    & \gamma_i \bigg( \sum_{j: \beta^{(\mathcal{S},l)}_{ji} \neq 0}  
    \norm*{ \underline{\Delta}^{(\mathcal{S},\cdot)}_{ji}}_2
    - \sum_{j: \beta^{(\mathcal{S},l)}_{ji} = 0}  \norm*{\underline\Delta^{(\mathcal{S},\cdot)}_{ji}}_2 \bigg) \\
    & \geq \left(C(\kappa_3) - C(\kappa_1,\kappa_3) d \sqrt{\frac{\log(mn_0pq)}{n_l p}}\right) \sum_{j:j \neq i} \norm*{\underline\Delta^{(\mathcal{S},\cdot)}_{ji}}_2^2
    - \gamma_i \norm*{\underline\Delta_{ji}^{(\mathcal{S},\cdot)}}_2 \\
\end{aligned}
\end{equation*}
uniformly over $i=1,\dots,d$ with the same probability. As a result, with a sufficiently large $n_0$ and the same probability, because $d \sqrt{\frac{\log(mn_0pq)}{n_o p}} \rightarrow 0$ as $n_0 \rightarrow \infty$ due to \cref{assmp:spatial_sample},
\begin{equation*}
\begin{aligned}
    & C(\kappa_3) \sum_{j:j \neq i} \norm*{\underline\Delta^{(\mathcal{S},\cdot)}_{ji}}_2^2 \\
    & \leq \gamma_i \bigg( \sum_{j: \beta^{(\mathcal{S},l)}_{ji} \neq 0}  
    \norm*{ \underline{\Delta}^{(\mathcal{S},\cdot)}_{ji}}_2
    - \sum_{j: \beta^{(\mathcal{S},l)}_{ji} = 0}  \norm*{\underline\Delta^{(\mathcal{S},\cdot)}_{ji}}_2 \bigg) 
    + \gamma_i
    ~ \sum_{j:j \neq i} \norm*{\underline\Delta_{ji}^{(\mathcal{S},\cdot)}}_2 \\
    & \leq C(\kappa_1, \kappa_3) ~ \sqrt{\frac{m + \log(mn_0pq)}{n_0 p}}
    ~ \sum_{j:j \neq i} \norm*{\underline\Delta_{ji}^{(\mathcal{S},\cdot)}}_2 \\
    & \leq C(\kappa_1, \kappa_3) ~ \sqrt{d \frac{m + \log(mn_0pq)}{n_0 p}} 
    ~ \sqrt{\sum_{j:j \neq i} \norm*{\underline\Delta_{ji}^{(\mathcal{S},\cdot)}}_2^2},
\end{aligned}
\end{equation*}
uniformly over $i=1,\dots,d$, and
\begin{equation*}
\begin{aligned}
    & \max_i \sum_{j:j \neq i} \norm*{\underline\Delta^{(\mathcal{S},\cdot)}_{ji}}_2^2 
    \leq C(\kappa_1, \kappa_3) ~d ~\frac{m + \log(mn_0pq)}{n_0 p}.
\end{aligned}
\end{equation*}
In addition, due to \cref{eq:delta_zero_vs_nonzero}, under the same event,
\begin{equation*}
\begin{aligned}
    & \max_i \sum_{j:j \neq i} \norm*{\underline\Delta^{(\mathcal{S},\cdot)}_{ji}}_2 
    \leq C \max_i \sum_{j:\beta_{ji} \neq 0} \norm*{\underline\Delta^{(\mathcal{S},\cdot)}_{ji}}_2 \\
    & \leq C \sqrt{d} \max_i \sqrt{\sum_{j:j \neq i} \norm*{\underline\Delta_{ji}^{(\mathcal{S},\cdot)}}_2^2} 
    \leq C(\kappa_1,\kappa_3) ~d ~\sqrt{\frac{m + \log(mn_0pq)}{n_0 p}}.
\end{aligned}
\end{equation*}
Last, it follows \cref{eq:basic_ineq,eq:basic_ineq_left} that
\begin{equation*}
\begin{aligned}
    & \max_i \frac{1}{2 n_0 p} \sum_{l=1}^m \norm*{X^{(\mathcal{S},l)}\Delta^{(\mathcal{S},l))}_{\cdot,i}}_2^2 \\
    & \leq \max_i \left[ 
    \gamma_i \sum_{j:j \neq i} \left( 
    \norm*{ \underline{\beta}^{(\mathcal{S},\cdot)}_{ji}}_2
    - \norm*{ \hat{\underline\beta}^{(\mathcal{S},\cdot)}_{ji}}_2 \right)
    + \frac{1}{n_0 p} \sum_{l=1}^m  (X^{(\mathcal{S},l)}\Delta^{(\mathcal{S},l)}_{\cdot,i})^\top \epsilon^{(\mathcal{S},l)}_{\cdot,i} \right] \\
    & \leq \max_i \left[ 
    \gamma_i \sum_{j: j \neq i}  
    \norm*{ \underline{\Delta}^{(\mathcal{S},\cdot)}_{ji}}_2
    + C(\kappa_1, \kappa_3) ~ \sqrt{\frac{m + \log(mn_0pq)}{n_0 p}} 
    ~ \sum_{j:j \neq i} \norm*{ \underline{\Delta}_{ji}^{(\mathcal{S},\cdot)}}_2
    \right] \\
    & \leq C(\kappa_1,\kappa_3) ~d ~\frac{m + \log(mn_0pq)}{n_0 p}
\end{aligned}
\end{equation*}
under the same event.

%%%%%%%%%%%%%%%%%%%%%%%%%%%%%%%%%%%%%%%%%%%%%%%%%%%%%%%
%%%%%%%%%%%%%%%%%%%%%%%%%%%%%%%%%%%%%%%%%%%%%%%%%%%%%%%
%%%%%%%%%%%%%%%%%%%%%%%%%%%%%%%%%%%%%%%%%%%%%%%%%%%%%%%
%%%%%%%%%%%%%%%%%%%%%%%%%%%%%%%%%%%%%%%%%%%%%%%%%%%%%%%

\subsection{Proof of Proposition~\ref{thm:inf_CLT}} \label{app:pf_inf_CLT}

Let $O^{(\mathcal{S},l)}_{ij}$ be the error of $\hat\rho_{ij}^{(\mathcal{S}, l)}$ not governed by $\Theta^{(\mathcal{S},l)}_{ij}$. 
That is, $O^{(\mathcal{S},l)}_{ij} := \hat{\rho}^{(\mathcal{S},l)}_{ij} - \rho^{(\mathcal{S},l)}_{ij} - \Theta^{(\mathcal{S},l)}_{ij}$. In the following lemma, we prove that $O^{(\mathcal{S},l)}_{ij}$ vanishes asymptotically. We provide the proof in \cref{app:pf_spatial_est_rho}.

\begin{lemma} \label{thm:spatial_est_rho}
Suppose that $\hat{\rho}^{(\mathcal{S},l)}$ is estimated based on $\gamma_i$'s given as in \cref{thm:spatial_regression}. Then, under \crefrange{assmp:balanced_sample}{assmp:temporal_sample},
\begin{equation*}
    \Pr\left[\max_{i,j: i \neq j} \sum_{l=1}^m  \abs*{O^{(\mathcal{S},l)}_{ij}} \geq C(\kappa_1,\kappa_3) ~d ~\frac{m + \log(mn_0pq)}{n_0 p} \right] \leq C(mn_0pq)^{-1/2},
\end{equation*}
for a sufficiently large $n_0$.
\end{lemma}

% It then immediately follows that $ \Delta \mat{P}_{\mathcal{S}} = \mat{P}_{n,d,\mathcal{S}}(\mat{\xi})-\mat{P}^*_{n,d,\mathcal{S}}(\mat{\xi}) =  \frac{1}{\sqrt{d}}\sum_{t=1}^d  \sqrt{n_tp } (\mat{\xi}_{t\mathcal{S}} \circ \mat{\Theta}_{t\mathcal{S}} + \mat{\xi}_{t\mathcal{S}} \circ \mat{O}_{t\mathcal{S}})$,
%with $\|\sum_{t=1}^d \frac{ \sqrt{n_t p}}{\sqrt{d}}\mat{\xi}_{t\mathcal{S}} \circ \mat{O}_{t\mathcal{S}}\|_{\infty}$ being a smaller term as shown in Proposition~\ref{prop:partial_corr}.

% \begin{remark} \label{remark:sample_size_requirement}
% % For single edge test, we need guarantee for $\sum_{l=1}^m \frac{\sqrt{n_l p}}{\sqrt{m}} O^{(\mathcal{S},l)}_{ij}$ to converge to $0$ in probability, which requires the condition that $\frac{d^2 (m + \log q)^2}{m} = o(n_0p)$; 
% For the simultaneous edge test,
% in order to make $\norm{\sum_{l=1}^m \frac{ \sqrt{n_l p}}{\sqrt{m}} O^{(\mathcal{S},l)}_E}_{\infty}$ converging to $0$ in probability, the sample size requirement is $n_0p = \omega(\frac{d^2 (m + \log(mn_0pq))^2}{m})$ as $n_0 \rightarrow \infty$. In comparison, one can also naively apply the estimation procedure for each graph separately as in \cite{chen2019}, and compute a similar test statistic to perform single edge and multiple edge tests following our procedure. However, such a naive method will require a much stronger sample size assumption, which is $n_0p = \omega(d^2 m \log^2(mn_0pq))$.
% \end{remark}

Let $\Theta_{E} = \frac{1}{\sqrt{m}} \sum_{l=1}^m  \sqrt{n_l p} \Theta^{(\mathcal{S},l)}_{E}$, and $O_E = \frac{1}{\sqrt{m}} \sum_{l=1}^m  \sqrt{n_l p} O^{(\mathcal{S},l)}_{E}$, where $\Theta^{(\mathcal{S},l)}_E$ and $O^{(\mathcal{S},l)}_E$ are the collections of $\Theta^{(\mathcal{S},l)}_{ij}$ and $O^{(\mathcal{S},l)}_{ij}$ for $(i,j) \in E$, respectively. For any $x>0$ and $\delta>0$,
\begin{equation*}
\begin{aligned}
    & \Pr[\norm{\hat{T}_E- T_E}_\infty > x] 
    \leq \Pr[\norm{\Theta_E}_\infty > x - \delta]
    + \Pr[\norm{O_E}_\infty > \delta] \\
    & \leq \Pr[\norm{Z}_\infty > x - \delta]
    + KS_\infty(\Theta_E, Z)
    + \Pr[\norm{O_E}_\infty > \delta] \\
    & = \Pr[\norm{Z}_\infty > x]+ \Pr[x \geq \norm{Z}_\infty > x - \delta]+ KS_\infty(\Theta_E, Z) + \Pr[\norm{O_E}_\infty > \delta],
\end{aligned}
\end{equation*}
where $KS_\infty(\Theta_E, Z) := \sup_{x>0} \abs{ \Pr[\norm{\Theta_E}_\infty>x] - \Pr[ \norm{Z}_\infty > x ] }$ is the Kolmogorov-Smirnov distance between $\norm{\Theta_E}_\infty$ and $\norm{Z}_\infty$. Similarly,
\begin{equation*}
\begin{aligned}
    &\Pr[\norm{\hat{T}_E - T_E}_\infty > x] 
    \geq \Pr[\norm{\Theta_E}_\infty > x + \delta] 
    - \Pr[\norm{O_E}_\infty > \delta] \\
    &\geq \Pr[\norm{Z}_\infty > x + \delta]
    - KS_\infty(\Theta_E, Z)
    - \Pr[\norm{O_E}_\infty > \delta]\\
    & = \Pr[\norm{Z}_\infty > x]
    - \Pr[x + \delta \geq \norm{Z}_\infty > x] 
    - KS_\infty(\Theta_E, Z)
    - \Pr[\norm{O_E}_\infty >\delta].
\end{aligned}
\end{equation*}
Therefore we conclude that 
\begin{equation*}
\begin{aligned}
    & \sup_{x>0} \abs*{ \Pr[\norm{\hat{T}_E - T_E}_\infty > x] 
        - \Pr[\norm{Z}_\infty >x]} \\
    & \leq KS_\infty(\Theta_E,Z) 
    + \sup_{x>0} \Pr[x+\delta \geq \norm{Z}_\infty \geq x-\delta]
    + \Pr[\norm{O_E}_\infty > \delta]. 
\end{aligned}
\end{equation*}
We use the anti-concentration inequality of the supremum norm of Gaussian random vectors with mean zero (Corollary 1, \citealp{Chernozhukov2015}) and the fact that
\begin{equation*} 
    S_{(i,j),(i,j)} = \frac{1}{m}\sum_{l=1}^m \frac{\norm{\Sigma^{(\mathcal{T},l)}}^2_F}{p} (1 - \rho^{(\mathcal{S},l)2}_{ij})^2 \geq C(\kappa_3)
\end{equation*} to control the second term:
\begin{equation} \label{eq:KS_decomp_second}
    \sup_{x>0} \Pr[x + \delta \geq \norm{Z}_\infty \geq x - \delta] \leq C(\kappa_3) \delta \sqrt{\log \abs{E}}.
\end{equation}
Setting $\delta = C(\kappa_1, \kappa_3) d \frac{m + \log (qmn_0 p)}{\sqrt{m n_0 p}}$, which converges to $0$ as $n \rightarrow \infty$ due to \cref{assmp:spatial_sample}, \cref{thm:spatial_est_rho} provides an upperbound over the size of the last term.

% and further notice that $\lambda_* = 
% O(\sqrt{\frac{d+\log q}{n_0 p}}) $, we have $\lVert \mat{O}^P_\mathcal{S} \rVert_{\infty} = O(\sqrt{\frac{n_0 p}{d}} s \lambda_*^2)$; by Assumption~\ref{as:sparse}
% ,  we have $\lVert \mat{O}^P_\mathcal{S} \rVert_{\infty} = o(\frac{1}{\sqrt{\log q}})$. Therefore, we can find $\delta = o(\frac{1}{\sqrt{\log q} })$, such that both $\sup_{x>0} \mathbb{P} ( x-\delta<\lVert \mat{\zeta} \rVert_{\infty} \leq x+\delta)$ and $\mathbb{P} (\lVert \mat{O}^P_\mathcal{S} \rVert_{\infty} >\delta)$ goes to 0. 

Now it suffices to derive a probabilistic upperbound for $KS_\infty(\Theta_E, Z)$. We recall that each element of $\Theta_E$ is
\begin{equation*}
\begin{aligned}
    \Theta_{ij} = & \frac{1}{\sqrt{m}} \sum_{l=1}^m \sqrt{n_l p} \Theta^{(\mathcal{S},l)}_{ij} \\
    = & \frac{1}{\sqrt{m}} \sum_{l=1}^m \sqrt{n_l p} 
    \left( \frac{\tilde\phi^{(\mathcal{S},l)}_{ij}}{\sqrt{\Phi^{(\mathcal{S},l)}_{ii} \Phi^{(\mathcal{S},l)}_{jj}}} 
    - \frac{\Phi^{(\mathcal{S},l)}_{ij} \tilde\phi^{(\mathcal{S},l)}_{jj}}{2 \Phi^{(\mathcal{S},l)}_{jj} \sqrt{\Phi^{(\mathcal{S},l)}_{ii} \Phi^{(\mathcal{S},l)}_{jj}}}
    - \frac{\Phi^{(\mathcal{S},l)}_{ij} \tilde\phi^{(\mathcal{S},l)}_{ii}}{2 \Phi^{(\mathcal{S},l)}_{ii} \sqrt{\Phi^{(\mathcal{S},l)}_{ii} \Phi^{(\mathcal{S},l)}_{jj}}} \right),
\end{aligned}
\end{equation*}
where $\tilde\phi^{(\mathcal{S},l)}_{ij} := \frac{1}{n_l p} \sum_{k=1}^{n_l} \sum_{t=1}^p \epsilon^{(\mathcal{S},k,l)\top}_{ti} \epsilon^{(\mathcal{S},k,l)}_{tj} - \Phi^{(\mathcal{S},l)}_{ij}$. We notice that $\epsilon^{(\mathcal{S},k,l)}_{ti}\epsilon^{(\mathcal{S},k,l)}_{tj}$ are not independent across $t = 1, \dots, p$ due to the temporal association, while the state-of-the-art theory of high-dimensional CLT elaborates only over independent random vectors yet. We use \cref{eq:iid_EX_EE} to convert $\phi^{(\mathcal{S},l)}$ into a summation of independent random variables:
\begin{equation*}
    \tilde\phi^{(\mathcal{S},l)}_{ij} 
    = \frac{1}{n_l p} \sum_{k=1}^{n_l} \sum_{t=1}^p \lambda^{(\mathcal{T},l)}_t \xi^{(k,l)}_{ti} \xi^{(k,l)}_{tj} 
    - \Phi^{(\mathcal{S},l)}_{ij} 
    = \frac{1}{n_l p} \sum_{k=1}^{n_l} \sum_{t=1}^p \lambda^{(\mathcal{T},l)}_t \left(\xi^{(k,l)}_{ti} \xi^{(k,l)}_{tj} - \Phi^{(\mathcal{S},l)}_{ij} \right),  
\end{equation*}
where $\xi^{(k,l)}_t \distiid \distNorm(0,\Phi^{(\mathcal{S},l)})$, and the last equation holds based on $\tr(\Sigma^{(\mathcal{T},l)}) = 1$.
Then, we rewrite 
\begin{equation*}
\begin{aligned}
    \Theta_{ij} = \frac{1}{\sqrt{N p}} \sum_{l=1}^m \sum_{k=1}^{n_l} 
    \sum_{t=1}^p \theta^{(k,l)}_{t,ij},
\end{aligned}
\end{equation*}
where $N = \sum_{l=1}^m n_l$, and
\begin{equation*}
\begin{aligned}
    \theta^{(k,l)}_{t,ij} := &
    \lambda^{(\mathcal{T},l)}_t \sqrt{\frac{N}{n_l m}} \left(
    \frac{\xi^{(k,l)}_{t,i} \xi^{(k,l)}_{t,j} 
        - \Phi^{(\mathcal{S},l)}_{ij}}
    {\sqrt{\Phi^{(\mathcal{S},l)}_{ii} \Phi^{(\mathcal{S},l)}_{jj}}} 
    - \frac{\Phi^{(\mathcal{S},l)}_{ij}
        (\xi^{(k,l)2}_{t,j} - \Phi^{(\mathcal{S},l)}_{jj})}
    {2\Phi^{(\mathcal{S},l)}_{jj} 
    \sqrt{\Phi^{(\mathcal{S},l)}_{ii} \Phi^{(\mathcal{S},l)}_{jj}}}
    - \frac{\Phi^{(\mathcal{S},l)}_{ij}
        (\xi^{(k,l)2}_{t,i}  - \Phi^{(\mathcal{S},l)}_{jj})}
    {2\Phi^{(\mathcal{S},l)}_{ii} 
    \sqrt{\Phi^{(\mathcal{S},l)}_{ii} \Phi^{(\mathcal{S},l)}_{jj}}}
    \right).
\end{aligned}
\end{equation*}

% \begin{align*}
%   \Theta^P_{\mathcal{S}}(i)
%   & = \frac{1}{\sqrt{Np}}\sum_{t=1}^d   \sum_{k=1}^{n_t} \sum_{l=1}^p \theta^{(k)}_{{\mathcal{S}},tl}(i),
% \end{align*}
% with 
% \begin{align*}
%   \theta^{(k)}_{{\mathcal{S}},tl}(i)
%   & = \frac{\lambda_{tl}\xi_{t,\chi_1{(i)},\chi_2{(i)} } \sqrt{N}}{\sqrt{n_t d}}   ( \frac{\epsilon^{(k)}_{tl\chi_1(i)} \epsilon^{(k)}_{tl\chi_2(i)} 
%   -r_{t,\chi_1{(i)},\chi_2{(i)}}  } {   \sqrt{r_{t,\chi_1(i),\chi_1(i)} r_{t,\chi_2(i),\chi_2(i)}} }\\
%   &- \frac{r_{t,\chi_1{(i)}, \chi_2{(i)}}}{2 r_{t, \chi_1{(i)}, \chi_1{(i)} }}\frac{({\epsilon}^{(k)}_{tl\chi_1(i)})^2  - r_{t,\chi_1{(i)}, \chi_1{(i)} } } {   \sqrt{r_{t,\chi_1(i),\chi_1(i)} r_{t,\chi_2(i),\chi_2(i)}} } -  \frac{r_{t,\chi_1{(i)}, \chi_2{(i)}}}{2 r_{t, \chi_2{(i)}, \chi_2{(i)} }}\frac{({\epsilon}^{(k)}_{tl\chi_2(i)})^2  - r^l_{t,\chi_2{(i)}, \chi_2{(i)} } } {   \sqrt{r_{t,\chi_1(i),\chi_1(i)} r_{t,\chi_2(i),\chi_2(i)}} }).
% \end{align*}
For fixed $i$ and $j$, because $\xi^{(k,l)}_{ti}$ and $\xi^{(k,l)}_{tj}$ are Gaussian random variables independent across $l$, $k$, and $t$, $\theta^{(k,l)}_{t,ij}$'s are independent sub-exponential random variable across $l$, $k$, and $t$. Due to \cref{assmp:eigenvalues}, $\theta^{(k,l)}_{t,ij}$'s has uniformly bounded $\psi_1$-Orlicz norms (see \cref{app:pf_sum_phi2} for the definition of the Orlicz norm). That is,
\begin{equation*}
    \max_{k,l,t,(i,j) \in E} \norm*{\theta^{(k,l)}_{t,ij}/\sqrt{\Var[\theta^{(k,l)}_{t,ij}]}}_{\psi_1} \leq C(\kappa_1,\kappa_3).
\end{equation*}
This automatically satisfies Condition (M) in \citet{chernozhukov2020nearly}. Although Condition (E.2), requiring bounded $\psi_2$-Orlicz norm, is not satisfied, the proof of Corollary 2.1 under Conditions (M) and (E.2) is easily modified under bounded $\psi_2$-Orlicz norms. Based on the facts that
\begin{equation*}
    \left(\Exp\left[
        \max_{k,l,t,(i,j) \in E} \abs*{\theta^{(k,l)}_{t,ij}/\sqrt{\Var[\theta^{(k,l)}_{t,ij}]}}^4
    \right]\right)^{1/4}
    \leq C(\kappa_1, \kappa_3) \log(\abs{E}mn_0p),
\end{equation*}
\begin{equation*}
    \Pr\left[
        \max_{(i,j) \in E} \norm*{\theta^{(k,l)}_{t,ij}/\sqrt{\Var[\theta^{(k,l)}_{t,ij}]}}_\infty > C B_n \log(dn)
    \right] \leq C,
\end{equation*}
%To obtain a non-asymptotic bound for $\sup_{x > 0} \abs{\Pr[\norm{T_E}_\infty > x] - \Pr[\norm{Z}_\infty > x]}$, we apply Corollary 2.1 in \citet{Chernozhukov2012} to obtain
Theorem 2.2 in \citet{chernozhukov2020nearly} implies
\begin{equation*}
    KS_\infty(\Theta_E, Z) \leq \frac{C(\kappa_1,\kappa_3)}{\sqrt{mn_0p}} \max\left\{     
        (\log \abs{E})^{2} \log(m n_0 p), (\log \abs{E})^{5/2} 
    \right\}.
\end{equation*}
Alongside the results in \cref{thm:spatial_est_rho,eq:KS_decomp_second}, it obtains the desired result.

% According to Corollary 2.1 in \cite{Chernozhukov2012}, our case can be fitted in (E.1): (1)Our random variables are sub-exponential random variables. (2)By Assumption~\ref{as:sparse}, the condition $(\log(n_0 dpq))^7/n_0 dp  = O((n_0pd)^{-c})$ with $c>0$ is satisfied. We conclude that Equation~\eqref{eqn:cherno} holds with $\sup_{x>0} |\mathbb{P} (\lVert \Delta \mat{P}_\mathcal{S} \rVert_{\infty}>x) - \mathbb{P}(\lVert {\mat{\zeta}} \rVert_{\infty}>x)| = O\left((n_0pd)^{-c}\right)$,  and we finish our proof.

%%%%%%%%%%%%%%%%%%%%%%%%%%%%%%%%%%%%%%%%%%%%%%%%%%%%%%%
%%%%%%%%%%%%%%%%%%%%%%%%%%%%%%%%%%%%%%%%%%%%%%%%%%%%%%%
%%%%%%%%%%%%%%%%%%%%%%%%%%%%%%%%%%%%%%%%%%%%%%%%%%%%%%%
%%%%%%%%%%%%%%%%%%%%%%%%%%%%%%%%%%%%%%%%%%%%%%%%%%%%%%%

\subsection{Proof of Theorem~\ref{thm:inf_bootstrap}} \label{app:pf_inf_bootstrap}

We denote the conditional Kolmogorov-Smirnov distance between $\norm{\hat{Z}}_\infty$ and $\norm{Z}_\infty$ given observed data $\mathcal{D}$ by $KS_\infty(\hat{Z},Z|\mathcal{D})$.
\begin{equation*}
    KS_\infty(\hat{Z},Z|\mathcal{D}) := \sup_{x>0} \abs{\Pr[\norm{\hat{Z}}_\infty > x | \mathcal{D}]
    - \Pr[\norm{Z}_\infty >x]}.
\end{equation*}
The error of the probability estimate is upperbounded by
\begin{equation*}
\begin{aligned}
    &\sup_{x>0} \abs{\Pr[\norm{\hat{T}_E}_\infty >x] - \Pr[\norm{\hat{Z}}_\infty > x | \mathcal{D}]} \\
    & \leq \sup_{x>0} \abs{\Pr[\norm{\hat{T}_E}_\infty >x] - \Pr[\norm{Z}_\infty > x]}
    + KS_\infty(\hat{Z},Z|\mathcal{D}).
\end{aligned}
\end{equation*}
\cref{thm:inf_CLT} provides an upperbound for the first term. Therefore, it suffices to derive an upperbound for $KS_\infty(\hat{Z},Z|\mathcal{D})$. According to Lemma 2.1 in \cite{chernozhukov2020nearly} and \cref{assmp:eigenvalues},
\begin{equation} \label{eq:KS_normal}
    KS_\infty(\hat{Z},Z|\mathcal{D})
    \leq C(\kappa_3) \norm{\hat{S}_{EE} - S_{EE}}_{\infty} \log\abs{E} (1 \vee \abs{\log\norm{\hat{S}_{EE} - S_{EE}}_\infty}).
\end{equation}
Therefore, it suffices to upperbound the size of $\norm{\hat{S}_{EE} - S_{EE}}_\infty$. In \cref{eq:inf_S}, $S_{(i_1,j_1),(i_2,j_2)}$ is given by
\begin{equation} \label{eq:S_rev}
\begin{aligned}
    & S_{(i_1,j_1),(i_2,j_2)} \\
    & := \frac{1}{m} \sum_{l=1}^m \frac{\norm{\Sigma^{(\mathcal{T},l)}}_F^2}{p} 
    \left[ \begin{aligned} 
    & \rho^{(\mathcal{S},l)}_{i_1 i_2} \rho^{(\mathcal{S},l)}_{j_1 j_2}
    + \rho^{(\mathcal{S},l)}_{i_1 j_2} \rho^{(\mathcal{S},l)}_{i_2 j_1}
    + \frac{1}{2} \rho^{(\mathcal{S},l)}_{i_1 j_1} \rho^{(\mathcal{S},l)}_{i_2 j_2} 
    \Big( \rho^{(\mathcal{S},l)2}_{i_1 i_2} + \rho^{(\mathcal{S},l)2}_{j_1 j_2}
    + \rho^{(\mathcal{S},l)2}_{i_1 j_2} + \rho^{(\mathcal{S},l)2}_{i_2 j_1} \Big) \\
    & - \rho^{(\mathcal{S},l)}_{i_1 i_2} 
    \rho^{(\mathcal{S},l)}_{i_2 j_2} \rho^{(\mathcal{S},l)}_{i_2 j_1}
    - \rho^{(\mathcal{S},l)}_{i_1 i_2}
    \rho^{(\mathcal{S},l)}_{i_1 j_1} \rho^{(\mathcal{S},l)}_{i_1 j_2}
    - \rho^{(\mathcal{S},l)}_{j_1 j_2} 
    \rho^{(\mathcal{S},l)}_{i_2 j_2} \rho^{(\mathcal{S},l)}_{i_1 j_2}
    - \rho^{(\mathcal{S},l)}_{j_1 j_2} \rho^{(\mathcal{S},l)}_{i_2 j_1} \rho^{(\mathcal{S},l)}_{i_1 j_1}
    \end{aligned} \right].
\end{aligned}
\end{equation}
\cref{thm:temporal_est_F} gives a bound on the error in $\norm{\hat{\Sigma}^{(\mathcal{T},l)}}_F^2$:
\begin{equation*}
    \max_{l=1,\dots,m} \frac{\norm{\hat{\Sigma}^{(\mathcal{T},l)}}_F^2 - \norm{\Sigma^{(\mathcal{T},l)}}_F^2}{p} 
    \leq C(\kappa_1, \kappa_3) \frac{\log(mn_0pq)}{(q n_0)^{1-\frac{1}{2(\alpha_0+1)}}}
\end{equation*}
with probability at least $1 - (mn_0pq)^{-1/2}$. On the other hand, \cref{thm:spatial_est_rho} implies that the error in $\hat{\rho}^{(\mathcal{S},l)}_{ij}$ is mainly driven by $\Theta^{(\mathcal{S},l)}_{ij}$, where the size of $\Theta^{(\mathcal{S},l)}_{ij}$ is bounded by the Berry-Esseen bound for sub-exponential random variables (for example, see \citealp{kuchibhotla2018moving}):
\begin{equation*}
    \max_{i,j} \frac{1}{m} \sum_{l=1}^m \abs{\Theta^{(\mathcal{S},l)}_{ij}} 
    \leq C(\kappa_1,\kappa_3) \sqrt{\frac{m+\log(mn_0pq)}{mn_0p}}
\end{equation*}
with probability at least $1 - (mn_0pq)^{-1/2}$. (Notice that, because of \cref{assmp:spatial_sample}, the above probabilistic bound is larger than $d \frac{m + \log(mn_0pq)}{n_0p}$, which is given as a probabilistic bound for $\max_{i,j} \sum_{l=1}^m \abs{O^{(\mathcal{S},l)}_{ij}}$ in \cref{thm:spatial_est_rho}.) Because \cref{assmp:eigenvalues} implies that both $\frac{\norm{\Sigma^{(\mathcal{T},l)}}_F^2}{p}$ and $\min_{i,j} \rho^{(\mathcal{S},l)}_{ij}$ are bounded away from both $0$ and $\infty$ by some constants dependent to $\kappa_3$,
\begin{equation*}
\begin{aligned}
    \norm{\hat{S}_{EE} - S_{EE}}_\infty
    & \leq C(\kappa_1,\kappa_3) \max\left\{ 
    \max_{l=1,\dots,m} \abs*{\frac{\norm{\hat{\Sigma}^{(\mathcal{T},l)}}_F^2 - \norm{\Sigma^{(\mathcal{T},l)}}_F^2}{p}},
    \max_{i,j} \frac{1}{m} \sum_{l=1}^m \abs{\Theta^{(\mathcal{S},l)}_{ij}} \right\} \\
    & \leq C(\kappa_1,\kappa_3,\kappa_5) \max\left\{
    \sqrt{\frac{\log(mn_0pq)}{(n_0q)^{1-\frac{1}{2(\alpha_0+1)}}}},
    \sqrt{\frac{m+\log(mn_0pq)}{mn_0p}}\right\}
\end{aligned}
\end{equation*}
with probability at least $1 - C(mn_0pq)^{-1/2}$ for a sufficiently large $n_0$. Plugging the result into \cref{eq:KS_normal}, we obtain the conclusion of the theorem.

\subsection{Proof of Theorem~\ref{thm:inf_power}} \label{app:pf_inf_power}

Based on the definition of $\mathcal{C}_E(1-\alpha)$ in \cref{eq:inf_CI},
\begin{equation*}
    \mathbf{0} \in \mathcal{C}_E(1-\alpha) \iff \norm{\hat{T}_E}_\infty \leq \hat{q}_{\norm{\hat{Z}}_\infty,1-\alpha}.
\end{equation*}
Under the null hypothesis, $T_E = \mathbf{0}$, and
\begin{equation*}
\begin{aligned}
    \Pr[\mathbf{0} \in \mathcal{C}_E(1-\alpha)] 
    & = \Pr[\norm{\hat{T}_E}_\infty \leq \hat{q}_{\norm{\hat{Z}}_\infty,1-\alpha}] \\ 
    & \leq \Pr[\norm{\hat{Z}}_\infty \leq \hat{q}_{\norm{\hat{Z}}_\infty,1-\alpha}|\mathcal{D}] 
    + \sup_{x>0} \abs{\Pr[\norm{\hat{T}_E}_\infty \leq x] - \Pr[\norm{\hat{Z}}_\infty \leq x|\mathcal{D}]} \\
    & = 1-\alpha
    + \sup_{x>0} \abs{\Pr[\norm{\hat{T}_E}_\infty \leq x] - \Pr[\norm{\hat{Z}}_\infty \leq x|\mathcal{D}]}
\end{aligned}
\end{equation*}
almost surely. Then, due to \cref{thm:inf_bootstrap},
\begin{equation*}
    \Pr[\mathbf{0} \in \mathcal{C}_E(1-\alpha)]
    \leq 1 - \alpha + \mathcal{M}
\end{equation*}
with probability at least $1 - C(mn_0pq)^{-1/2}$, where $\mathcal{M}$ is the probabilistic upper bound for $\sup_{x>0} \abs{\Pr[\norm{\hat{T}_E}_\infty \leq x] - \Pr[\norm{\hat{Z}}_\infty \leq x|\mathcal{D}]}$ in \cref{thm:inf_bootstrap}. Similarly,
\begin{equation*}
    \Pr[\mathbf{0} \in \mathcal{C}_E(1-\alpha)]
    \geq 1 - \alpha - \mathcal{M}
\end{equation*}
with the same probability.
Because the assumed sample complexity implies that $\mathcal{M} \rightarrow 0$, $\Pr[\mathbf{0} \notin \mathcal{C}_E(1-\alpha)] \overset{p}{\rightarrow} \alpha$, which verifies the validity of the testing procedure.

% Let's denote by $\mathcal{M}$ the probabilistic upper bound for $\sup_{x>0} \abs{\Pr[\norm{\hat{T}_E}_\infty \leq x] - \Pr[\norm{\hat{Z}}_\infty \leq x|\mathcal{D}]} \rightarrow 0$ due to \cref{thm:inf_bootstrap}. Then, the assumed sample complexity implies that $\mathcal{M} \rightarrow 0$ as $n_0 \rightarrow \infty$. Hence, given any fixed $M > 0$, 
% \begin{equation*}
% \begin{aligned}
%     \Pr[\mathbf{0} \in \mathcal{C}_E(1-\alpha)] 
%     & \leq 1 - \alpha
% \end{aligned}
% \end{equation*}
% Under the assumed sample complexity, The first claim directly follows from \cref{thm:multiple_graph}, and we know that $\mathbb{P}_{H_0}(\mat{c} \notin C_S(1-\alpha)) \rightarrow \alpha$.

For the second claim, we use existing results in the extreme value theory of Gaussian random vectors (e.g., see Exercise 5.10, \citealp{wainwright2019high}) that 
\begin{equation*}
    \Pr\left[\norm{\hat{Z}}_\infty 
    \geq \Exp[\norm{\hat{Z}}_\infty|\mathcal{D}] + \nu \Big|\mathcal{D} \right] 
    \leq \exp\left(-\frac{\nu^2}{2\max_{(i,j) \in E} \hat{S}_{(i,j),(i,j)}} \right).
\end{equation*}
and
\begin{equation*}
    \Exp[\norm{\hat{Z}}_\infty|\mathcal{D}]
    \leq C \sqrt{\log q} \max_{(i,j) \in E} \hat{S}_{(i,j),(i,j)},
\end{equation*}
almost surely. With $\nu = \sqrt{-2 \log \alpha \max_{(i,j) \in E} \hat{S}_{(i,j),(i,j)}}$,
\begin{equation*}
    \hat{q}_{\norm{\hat{Z}}_\infty, 1-\alpha} \leq (C \sqrt{\log q}+\sqrt{-2\log \alpha}) \max_{(i,j) \in E} \hat{S}_{(i,j),(i,j)},
\end{equation*}
almost surely. As we saw in \cref{app:pf_inf_bootstrap}, 
\begin{equation*}
\begin{aligned}
    \norm{\hat{S}_{EE} - S_{EE}}_\infty
    \leq C(\kappa_1,\kappa_3,\kappa_5) \max\left\{
    \frac{\log(mn_0pq)}{(n_0q)^{1-\frac{1}{2(\alpha_0+1)}}},
    \sqrt{\frac{m+\log(mn_0pq)}{mn_0p}}\right\}
\end{aligned}
\end{equation*}
with probability at least $1 - C(mn_0pq)^{-1/2}$. Due to \cref{assmp:spatial_sample,assmp:temporal_sample}, the right hand side converges to $0$ as $n_0 \rightarrow \infty$, and therefore
\begin{equation*}
\begin{aligned}
    \hat{q}_{\norm{\hat{Z}}_\infty, 1-\alpha} 
    % \leq & ~C(\kappa_1,\kappa_3,\kappa_5) (\sqrt{\log q}+\sqrt{-2\log \alpha}) \\
    % & \times \left( \norm{S_{EE}}_\infty
    % + \max\left\{ \frac{\log(mn_0pq)}{(n_0q)^{1-\frac{1}{2(\alpha_0+1)}}},
    % \sqrt{\frac{m+\log(mn_0pq)}{mn_0p}}\right\} \right) \\
    \leq & ~C(\kappa_1,\kappa_3,\kappa_5) (\sqrt{\log q}+\sqrt{-2\log \alpha}) 
    \max_{(i,j) \in E} S_{(i,j),(i,j)}
\end{aligned}
\end{equation*}
with the same probability for sufficiently large $n_0$. Suppose that the assumed alternative hypothesis implies that
\begin{equation*}
    \norm{T_E}_\infty 
    \geq C(\kappa_1,\kappa_3,\kappa_5) \sqrt{\log q} \max_{(i,j) \in E} S_{(i,j),(i,j)}
    \geq \hat{q}_{\norm{\hat{Z}}_\infty,1-\alpha} + \hat{q}_{\norm{\hat{Z}}_\infty,1/q}.
\end{equation*}
Then, 
\begin{equation*}
\begin{aligned}
    \Pr[\mathbf{0} \in \mathcal{C}_E(1-\alpha)] 
    & = \Pr[\norm{\hat{T}_E}_\infty \leq \hat{q}_{\norm{\hat{Z}}_\infty,1-\alpha}] \\ 
    & \leq \Pr[\norm{T_E}_\infty - \norm{\hat{T}_E - T_E}_\infty 
    \leq \hat{q}_{\norm{\hat{Z}}_\infty,1-\alpha}] \\
    & = \Pr[\norm{\hat{T}_E - T_E}_\infty 
    \geq \hat{q}_{\norm{\hat{Z}}_\infty,1/q}] \\
    & \leq \Pr[\norm{\hat{Z}}_\infty 
    \geq \hat{q}_{\norm{\hat{Z}}_\infty,1/q}|\mathcal{D}] 
    + \sup_{x>0} \abs*{\Pr[\norm{\hat{T}_E}_\infty \leq x] - \Pr[\norm{\hat{Z}}_\infty \leq x|\mathcal{D}]} \\
    & = 1/q
    + \sup_{x>0} \abs*{\Pr[\norm{\hat{T}_E}_\infty \leq x] - \Pr[\norm{\hat{Z}}_\infty \leq x|\mathcal{D}]},
\end{aligned}
\end{equation*}
which converges to $0$ as $n_0 \rightarrow \infty$, due to the assumed sample complexity. It verifies the second conclusion about the power analysis.

\subsection{Proof of Theorem~\ref{thm:temporal_regression}} \label{app:pf_temporal_regression}

Let $\tilde{\beta}^{(\mathcal{T},l)}$ be the projected parameter of $\beta^{(\mathcal{T},l)}$ onto the parameter space under the banded Cholesky factor assumption. That is, for $t = 1, \dots, p$,
\begin{equation*}
\begin{aligned}
    & \tilde{\beta}^{(\mathcal{T},l)}_{\cdot,t}
    := \argmin_{b \in \reals^p}
     \Exp\left[\norm{X^{(\mathcal{T},l)}_{t,\cdot} - X^{(\mathcal{T},l)\top} b}_2^2\right] \\
    & \text{w.r.t} ~~ b_s = 0 ~~ \text{where} ~~ s < t-h_l ~~\text{or} ~~ s \geq t.
\end{aligned}
\end{equation*}
According to the regression theory, $\tilde{\beta}^{(\mathcal{T},l)}_{\cdot, t}$ can be rewritten in terms of $\Sigma^{(\mathcal{T},l)}$ only:
\begin{equation*}
    \tilde{\beta}^{(\mathcal{T},l)}_{t-h:t-1, t} = - \tilde{\Omega}^{(\mathcal{T},l)}_{t,h_l+1,1:h_l} / \tilde{\Omega}^{(\mathcal{T},l)}_{t,h_l+1,h_l+1},
\end{equation*}
where $\tilde{\Omega}^{(\mathcal{T},l)}_t = \left(\Sigma^{(\mathcal{T},l)}_{t-h_l:t,t-h_l:t}\right)^{-1} \in \reals^{h_l+1 \times h_l+1}$. 
We denote the regression error by
\begin{equation*}
    \tilde{\epsilon}^{(\mathcal{T},l)}_{t,i} := X^{(\mathcal{T},l)}_{t,i} - X^{(\mathcal{T},l)\top}_{\cdot,i} \tilde{\beta}^{(\mathcal{T},l)}_{\cdot,t},
\end{equation*}
and $\tilde{\Phi}^{(\mathcal{T},l)}_{tt} := \frac{1}{n_l q} \Exp[\norm{ \tilde{\epsilon}^{(\mathcal{T},l)}_{t,i}}_2^2]$.
Then, 
\begin{equation*}
    \tilde{\Phi}^{(\mathcal{T},l)}_{tt} = \frac{\tr(\Sigma^{(\mathcal{S},l)})}{q} \frac{1}{\tilde{\Omega}^{(\mathcal{T},l)}_{t,h_l+1,h_l+1}}.
\end{equation*}
We notice that $\tilde{\Omega}^{(\mathcal{T},l)}_{t,h_l+1,1:h_l} / \tilde{\Omega}^{(\mathcal{T},l)}_{t,h_l+1,h_l+1}$ and $1 / \tilde{\Omega}^{(\mathcal{T},l)}_{t,h_l+1,h_l+1}$ are the projected parameter and prediction error of the $AR(h_l)$ regression on vector-variate observations, so the bias analyses in the existing literature (e.g., \citet{liu2017}) is useful here.
For example, Lemma B.3 in \citet{liu2017} implies that under \cref{assmp:temporal_sample}
\begin{equation*}
\begin{aligned} \label{eq:bias_beta_Phi_T}
    \norm*{\tilde{\beta}^{(\mathcal{T},l)}_{\cdot,t}
    - \beta^{(\mathcal{T},l)}_{\cdot,t}}_2 
    & \leq C(\kappa_3, \kappa_5) (h_l-1)^{-\alpha_l-1/2}, \\
    \abs*{\tilde{\Phi}^{(\mathcal{T},l)}_{tt} - \frac{\tr(\Sigma^{(\mathcal{S},l)})}{q} \Phi^{(\mathcal{T},l)}_{tt}} 
    & \leq C(\kappa_3, \kappa_5) (h_l-1)^{-\alpha_l-1/2}.
\end{aligned}
\end{equation*}
Now, it suffices to establish the probabilistic error bound of $\hat{\beta}^{(\mathcal{T},l)}$ and $\hat{\Phi}^{(\mathcal{T},l)}$ against $\tilde{\beta}^{(\mathcal{T},l)}$ and $\tilde{\Phi}^{(\mathcal{T},l)}$, respectively.
We take a similar approach with the proof of \cref{thm:spatial_regression}. First, we notice that
\begin{equation*}
    \hat{\beta}^{(\mathcal{T},l)}_{t-h_l:t-1,t} = 
    \left(X^{(\mathcal{T},l)}_{t-h_l:t-1,\cdot} X^{(\mathcal{T},l)\top}_{t-h_l:t-1,\cdot} \right)^{-1} 
    \left( X^{(\mathcal{T},l)}_{t-h_l:t-1,\cdot} X^{(\mathcal{T},l)\top}_{t,\cdot} \right).
\end{equation*}
Then,
\begin{equation*}
\begin{aligned}
    & \left(\frac{1}{n_l q} X^{(\mathcal{T},l)}_{t-h_l:t-1,\cdot} X^{(\mathcal{T},l)\top}_{t-h_l:t-1,\cdot} \right) (\hat{\beta}^{(\mathcal{T},l)}_{\cdot,t} - \tilde{\beta}^{(\mathcal{T},l)}_{\cdot,t}) \\
    & = \frac{1}{n_l q} X^{(\mathcal{T},l)}_{t-h_l:t-1,\cdot} X^{(\mathcal{T},l)\top}_{t,\cdot}
    - \left(\frac{1}{n_l q} X^{(\mathcal{T},l)}_{t-h_l:t-1,\cdot} X^{(\mathcal{T},l)\top}_{t-h_l:t-1,\cdot} \right) \tilde{\beta}^{(\mathcal{T},l)}_{t-h_l:t-1,t} \\
    & = \frac{1}{n_l q} X^{(\mathcal{T},l)}_{t-h_l:t-1,\cdot} \tilde{\epsilon}^{(\mathcal{T},l)\top}_{t,\cdot}.
\end{aligned}
\end{equation*}
By a similar argument with the proof of \cref{lem:l_inf_EX},
\begin{equation*} 
\begin{aligned}
    & \Pr\left[ \max_{l=1,\dots,m} \max_{t=1,\dots,p} 
    \norm*{\left(\frac{1}{n_l q} X^{(\mathcal{T},l)}_{t-h_l:t-1,\cdot} X^{(\mathcal{T},l)\top}_{t-h_l:t-1,\cdot} \right)
    \tilde\Delta^{(\mathcal{T},l)}_{\cdot,t}}_\infty 
    \geq C(\kappa_1,\kappa_3) \sqrt{\frac{\log(mn_0pq)}{n_0q}} \right] \\
    & = \Pr\left[ \max_{l=1,\dots,m} \max_{t=1,\dots,p} 
    \norm*{\frac{1}{n_l q} X^{(\mathcal{T},l)}_{t-h_l:t-1,\cdot} \tilde{\epsilon}^{(\mathcal{T},l)\top}_{t,\cdot}}_\infty
    \geq C(\kappa_1,\kappa_3) \sqrt{\frac{\log(mn_0pq)}{n_0q}} \right] \\
    & \leq C (mn_0pq)^{-1/2}
\end{aligned}
\end{equation*}
for some positive constants $C(\kappa_3)$ and $C$, where $\tilde\Delta^{(\mathcal{T},l)} := \hat{\beta}^{(\mathcal{T},l)} - \tilde\beta^{(\mathcal{T},l)}$. Under the event, 
\begin{equation} \label{eq:var_beta_T_upper}
\begin{aligned}
    \abs*{ \tilde\Delta^{(\mathcal{T},l)\top}_{\cdot,t} 
    \left(\frac{1}{n_l q} X^{(\mathcal{T},l)}_{t-h_l:t-1,\cdot} X^{(\mathcal{T},l)\top}_{t-h_l:t-1,\cdot} \right)
    \tilde\Delta^{(\mathcal{T},l)}_{\cdot,t} }
    & \leq \norm*{\left(\frac{1}{n_l q} X^{(\mathcal{T},l)}_{t-h_l:t-1,\cdot} X^{(\mathcal{T},l)\top}_{t-h_l:t-1,\cdot} \right)
    \tilde\Delta^{(\mathcal{T},l)}_{\cdot,t}}_\infty
    \norm{\tilde\Delta^{(\mathcal{T},l)}_{\cdot,t}}_1  \\
    & \leq C(\kappa_1,\kappa_3) \sqrt{\frac{\log(mn_0pq)}{n_0q}} \norm{\tilde\Delta^{(\mathcal{T},l)}_{\cdot,t}}_1 \\
    & \leq C(\kappa_1,\kappa_3) \sqrt{h_l \frac{\log(mn_0pq)}{n_0q}} \norm{\tilde\Delta^{(\mathcal{T},l)}_{\cdot,t}}_2, \\
\end{aligned}
\end{equation}
for any $t=1,\dots,T$ and $l=1,\dots,m$. On the other hand, by a similar argument with the proof of \cref{lem:l_inf_XX},
\begin{equation*}
\begin{aligned}
    & \Pr\left[ \begin{aligned}
        \max_{l=1,\dots,m} \max_{s,t} 
        \abs*{ \frac{1}{n_l q} X^{(\mathcal{T},l)}_{s,\cdot} X^{(\mathcal{T},l)\top}_{t,\cdot} 
        - \frac{\tr(\Sigma^{(\mathcal{S},l)})}{q} \Sigma^{(\mathcal{T},l)}_{st} } 
        \geq C(\kappa_1,\kappa_3) \sqrt{\frac{\log(mn_0pq)}{n_0q}}
    \end{aligned} \right] \\
    & \leq C (mn_0pq)^{-1/2}
\end{aligned}
\end{equation*}
for some positive constants $C(\kappa_3)$ and $C$. Under the event,
\begin{equation} \label{eq:l_2_XX}
\begin{aligned}
    & \abs*{ \tilde\Delta^{(\mathcal{T},l)\top}_{\cdot,t} 
    \left(\frac{1}{n_l q} X^{(\mathcal{T},l)}_{t-h_l:t-1,\cdot} X^{(\mathcal{T},l)\top}_{t-h_l:t-1,\cdot} 
    - \frac{\tr(\Sigma^{(\mathcal{S},l)})}{q} \Sigma^{(\mathcal{T},l)}_{t-h_l:t-1, t-h_l:t-1} \right)
    \tilde\Delta^{(\mathcal{T},l)}_{\cdot,t} } \\
    & \leq \norm*{\frac{1}{n_l q} X^{(\mathcal{T},l)}_{t-h_l:t-1,\cdot} X^{(\mathcal{T},l)\top}_{t-h_l:t-1,\cdot} 
    - \frac{\tr(\Sigma^{(\mathcal{S},l)})}{q} \Sigma^{(\mathcal{T},l)}_{t-h_l:t-1, t-h_l:t-1}}_\infty
    \norm{\tilde\Delta^{(\mathcal{T},l)}_{\cdot,t}}_1^2  \\
    & \leq C(\kappa_1,\kappa_3) \sqrt{\frac{\log(mn_0pq)}{n_0q}} \norm{\tilde\Delta^{(\mathcal{T},l)}_{\cdot,t}}_1^2 
    \leq C(\kappa_1,\kappa_3) h_l \sqrt{\frac{\log(mn_0pq)}{n_0q}} \norm{\tilde\Delta^{(\mathcal{T},l)}_{\cdot,t}}_2^2 \\
    & \leq \frac{1}{2\kappa_3} \norm{\tilde\Delta^{(\mathcal{T},l)}_{\cdot,t}}_2^2,
\end{aligned}
\end{equation}
for any $t=1,\dots,T$, $l=1,\dots,m$, and a sufficiently large $n_0$ due to \cref{assmp:temporal_sample}. Moreover with \cref{assmp:eigenvalues},
\begin{equation} \label{eq:var_beta_T_lower}
    \abs*{ \tilde\Delta^{(\mathcal{T},l)\top}_{\cdot,t} 
    \left(\frac{1}{n_l q} X^{(\mathcal{T},l)}_{t-h_l:t-1,\cdot} X^{(\mathcal{T},l)\top}_{t-h_l:t-1,\cdot} \right)
    \tilde\Delta^{(\mathcal{T},l)}_{\cdot,t} } \geq \frac{1}{ 2 \kappa_3} \norm{\tilde\Delta^{(\mathcal{T},l)}_{\cdot,t}}_2^2
\end{equation}
with probability at least $1 - C(mn_0pq)^{-1/2}$. \cref{eq:var_beta_T_upper,eq:var_beta_T_lower} imply
\begin{equation*}
    \frac{1}{2 \kappa_3} \norm{\tilde\Delta^{(\mathcal{T},l)}_{\cdot,t}}_2^2 
    \leq C(\kappa_1, \kappa_3) \sqrt{h_l \frac{\log(mn_0pq)}{n_0q}} \norm{\tilde\Delta^{(\mathcal{T},l)}_{\cdot,t}}_2,
\end{equation*}
and
\begin{equation} \label{eq:var_beta_T}
    \norm{\tilde\Delta^{(\mathcal{T},l)}_{\cdot,t}}_2 
    \leq C(\kappa_1,\kappa_3) \sqrt{h_l \frac{\log(mn_0pq)}{n_0q}}
    \textand
    \norm{\tilde\Delta^{(\mathcal{T},l)}_{\cdot,t}}_1 
    \leq C(\kappa_1,\kappa_3) h_l \sqrt{\frac{\log(mn_0pq)}{n_0q}}.
\end{equation}
for any $t=1,\dots,T$, $l=1,\dots,m$ and a sufficiently large $n_0$, with probability at least $1 - C(mn_0pq)^{-1/2}$ . Integrating \cref{eq:bias_beta_Phi_T,eq:var_beta_T}, we obtain the first result that, with the same probability,
\begin{equation*}
\begin{aligned}
    \norm{\hat\beta^{(\mathcal{T},l)}_{\cdot,t} - \beta^{(\mathcal{T},l)}_{\cdot,t}}_2
    & \leq C(\kappa_3,\kappa_5)(h_l-1)^{-\alpha_l-1/2} + C(\kappa_1,\kappa_3) h_l \sqrt{\frac{\log(mn_0pq)}{n_0q}} \\
    & \leq C(\kappa_1,\kappa_3,\kappa_5) \sqrt{\frac{\log(mn_0pq)}{(n_0q)^{1-1/2(\alpha_0+1)}}},
\end{aligned}
\end{equation*}
for any $t=1,\dots,T$, $l=1,\dots,m$ and a sufficiently large $n_0$, given that $h_l = \floor{(n_l q)^{1/2(\alpha_l+1)}}$.

For the second result, we observe that
\begin{equation*}
\begin{aligned}
    \abs*{\hat{\Phi}^{(\mathcal{T},l)}_{tt} - \tilde{\Phi}^{(\mathcal{T},l)}_{tt}}
    \leq \frac{1}{n_l q} \abs*{ \norm{\hat\epsilon^{(\mathcal{T},l)}_{t,\cdot}}_2^2 - \norm{\tilde\epsilon^{(\mathcal{T},l)}_{t,\cdot}}_2^2 } 
    + \abs*{ \frac{1}{n_l q} \norm{\tilde{\epsilon}^{(\mathcal{T},l)}_{t,\cdot}}_2^2 - \tilde{\Phi}^{(\mathcal{T},l)}_{tt}}.
\end{aligned}
\end{equation*}
For the first term,
\begin{equation*}
\begin{aligned}
    & \frac{1}{n_l q} \abs*{ \norm{\hat\epsilon^{(\mathcal{T},l)}_{t,\cdot}}_2^2 - \norm{\tilde\epsilon^{(\mathcal{T},l)}_{t,\cdot}}_2^2 } \\
    & = \frac{2}{n_l q} \abs*{\tilde\Delta^{(\mathcal{T},l)\top}_{\cdot,t}
    X^{(\mathcal{T},l)}_{t-h:t-1, \cdot} \tilde{\epsilon}^{(\mathcal{T},l)\top}_{t,\cdot}}
    + \frac{1}{n_l q} \abs*{\tilde\Delta^{(\mathcal{T},l)\top}_{\cdot,t} X^{(\mathcal{T},l)}_{t-h:t-1} X^{(\mathcal{T},l)\top}_{t-h:t-1} \tilde\Delta^{(\mathcal{T},l)}_{\cdot,t} } \\
    & \leq \norm*{\tilde\Delta^{(\mathcal{T},l)\top}_{\cdot,t}}_1 
    \norm*{\frac{2}{n_l q} X^{(\mathcal{T},l)}_{t-h:t-1, \cdot} \tilde{\epsilon}^{(\mathcal{T},l)\top}_{t,\cdot}}_\infty
    + \norm*{\tilde\Delta^{(\mathcal{T},l)\top}_{\cdot,t}}_2^2 \norm*{\frac{1}{n_l q} X^{(\mathcal{T},l)}_{t-h:t-1} X^{(\mathcal{T},l)\top}_{t-h:t-1}}_2 \\
    & \leq C(\kappa_1, \kappa_3) h \frac{\log p}{n_l q}
\end{aligned}
\end{equation*}
for any $t=1,\dots,T$, $l=1,\dots,m$, and a sufficiently large $n_0$, with probability at least $1 - C(mn_0pq)^{-1/2}$, based on \cref{eq:var_beta_T_upper,eq:l_2_XX,assmp:eigenvalues}. For the second term, a similar argument with the proof of \cref{lem:l_inf_EE} implies that
\begin{equation*}
\begin{aligned}
    & \Pr\left[ \begin{aligned}
        \max_{l=1,\dots,m} \max_{s,t} 
        \abs*{ \frac{1}{n_l q} \tilde{\epsilon}^{(\mathcal{T},l)}_{s,\cdot} \tilde{\epsilon}^{(\mathcal{T},l)\top}_{t,\cdot} 
        - \frac{\tr(\Sigma^{(\mathcal{S},l)})}{q} \tilde{\Phi}^{(\mathcal{T},l)}_{st} } 
        \geq C(\kappa_1,\kappa_3) \sqrt{\frac{\log(mn_0pq)}{n_l q}}
    \end{aligned} \right] \\
    & \leq C (mn_0pq)^{-1/2}.
\end{aligned}
\end{equation*}
In sum,
\begin{equation*}
    \abs*{\hat{\Phi}^{(\mathcal{T},l)}_{tt} - \tilde{\Phi}^{(\mathcal{T},l)}_{tt}}
    \leq C(\kappa_1,\kappa_3) h \frac{\log p}{n_0q} 
    + C(\kappa_1,\kappa_3) \sqrt{\frac{\log(mn_0pq)}{n_l q}}
    \leq C(\kappa_1,\kappa_3) \sqrt{h \frac{\log p}{n_0q}}
\end{equation*}
for any $t=1,\dots,T$, $l=1,\dots,m$ and a sufficiently large $n_0$, with probability at least $1 - C(mn_0pq)^{-1/2}$, which implies the second conclusion where $h = \floor{(n_lq)^{1/2(\alpha_l+1)}}$.

%%%%%%%%%%%%%%%%%%%%%%%%%%%%%%%%%%%%%%%%%%%%%%%%%%%%%%%
%%%%%%%%%%%%%%%%%%%%%%%%%%%%%%%%%%%%%%%%%%%%%%%%%%%%%%%
%%%%%%%%%%%%%%%%%%%%%%%%%%%%%%%%%%%%%%%%%%%%%%%%%%%%%%%
%%%%%%%%%%%%%%%%%%%%%%%%%%%%%%%%%%%%%%%%%%%%%%%%%%%%%%%

\subsection{Proof of Theorem~\ref{thm:temporal_est_Sigma}} \label{app:pf_temporal_est_Sigma}

We recall the Cholesky decomposition of the temporal precision matrix:
\begin{equation*}
\begin{aligned}
    \Omega^{(\mathcal{T},l)} & = \frac{\tr(\Sigma^{(\mathcal{S},l)})}{q} 
    (I - \beta^{(\mathcal{T},l)\top}) 
    \Phi^{(\mathcal{T},l)-1} (I - \beta^{(\mathcal{T},l)}) \\
    \Sigma^{(\mathcal{T},l)} & = \frac{q}{\tr(\Sigma^{(\mathcal{S},l)})} 
    (I - \beta^{(\mathcal{T},l)\top})^{-1} 
    \Phi^{(\mathcal{T},l)} (I - \beta^{(\mathcal{T},l)})^{-1}
\end{aligned}
\end{equation*}
Our temporal covariance matrix estimate is based on the estimated Cholesky factor, $\hat{\beta}^{(\mathcal{T},l)}$, and noise variance, $\hat{\Phi}^{(\mathcal{T},l)}$, in \cref{eq:temporal_est_beta,eq:temporal_est_Phi}. In regard to their estimate errors, we use the following lemma as a corollary of \cref{thm:temporal_regression}. See \cref{app:pf_temporal_beta_Phi_F} for the proof.

\begin{lemma} \label{lem:temporal_beta_Phi_F}
    Suppose that $h_l = \floor{(n_l q)^{1/(1+\alpha)}}$ and $\eta = C(\kappa_3)$ satisfies $\eta \leq \lambda_1(I - \beta^{(\mathcal{T},l)})$ for $l = 1, \dots, m$. Then, following the procedure defined in \cref{sec:temporal_est}, 
    \begin{equation*}
    \begin{aligned}
        \mathbb{P} \left[ \begin{aligned}
            \max_l \frac{1}{p} \norm*{\hat{\Phi}^{(\mathcal{T},l)} - \frac{\tr(\Sigma^{(\mathcal{S},l)})}{q} \Phi^{(\mathcal{T},l)}}_F^2
            & \geq C(\kappa_1, \kappa_3, \kappa_5) {\frac{\log (mn_0pq)}{(n_0q)^{1-1/(2\alpha_0+2)}}}, \\
            \max_l \frac{1}{p} \norm*{\hat{\Phi}^{(\mathcal{T},l)-1} - \frac{q}{\tr(\Sigma^{(\mathcal{S},l)})} \Phi^{(\mathcal{T},l)-1}}_F^2
            & \geq C(\kappa_1, \kappa_3, \kappa_5) {\frac{\log (mn_0pq)}{(n_0q)^{1-1/(2\alpha_0+2)}}} \\
        \end{aligned} \right] & \\
        \leq C(mn_0pq)^{-1/2}, & \\
    \end{aligned}
    \end{equation*}
    \begin{equation*}
    \begin{aligned}
        \mathbb{P} \left[ \begin{aligned}
            \max_l \frac{1}{p} \norm{P_\eta(I-\hat{\beta}^{(\mathcal{T},l)})
            - (I - \beta^{(\mathcal{T},l)})}_F^2
            & \geq C(\kappa_1, \kappa_3, \kappa_5) 
            {\frac{\log (mn_0pq)}{(n_0q)^{1-1/(2\alpha_0+2)}}}, \\
            \max_l \frac{1}{p} \norm{P_\eta(I-\hat{\beta}^{(\mathcal{T},l)})^{-1} 
            - (I - \beta^{(\mathcal{T},l)})^{-1}}_F^2
            & \geq C(\kappa_1, \kappa_3, \kappa_5) 
            {\frac{\log (mn_0pq)}{(n_0q)^{1-1/(2\alpha_0+2)}}} \\
        \end{aligned} \right] & \\
        \leq C (mn_0pq)^{-1/2}, &
    \end{aligned}
    \end{equation*}
    for sufficiently large $n_0$.
\end{lemma}
We note that \cref{assmp:eigenvalues} implies that the operator norms $I-\beta^{(\mathcal{T},l)}$, $(I-\beta^{(\mathcal{T},l)})^{-1}$, $\Phi^{(\mathcal{T},l)}$, and $\Phi^{(\mathcal{T},l)-1}$ are bounded; see Lemma B.2 in \citet{liu2017}. The eigenvalue truncation in \cref{eq:temporal_eig_cut} implies the bounded operator norm for $P_\eta(I-\hat{\beta}^{(\mathcal{T},l)})$ and $P_\eta(I-\hat{\beta}^{(\mathcal{T},l)})^{-1}$. Finally, \cref{thm:temporal_est_Sigma} implies the bounded operator norm for $\hat{\Phi}^{(\mathcal{T},l)}$ and $\hat{\Phi}^{(\mathcal{T},l)-1}$. Namely,
\begin{equation*}
\begin{aligned}
    \norm{I-\beta^{(\mathcal{T},l)}}_\mathrm{op},
    \norm{P_\eta(I-\hat{\beta}^{(\mathcal{T},l)})}_\mathrm{op} & \leq C(\kappa_3) \\
    \norm{(I-\beta^{(\mathcal{T},l)})^{-1}}_\mathrm{op},
    \norm{P_\eta(I-\hat{\beta}^{(\mathcal{T},l)})^{-1}}_\mathrm{op} & \leq C(\kappa_3) \\
    \norm{\Phi^{(\mathcal{T},l)}}_\mathrm{op},
    \norm*{\frac{\tr(\Sigma^{(\mathcal{S},l)})}{q} \hat{\Phi}^{(\mathcal{T},l)}}_\mathrm{op} & \leq C(\kappa_3) \\
    \norm{\Phi^{(\mathcal{T},l)-1}}_\mathrm{op},
    \norm*{\frac{q}{\tr(\Sigma^{(\mathcal{S},l)})} \hat{\Phi}^{(\mathcal{T},l)-1}}_\mathrm{op} & \leq C(\kappa_3) \\
\end{aligned}
\end{equation*}
for any $l=1,\dots,m$ and a sufficiently large $n_0$, with probability at least $1 - C(mn_0pq)^{-1/2}$.

Now, we move on to connect the error in the temporal covariance matrix estimate with the Frobenius norm bounds in \cref{lem:temporal_beta_Phi_F}. We use a well known inequality about the Frobenius and operator norms:
\begin{equation*}
    \norm{AB}_F \leq \norm{A}_\mathrm{op} \norm{B}_F
\end{equation*}
for any compatible matrices $A$ and $B$. For the temporal covariance matrix,
\begin{equation*}
\begin{aligned}
    & \norm*{\bar{\Sigma}^{(\mathcal{T},l)} - \frac{\tr(\Sigma^{(\mathcal{S},l)})}{q} \Sigma^{(\mathcal{T},l)}}_F \\
    & \leq \norm{(I - \beta^{(\mathcal{T},l)})^{-1}}_\mathrm{op}
    \norm*{\frac{\tr(\Sigma^{(\mathcal{S},l)})}{q} \Phi^{(\mathcal{T},l)}}_\mathrm{op}
    \norm{P_\eta(I - \hat{\beta}^{(\mathcal{T},l)})^{-1} 
    - (I - \beta^{(\mathcal{T},l)})^{-1}}_F \\
    & ~~ + \norm{(I - \beta^{(\mathcal{T},l)})^{-1}}_\mathrm{op}
    \norm*{\hat{\Phi}^{(\mathcal{T},l)} 
    - \frac{\tr(\Sigma^{(\mathcal{S},l)})}{q} \Phi^{(\mathcal{T},l)}}_F
    \norm{P_\eta(I - \hat\beta^{(\mathcal{T},l)})^{-1}}_\mathrm{op} \\
    & ~~ + \norm{P_\eta(I - \hat{\beta}^{(\mathcal{T},l)})^{-1} 
    - (I - \beta^{(\mathcal{T},l)})^{-1}}_F
    \norm*{\frac{\tr(\Sigma^{(\mathcal{S},l)})}{q} \Phi^{(\mathcal{T},l)}}_\mathrm{op}
    \norm{P_\eta(I - \hat\beta^{(\mathcal{T},l)})^{-1}}_\mathrm{op} \\
\end{aligned}
\end{equation*}
and hence, plugging in the previously obtained bounds,  
\begin{equation*}
    \norm*{\bar{\Sigma}^{(\mathcal{T},l)} - \frac{\tr(\Sigma^{(\mathcal{S},l)})}{q} \Sigma^{(\mathcal{T},l)}}_F 
    \leq C(\kappa_1, \kappa_3, \kappa_5) \sqrt{p} \sqrt{\frac{\log(mn_0pq)}{(n_0q)^{1-1/(2\alpha_0 + 2)}}}
\end{equation*}
for any $l=1,\dots,m$ with probability at least $1 - C(mn_0pq)^{-1/2}$. The second conclusion about $\bar{\Omega}^{(\mathcal{T},l)}$ is similarly obtained.

%%%%%%%%%%%%%%%%%%%%%%%%%%%%%%%%%%%%%%%%%%%%%%%%%%%%%%%
%%%%%%%%%%%%%%%%%%%%%%%%%%%%%%%%%%%%%%%%%%%%%%%%%%%%%%%
%%%%%%%%%%%%%%%%%%%%%%%%%%%%%%%%%%%%%%%%%%%%%%%%%%%%%%%
%%%%%%%%%%%%%%%%%%%%%%%%%%%%%%%%%%%%%%%%%%%%%%%%%%%%%%%

\subsection{Proof of Corollary~\ref{thm:temporal_est_F}} \label{app:pf_temporal_est_F}

The error in the Frobenius norm estimate can be upperbounded by a function of the error in the covariance matrix estimate:
\begin{equation*}
\begin{aligned}
    & \frac{1}{p} \abs*{\norm{\hat{\Sigma}^{(\mathcal{T},l)}}_F^2 - \norm{\Sigma^{(\mathcal{T},l)}}_F^2} \\
    & = \frac{1}{p} \left( \norm{\hat{\Sigma}^{(\mathcal{T},l)}}_F + \norm{\Sigma^{(\mathcal{T},l)}}_F \right)
    \abs*{\norm{\hat{\Sigma}^{(\mathcal{T},l)}}_F - \norm{\Sigma^{(\mathcal{T},l)}}_F} \\
    & \leq \left( \frac{2}{\sqrt{p}} \norm{\Sigma^{(\mathcal{T},l)}}_F + \frac{1}{\sqrt{p}} \norm{\hat{\Sigma}^{(\mathcal{T},l)} - \Sigma^{(\mathcal{T},l)}}_F \right) \frac{1}{\sqrt{p}} \norm{ \hat{\Sigma}^{(\mathcal{T},l)} - \Sigma^{(\mathcal{T},l)}}_F.
\end{aligned}
\end{equation*}
Because $\frac{1}{\sqrt{p}} \norm{\Sigma^{(\mathcal{T},l)}}_F$ is bounded away from both $0$ and $\infty$, and \cref{thm:temporal_est_Sigma} implies that
\begin{equation*}
    \frac{1}{\sqrt{p}} \norm{\hat{\Sigma}^{(\mathcal{T},l)} - \Sigma^{(\mathcal{T},l)}}_F 
    \leq C(\kappa_1,\kappa_3,\kappa_5) \sqrt{\frac{\log(mn_0pq)}{p (n_0q)^{1-1/(2\alpha_0+2)}}} 
\end{equation*} 
with probability at least $1 - C(mn_0pq)^{-1/2}$,
\begin{equation*}
\begin{aligned}
    \frac{1}{p} \abs*{\norm{\hat{\Sigma}^{(\mathcal{T},l)}}_F^2 - \norm{\Sigma^{(\mathcal{T},l)}}_F^2}
    & \leq C(\kappa_1,\kappa_3,\kappa_5) \frac{1}{\sqrt{p}} \norm{\hat{\Sigma}^{(\mathcal{T},l)} - \Sigma^{(\mathcal{T},l)}}_F \\
    & \leq C(\kappa_1,\kappa_3,\kappa_5) \sqrt{\frac{\log(mn_0pq)}{p (n_0q)^{1-1/(2\alpha_0+2)}}}
\end{aligned}
\end{equation*}
with probability at least $1 - C(mn_0pq)^{-1/2}$ for sufficiently large $n_0$.

\section{Proof of the Lemmas}

\subsection{Proof of Lemma~\ref{lem:l_inf_XX}} \label{app:pf_l_inf_XX}

In \cref{eq:iid_XX}, $W^{(\mathcal{S},l)}_{ti} W^{(\mathcal{S},l)}_{tj}$ is a sub-exponential variable with
\begin{equation*}
    \Exp\left[\exp\left(\nu(W^{(\mathcal{S},l)}_{ti}W^{(\mathcal{S},l)}_{tj} - \Sigma^{(\mathcal{T},l)}_{tt} \Sigma^{(\mathcal{S},l)}_{ij})\right)\right] \leq e^{C(\kappa_3)\nu^2},
\end{equation*}
for all $\nu: \abs{\nu} < C(\kappa_3)$ due to \cref{assmp:eigenvalues}, and independent across $t = 1,\dots, n_l p$. According to Theorem 2.8.2 in \cite{vershynin2018} and \cref{assmp:balanced_sample,assmp:temporal_trace}, our main claim is a result of maximal inequality on sub-exponential random variables.

%%%%%%%%%%%%%%%%%%%%%%%%%%%%%%%%%%
%%%%%%%%%%%%%%%%%%%%%%%%%%%%%%%%%%
%%%%%%%%%%%%%%%%%%%%%%%%%%%%%%%%%%

\subsection{Proof of Lemma~\ref{lem:l_2_EX}} \label{app:pf_l_2_EX}

In \cref{eq:iid_EX_EE}, 
\begin{equation*}
    (\xi^{(\mathcal{S},l)}_{\cdot,i}, W^{(\mathcal{S},l)}_{\cdot,j}) \distiid N\left(0, I \otimes \diag\left(\frac{1}{\Omega^{(\mathcal{S},l)}_{ii}}, \Sigma^{(\mathcal{S},l)}_{jj}\right) \right).
\end{equation*}
Because $W^{(\mathcal{S},l)}_{\cdot,i}$ is independent to $\xi^{(\mathcal{S},l)}_{\cdot,j}$,
\begin{equation*}
    \left\{ \xi^{(\mathcal{S},l)}_{\cdot,i} W^{(\mathcal{S},l)\top}_{\cdot, j} | W^{(\mathcal{S},l)}_{\cdot, j} \right\} \dist N\left(0, \frac{\norm{W^{(\mathcal{S},l)}_{\cdot, j}}_2^2}{\Omega^{(\mathcal{S},l)}_{ii}}\right) ~~ a.s.,
\end{equation*}
and therefore
\begin{equation*}
\begin{aligned}
    \left\{\left(\frac{\sqrt{n_l p}}{\norm{X^{(\mathcal{S},l)}_{\cdot,j}}_2} \epsilon^{(\mathcal{S},l)\top}_{\cdot,i} X^{(\mathcal{S},l)}_{\cdot,j}\right)^2 \bigg| X^{\mathcal{S},l)}_{\cdot,j} \right\}
    & = \left\{\left(\frac{\sqrt{n_l p}}{\norm{X^{(\mathcal{S},l)}_{\cdot,j}}_2} \xi^{(\mathcal{S},l)\top}_{\cdot,i} W^{(\mathcal{S},l)}_{\cdot,j}\right)^2 \bigg| W^{\mathcal{S},l)}_{\cdot,j} \right\} \\
    & \dist n_l p \frac{\norm{W^{(\mathcal{S},l)}_{\cdot,j}}_2^2}
    {\norm{X^{(\mathcal{S},l)}_{\cdot,j}}_2^2} \chi^2(1) ~~ a.s.
\end{aligned}
\end{equation*}
Due to \cref{assmp:eigenvalues}, $\frac{\norm{W^{(\mathcal{S},l)}_{\cdot,j}}_2^2} {\norm{X^{(\mathcal{S},l)}_{\cdot,j}}_2^2} \leq C(\kappa_3)$.
Hence, the sub-exponential inequality in Theorem 2.8.2, \citet{vershynin2018} implies that, for any $\nu > 0$,
\begin{equation*}
\begin{aligned}
    & \Pr\left[\sum_{l=1}^m \left(\frac{\sqrt{n_l p}}{\norm{X^{(\mathcal{S},l)}_{\cdot,j}}_2} \epsilon^{(\mathcal{S},l)\top}_{\cdot,i} X^{(\mathcal{S},l)}_{\cdot,j}\right)^2
    - \sum_{l=1}^m \Exp \left(\frac{\sqrt{n_l p}}{\norm{X^{(\mathcal{S},l)}_{\cdot,j}}_2} \epsilon^{(\mathcal{S},l)\top}_{\cdot,i} X^{(\mathcal{S},l)}_{\cdot,j}\right)^2 \geq \eta \right] \\
    & \leq \exp \left[- C(\kappa_3) \min\left\{ 
    \left(\frac{\eta}{n_lp}\right)^2 \frac{1}{m},
    ~\frac{\eta}{n_lp}
    \right\} \right].
\end{aligned}
\end{equation*}
\cref{assmp:eigenvalues} implies that $\sum_{l=1}^m \Exp \left(\frac{\sqrt{n_l p}}{\norm{X^{(\mathcal{S},l)}_{\cdot,j}}_2} \epsilon^{(\mathcal{S},l)\top}_{\cdot,i} X^{(\mathcal{S},l)}_{\cdot,j}\right)^2 \leq C(\kappa_1,\kappa_3) ~mn_0p$. 
Plugging-in $\eta = C(\kappa_1,\kappa_3) \cdot n_0 p \cdot \max\left\{\sqrt{m \nu}, ~\nu \right\}$
\begin{equation*}
\begin{aligned}
    & \Pr\left[ \frac{1}{n_0 p} \sum_{l=1}^m \left(\frac{\sqrt{n_l p}}{\norm{X^{(\mathcal{S},l)}_{\cdot,j}}_2} \epsilon^{(\mathcal{S},l)\top}_{\cdot,i} X^{(\mathcal{S},l)}_{\cdot,j}\right)^2 
    \geq C(\kappa_1, \kappa_3) \left(m + \nu \right) \right] 
    \leq e^{- \nu}.
\end{aligned}
\end{equation*}
% The first conclusion follows the maximal inequality applied to the above probability bound. 
The second conclusion follows a similar argument with 
\begin{equation*}
    (\xi^{(\mathcal{S},l)}_{\cdot,i}, W^{(\mathcal{S},l)} \beta^{(\mathcal{S},l)}_{\cdot, j}) \distiid N\left(0, I \otimes \diag\left(\frac{1}{\Omega^{(\mathcal{S},l)}_{ii}},\frac{\Omega^{(\mathcal{S},l)}_{ii}\Sigma^{(\mathcal{S},l)}_{ii}-1}{\Omega^{(\mathcal{S},l)}_{ii}}\right)\right). 
\end{equation*}

% Plugging-in $\nu = C(\kappa_1,\kappa_3) \cdot n_0 p \cdot \max\left\{\sqrt{m \log(mn_0pq)}, ~\log(mn_0pq) \right\}$,
% \begin{equation*}
% \begin{aligned}
%     & \Pr\left[ \frac{1}{n_0 p} \sum_{l=1}^m \left(\frac{\sqrt{n_l p}}{\norm{X^{(\mathcal{S},l)}_{\cdot,j}}_2} \epsilon^{(\mathcal{S},l)\top}_{\cdot,i} X^{(\mathcal{S},l)}_{\cdot,j}\right)^2 
%     \geq C(\kappa_1, \kappa_3) \left(m + \log(mn_0pq) \right) \right] \\
%     & \leq \frac{2}{m(m-1)} (mn_0pq)^{-1/2}.
% \end{aligned}
% \end{equation*}

%%%%%%%%%%%%%%%%%%%%%%%%%%%%%%%%%%
%%%%%%%%%%%%%%%%%%%%%%%%%%%%%%%%%%
%%%%%%%%%%%%%%%%%%%%%%%%%%%%%%%%%%

\subsection{Proof of Lemma~\ref{lem:l_inf_EX}} \label{app:pf_l_inf_EX}

In \cref{eq:iid_EX_EE}, 
\begin{equation*}
    (\xi^{(\mathcal{S},l)}_{\cdot,i}, W^{(\mathcal{S},l)}_{\cdot,j}) \distiid N\left(0, I \otimes \diag\left(\frac{1}{\Omega^{(\mathcal{S},l)}_{ii}}, \Sigma^{(\mathcal{S},l)}_{jj}\right) \right),
\end{equation*}
\begin{equation*}
    (\xi^{(\mathcal{S},l)}_{\cdot,i}, W^{(\mathcal{S},l)} \beta^{(\mathcal{S},l)}_{\cdot, j}) \distiid N\left(0, I \otimes \diag\left(\frac{1}{\Omega^{(\mathcal{S},l)}_{ii}},\frac{\Omega^{(\mathcal{S},l)}_{ii}\Sigma^{(\mathcal{S},l)}_{ii}-1}{\Omega^{(\mathcal{S},l)}_{ii}}\right)\right). 
\end{equation*}
In other words, $\xi^{(\mathcal{S},l)}_{ti} W^{(\mathcal{S},l)}_{tj}$ and $\xi^{(\mathcal{S},l)}_{ti} W^{(\mathcal{S},l)}_{t,\cdot} \beta^{(\mathcal{S},l)}_{\cdot,i}$ are sub-exponential random variables with mean zero and constant bounded by $C(\kappa_3)$:
\begin{equation*}
    \Exp\left[\exp\left(\nu \xi^{(\mathcal{S},l)}_{ti} W^{(\mathcal{S},l)}_{tj}\right)\right] \leq e^{C(\kappa_3)\nu^2},
\end{equation*}
\begin{equation*}
    \Exp\left[\exp\left(\nu \xi^{(\mathcal{S},l)}_{ti} W^{(\mathcal{S},l)}_{t,\cdot} \beta^{(\mathcal{S},l)}_{\cdot, i}\right)\right] \leq e^{C(\kappa_3)\nu^2},
\end{equation*}
for all $\nu: \abs{\nu} < C(\kappa_3)$ due to \cref{assmp:eigenvalues}, and independent across $t = 1,\dots, n_l p$. Then, similar arguments as in \cref{lem:l_inf_XX} induce the desired conclusions.

%%%%%%%%%%%%%%%%%%%%%%%%%%%%%%%%%%
%%%%%%%%%%%%%%%%%%%%%%%%%%%%%%%%%%
%%%%%%%%%%%%%%%%%%%%%%%%%%%%%%%%%%

\subsection{Proof of Lemma~\ref{lem:l_inf_EE}} \label{app:pf_l_inf_EE}

In \cref{eq:iid_EX_EE}, $\xi^{(\mathcal{S},l)}_{ti} \xi^{(\mathcal{S},l)}_{tj}$ is a sub-exponential variable with
\begin{equation*}
    \Exp\left[\exp\left(\nu (\xi^{(\mathcal{S},l)}_{ti}\xi^{(\mathcal{S},l)}_{tj} - \Sigma^{(\mathcal{T},l)}_{tt} \Phi^{(\mathcal{S},l)}_{ij})\right)\right] \leq e^{C(\kappa_3)\nu^2},
\end{equation*}
for all $\nu: \abs{\nu} < C(\kappa_3)$ due to \cref{assmp:eigenvalues}, and independent across $t = 1,\dots, n_l p$. Following the same proof as in \cref{lem:l_inf_XX}, we prove the claim.

%%%%%%%%%%%%%%%%%%%%%%%%%%%%%%%%%%%%
%%%%%%%%%%%%%%%%%%%%%%%%%%%%%%%%%%%%
%%%%%%%%%%%%%%%%%%%%%%%%%%%%%%%%%%%%

\subsection{Proof of Lemma~\ref{thm:spatial_est_rho}} \label{app:pf_spatial_est_rho}

% Let $\hat{\delta}^{(\mathcal{S},l)}_{ij} := \frac{1}{n_l p} \left[ \hat\epsilon^{(\mathcal{S},l)\top}_{\cdot,i} \hat\epsilon^{(\mathcal{S},l)}_{\cdot,j}
%     - \epsilon^{(\mathcal{S},l)\top}_{\cdot,i} \epsilon^{(\mathcal{S},l)}_{\cdot,j} \right].$

For $i, j \in \{1, \dots, q\}$, let 
\begin{equation*}
    \hat\phi^{(\mathcal{S},l)}_{ij} := \frac{1}{n_l p} \left[ \hat\epsilon^{(\mathcal{S},l)\top}_{\cdot,i} \hat\epsilon^{(\mathcal{S},l)}_{\cdot,j}
    - \epsilon^{(\mathcal{S},l)\top}_{\cdot,i} \epsilon^{(\mathcal{S},l)}_{\cdot,j} \right] + \frac{1}{n_l p} \left(
        \norm{\epsilon^{(\mathcal{S},l)}_{\cdot,i}}_2^2 \Delta^{(\mathcal{S},l)}_{ij}
        + \norm{\epsilon^{(\mathcal{S},l)}_{\cdot,j}}_2^2 \Delta^{(\mathcal{S},l)}_{ji}
    \right) \mathbb{I}(i \neq j)
\end{equation*}
and
\begin{equation*}
    \tilde\phi^{(\mathcal{S},l)}_{ij} := \frac{1}{n_l p} \epsilon^{(\mathcal{S},l)\top}_{\cdot,i} \epsilon^{(\mathcal{S},l)}_{\cdot,j} - \Phi^{(\mathcal{S},l)}_{ij}.
\end{equation*}
Then, %we can represent $\hat{\Phi}_{ij} - \Phi_{ij}$ by $\hat\phi^{(\mathcal{S},l)}$ and $\tilde\phi^{(\mathcal{S},l)}_{ij}$. 
for $i = j$, 
\begin{equation*}
    \hat{\Phi}^{(\mathcal{S},l)}_{ii} - \Phi^{(\mathcal{S},l)}_{ii} 
    = \frac{1}{n_l p}
    \hat\epsilon^{(\mathcal{S},l)\top}_{\cdot,i} \hat\epsilon^{(\mathcal{S},l)}_{\cdot,i} - \Phi^{(\mathcal{S},l)}_{ii} = \hat\phi^{(\mathcal{S},l)}_{ii} + \tilde\phi^{(\mathcal{S},l)}_{ii},
\end{equation*}
and, for $i \neq j$, because $\beta^{(\mathcal{S},l)}_{ij} = - \Phi^{(\mathcal{S},l)}_{ij} / \Phi^{(\mathcal{S},l)}_{ii}$,
\begin{equation*}
\begin{aligned}
    \Phi^{(\mathcal{S},l)}_{ij} - \hat{\Phi}^{(\mathcal{S},l)}_{ij} 
    = & - \left( \Phi^{(\mathcal{S},l)}_{ij} + \Phi^{(\mathcal{S},l)}_{ii} \beta^{(\mathcal{S},l)}_{ij} + \Phi^{(\mathcal{S},l)}_{jj} \beta^{(\mathcal{S},l)}_{ji} \right) \\
    & + \frac{1}{n_l p} \left(
    \hat\epsilon^{(\mathcal{S},l)\top}_{\cdot,i} \hat\epsilon^{(\mathcal{S},l)}_{\cdot,j}
    + \norm{\hat\epsilon^{(\mathcal{S},l)}_{\cdot,i}}^2_2 \hat\beta^{(\mathcal{S},l)}_{ij}
    + \norm{\hat\epsilon^{(\mathcal{S},l)}_{\cdot,j}}^2_2 \hat\beta^{(\mathcal{S},l)}_{ji}
    \right) \\
    = & ~~~\hat\phi^{(\mathcal{S},l)}_{ij}
    % +  \frac{1}{n_l p} \left(\norm{\epsilon^{(\mathcal{S},l)}_{\cdot,i}}_2^2 \Delta^{(\mathcal{S},l)}_{ij}
    % + \norm{\epsilon^{(\mathcal{S},l)}_{\cdot,j}}_2^2 \Delta^{(\mathcal{S},l)}_{ji}
    % \right)
    + \hat\phi^{(\mathcal{S},l)}_{ii} \beta^{(\mathcal{S},l)}_{ij}
    + \hat\phi^{(\mathcal{S},l)}_{jj} \beta^{(\mathcal{S},l)}_{ji} \\
    & + \tilde\phi^{(\mathcal{S},l)}_{ij}
    + \tilde\phi^{(\mathcal{S},l)}_{ii} \beta^{(\mathcal{S},l)}_{ij} 
    + \tilde\phi^{(\mathcal{S},l)}_{jj} \beta^{(\mathcal{S},l)}_{ji}
\end{aligned}
\end{equation*}
These error terms drive $\hat\rho^{(\mathcal{S},l)}_{ij} - \rho^{(\mathcal{S},l)}_{ij}$ for $i,j: i \neq j$ by
\begin{equation*}
\begin{aligned}
    \hat\rho^{(\mathcal{S},l)}_{ij} - \rho^{(\mathcal{S},l)}_{ij} 
    = & - \frac{\hat\Phi^{(\mathcal{S},l)}_{ij}}
    {\sqrt{\hat\Phi^{(\mathcal{S},l)}_{ii} \hat\Phi^{(\mathcal{S},l)}_{jj}}} 
    + \frac{\Phi^{(\mathcal{S},l)}_{ij}}
    {\sqrt{\Phi^{(\mathcal{S},l)}_{ii} \Phi^{(\mathcal{S},l)}_{jj}}} \\
    = & \frac{\Phi^{(\mathcal{S},l)}_{ij} - \hat\Phi^{(\mathcal{S},l)}_{ij}}
    {\sqrt{\hat\Phi^{(\mathcal{S},l)}_{ii} \hat\Phi^{(\mathcal{S},l)}_{jj}}} 
    + \Phi^{(\mathcal{S},l)}_{ij} \left(
    \frac{1}{\sqrt{\Phi^{(\mathcal{S},l)}_{ii} \Phi^{(\mathcal{S},l)}_{jj}}} - \frac{1}{\sqrt{\hat\Phi^{(\mathcal{S},l)}_{ii} \hat\Phi^{(\mathcal{S},l)}_{jj}}}
    \right) \\
    = & \frac{\hat\phi^{(\mathcal{S},l)}_{ij}
        + \hat\phi^{(\mathcal{S},l)}_{ii} \beta^{(\mathcal{S},l)}_{ij}
        + \hat\phi^{(\mathcal{S},l)}_{jj} \beta^{(\mathcal{S},l)}_{ji}
        + \tilde\phi^{(\mathcal{S},l)}_{ij}
        + \tilde\phi^{(\mathcal{S},l)}_{ii} \beta^{(\mathcal{S},l)}_{ij} 
        + \tilde\phi^{(\mathcal{S},l)}_{jj} \beta^{(\mathcal{S},l)}_{ji}}
    {\sqrt{\hat\Phi^{(\mathcal{S},l)}_{ii} \hat\Phi^{(\mathcal{S},l)}_{jj}}} \\
    & + \Phi^{(\mathcal{S},l)}_{ij} \left(
    \frac{1}{\sqrt{\Phi^{(\mathcal{S},l)}_{ii} \Phi^{(\mathcal{S},l)}_{jj}}} - \frac{1}{\sqrt{\hat\Phi^{(\mathcal{S},l)}_{ii} \hat\Phi^{(\mathcal{S},l)}_{jj}}}
    \right) \\
    % & + \frac{ \hat\Phi^{(\mathcal{S},l)}_{ij} 
    % \left( 
    %   {\hat\Phi^{(\mathcal{S},l)}_{ii} \hat\Phi^{(\mathcal{S},l)}_{jj}} 
    %     - {\Phi^{(\mathcal{S},l)}_{ii} \Phi^{(\mathcal{S},l)}_{jj}} 
    % \right)}
    % {\sqrt{\Phi^{(\mathcal{S},l)}_{ii} \Phi^{(\mathcal{S},l)}_{jj} 
    % \hat\Phi^{(\mathcal{S},l)}_{ii} \hat\Phi^{(\mathcal{S},l)}_{jj}}
    % \left( 
    %     \sqrt{\hat\Phi^{(\mathcal{S},l)}_{ii} \hat\Phi^{(\mathcal{S},l)}_{jj}} 
    %     + \sqrt{\Phi^{(\mathcal{S},l)}_{ii} \Phi^{(\mathcal{S},l)}_{jj}} 
    % \right)} \\
    % = & \frac{\tilde\phi^{(\mathcal{S},l)}_{ij}
    % + \tilde\phi^{(\mathcal{S},l)}_{ii} \beta^{(\mathcal{S},l)}_{ij} 
    % + \tilde\phi^{(\mathcal{S},l)}_{jj} \beta^{(\mathcal{S},l)}_{ji}}
    % {\sqrt{\Phi^{(\mathcal{S},l)}_{ii} \Phi^{(\mathcal{S},l)}_{jj}}}
    % + \frac{\Phi^{(\mathcal{S},l)}_{ij} \left(
    % \Phi^{(\mathcal{S},l)}_{ii} \tilde\phi^{(\mathcal{S},l)}_{jj} 
    % + \Phi^{(\mathcal{S},l)}_{jj} \tilde\phi^{(\mathcal{S},l)}_{ii} \right)}
    % {2\Phi^{(\mathcal{S},l)}_{ii} \Phi^{(\mathcal{S},l)}_{jj}
    % \sqrt{\Phi^{(\mathcal{S},l)}_{ii} \Phi^{(\mathcal{S},l)}_{jj}}} + (\dots)
\end{aligned}
\end{equation*}
We notice that
\begin{equation*}
\begin{aligned}
    \Theta^{(\mathcal{S},l)}_{ij} 
    = & \frac{\tilde\phi^{(\mathcal{S},l)}_{ij}
    + \tilde\phi^{(\mathcal{S},l)}_{ii} \beta^{(\mathcal{S},l)}_{ij} / 2
    + \tilde\phi^{(\mathcal{S},l)}_{jj} \beta^{(\mathcal{S},l)}_{ji} / 2}
    {\sqrt{\Phi^{(\mathcal{S},l)}_{ii} \Phi^{(\mathcal{S},l)}_{jj}}} \\
    = & \frac{\tilde\phi^{(\mathcal{S},l)}_{ij}
    + \tilde\phi^{(\mathcal{S},l)}_{ii} \beta^{(\mathcal{S},l)}_{ij} 
    + \tilde\phi^{(\mathcal{S},l)}_{jj} \beta^{(\mathcal{S},l)}_{ji}}
    {\sqrt{\Phi^{(\mathcal{S},l)}_{ii} \Phi^{(\mathcal{S},l)}_{jj}}}
    + \frac{\Phi^{(\mathcal{S},l)}_{ij} 
    \left( \Phi^{(\mathcal{S},l)}_{ii} \tilde\phi^{(\mathcal{S},l)}_{jj} 
    + \Phi^{(\mathcal{S},l)}_{jj} \tilde\phi^{(\mathcal{S},l)}_{ii} \right)}
    {2\Phi^{(\mathcal{S},l)}_{ii} \Phi^{(\mathcal{S},l)}_{jj}
    \sqrt{\Phi^{(\mathcal{S},l)}_{ii} \Phi^{(\mathcal{S},l)}_{jj}}},
\end{aligned}
\end{equation*}
because $\beta^{(\mathcal{S},l)}_{ij} = - \Phi^{(\mathcal{S},l)}_{ij} / \Phi^{(\mathcal{S},l)}_{ii}$. Hence, the remainder term turns out to be
\begin{equation*}
\begin{aligned}
    O^{(\mathcal{S},l)}_{ij}
    = & \left[\Phi^{(\mathcal{S},l)}_{ij} \left(
    \frac{1}{\sqrt{\Phi^{(\mathcal{S},l)}_{ii} \Phi^{(\mathcal{S},l)}_{jj}}} - \frac{1}{\sqrt{\hat\Phi^{(\mathcal{S},l)}_{ii} \hat\Phi^{(\mathcal{S},l)}_{jj}}}
    \right) - \frac{\Phi^{(\mathcal{S},l)}_{ij} 
    \left( \Phi^{(\mathcal{S},l)}_{ii} \tilde\phi^{(\mathcal{S},l)}_{jj} 
    + \Phi^{(\mathcal{S},l)}_{jj} \tilde\phi^{(\mathcal{S},l)}_{ii} \right)}
    {2\Phi^{(\mathcal{S},l)}_{ii} \Phi^{(\mathcal{S},l)}_{jj}
    \sqrt{\Phi^{(\mathcal{S},l)}_{ii} \Phi^{(\mathcal{S},l)}_{jj}}}\right] \\
    & + \left( \tilde\phi^{(\mathcal{S},l)}_{ij}
        + \tilde\phi^{(\mathcal{S},l)}_{ii} \beta^{(\mathcal{S},l)}_{ij}
        + \tilde\phi^{(\mathcal{S},l)}_{jj} \beta^{(\mathcal{S},l)}_{ji}
    \right)
    \left(
    \frac{1}{\sqrt{\hat\Phi^{(\mathcal{S},l)}_{ii} \hat\Phi^{(\mathcal{S},l)}_{jj}}}
    - \frac{1}{\sqrt{\Phi^{(\mathcal{S},l)}_{ii} \Phi^{(\mathcal{S},l)}_{jj}}} \right)\\
    & + \frac{\hat\phi^{(\mathcal{S},l)}_{ij}
    + \hat\phi^{(\mathcal{S},l)}_{ii} \beta^{(\mathcal{S},l)}_{ij}
    + \hat\phi^{(\mathcal{S},l)}_{jj} \beta^{(\mathcal{S},l)}_{ji}}
    {\sqrt{\hat\Phi^{(\mathcal{S},l)}_{ii} \hat\Phi^{(\mathcal{S},l)}_{jj}}} \\
\end{aligned}
\end{equation*}

For $\hat\phi^{(\mathcal{S},l)}_{ii}$'s,
% \cref{thm:regression} implies that 
% \begin{equation*}
%     \max_i \sum_{l=1}^m \frac{1}{2 n_0 p} \norm*{\hat\epsilon^{(\mathcal{S},l)}_{\cdot,i} - \epsilon^{(\mathcal{S},l)}_{\cdot,i}}_2^2 \leq C(\kappa_1, \kappa_3, \nu) ~s ~\frac{m+\log q}{n_0 p}
% \end{equation*}
% with probability at least $1 - 2(mn_0pq)^{-1/2}$, so with the same probability
\begin{equation*}
\begin{aligned}
    \max_i \sum_{l=1}^m \abs{\hat\phi^{(\mathcal{S},l)}_{ii}}
    \leq & \max_i \sum_{l=1}^m \abs{\norm{\hat\epsilon^{(\mathcal{S},l)}_{\cdot,i}}_2^2 - \norm{\epsilon^{(\mathcal{S},l)}_{\cdot,i}}_2^2} \\
    \leq & C(\kappa_3) \max_i \sum_{l=1}^m \frac{1}{2 n_0 p} \norm*{\hat\epsilon^{(\mathcal{S},l)}_{\cdot,i} - \epsilon^{(\mathcal{S},l)}_{\cdot,i}}_2^2 \\
    \leq & C(\kappa_1, \kappa_3) ~d ~\frac{m+\log(mn_0pq)}{n_0 p}
\end{aligned}
\end{equation*}
with probability at least $1- C(mn_0pq)^{-1/2}$ for a sufficienty large $n_0$, where the probability inequality at the last line is the result of \cref{thm:spatial_regression}. For $\tilde\phi^{(\mathcal{S},l)}_{ii}$'s,
\begin{equation*}
\begin{aligned}
    \max_{i} \max_l \abs{\tilde\phi^{(\mathcal{S},l)}_{ii}} \leq C(\kappa_1, \kappa_3) \sqrt{\frac{\log (mn_0pq)}{n_0 p}}
\end{aligned}
\end{equation*}
with probability at least $1-(mn_0pq)^{-1/2}$ as a result of \cref{lem:l_inf_EE}. \cref{assmp:spatial_sample} implies that
$\max_i \hat\Phi^{(\mathcal{S},l)}_{ii} \leq C(\kappa_1, \kappa_3)$, so both $\hat\phi^{(\mathcal{S},l)}$ and $\tilde\phi^{(\mathcal{S},l)}$ are bounded. As a result, it is easily seen that, with the same probability,
\begin{equation} \label{eq:sum_O_leq_max}
    \abs{\sum_{l=1}^m O^{(\mathcal{S},l)}_{ij}} \leq C(\kappa_1, \kappa_3)  \max\left\{ \begin{aligned}
        & \sum_l \abs{\hat\phi^{(\mathcal{S},l)}_{ii}}, 
        \sum_l \abs{\hat\phi^{(\mathcal{S},l)}_{ij}},
        \sum_l \abs{\hat\phi^{(\mathcal{S},l)}_{jj}}, \\
        & \sum_l \abs{\tilde\phi^{(\mathcal{S},l)2}_{ii}},
        \sum_l \abs{\tilde\phi^{(\mathcal{S},l)}_{ii}\tilde\phi^{(\mathcal{S},l)}_{jj}},
        \sum_l \abs{\tilde\phi^{(\mathcal{S},l)2}_{jj}}, \\
        & \sum_l \abs{\tilde\phi^{(\mathcal{S},l)}_{ii}\tilde\phi^{(\mathcal{S},l)}_{ij}},
        \sum_l \abs{\tilde\phi^{(\mathcal{S},l)}_{jj}\tilde\phi^{(\mathcal{S},l)}_{ij}}.
    \end{aligned} \right\}
\end{equation}
For example, the Taylor's theorem gives
\begin{equation*}
\begin{aligned}
    & \Phi^{(\mathcal{S},l)}_{ij} \left(
    \frac{1}{\sqrt{\Phi^{(\mathcal{S},l)}_{ii} \Phi^{(\mathcal{S},l)}_{jj}}} - \frac{1}{\sqrt{\hat\Phi^{(\mathcal{S},l)}_{ii} \hat\Phi^{(\mathcal{S},l)}_{jj}}}
    \right) 
    - \frac{\Phi^{(\mathcal{S},l)}_{ij} \left( 
        \Phi^{(\mathcal{S},l)}_{ii} \tilde\phi^{(\mathcal{S},l)}_{jj} 
        + \Phi^{(\mathcal{S},l)}_{jj} \tilde\phi^{(\mathcal{S},l)}_{ii} \right)}
    {2\Phi^{(\mathcal{S},l)}_{ii} \Phi^{(\mathcal{S},l)}_{jj}
    \sqrt{\Phi^{(\mathcal{S},l)}_{ii} \Phi^{(\mathcal{S},l)}_{jj}}} \\
    = & ~ \frac{\Phi^{(\mathcal{S},l)}_{ij} \left(
        \hat\Phi^{(\mathcal{S},l)}_{ii} \hat\Phi^{(\mathcal{S},l)}_{jj}
        - \Phi^{(\mathcal{S},l)}_{ii} \Phi^{(\mathcal{S},l)}_{jj}
        - \Phi^{(\mathcal{S},l)}_{ii} \tilde\phi^{(\mathcal{S},l)}_{jj} 
        - \Phi^{(\mathcal{S},l)}_{jj} \tilde\phi^{(\mathcal{S},l)}_{ii} \right)}
    {2\Phi^{(\mathcal{S},l)}_{ii} \Phi^{(\mathcal{S},l)}_{jj}
    \sqrt{\Phi^{(\mathcal{S},l)}_{ii} \Phi^{(\mathcal{S},l)}_{jj}}} \\
    & + \Phi^{(\mathcal{S},l)}_{ij} \int_{\hat{\Phi}^{(\mathcal{S},l)}_{ii}\hat{\Phi}^{(\mathcal{S},l)}_{jj}}
    ^{\Phi^{(\mathcal{S},l)}_{ii}\Phi^{(\mathcal{S},l)}_{jj}} 
    \frac{3}{4t^{5/2}} 
    (\Phi^{(\mathcal{S},l)}_{ii} \Phi^{(\mathcal{S},l)}_{jj} - t)^2 dt \\
    \leq & C(\kappa_1,\kappa_3,\kappa_4) \max\left\{ \begin{aligned}
        \hat\Phi^{(\mathcal{S},l)}_{ii} \hat\Phi^{(\mathcal{S},l)}_{jj}
        - \Phi^{(\mathcal{S},l)}_{ii} \Phi^{(\mathcal{S},l)}_{jj}
        - \Phi^{(\mathcal{S},l)}_{ii} \tilde\phi^{(\mathcal{S},l)}_{jj} 
        - \Phi^{(\mathcal{S},l)}_{jj} \tilde\phi^{(\mathcal{S},l)}_{ii}, & \\
        \left(\Phi^{(\mathcal{S},l)}_{ii} \Phi^{(\mathcal{S},l)}_{jj}
        - \hat\Phi^{(\mathcal{S},l)}_{ii} \hat\Phi^{(\mathcal{S},l)}_{jj}\right)^2 &
    \end{aligned} \right\},
\end{aligned}
\end{equation*}
which is easily fit into \cref{eq:sum_O_leq_max}.

For the second order terms of $\tilde\phi^{(\mathcal{S},l)}$ in \cref{eq:sum_O_leq_max}, we recognize that they are random variables exhibiting a mixture of sub-Gaussian, sub-exponential, and sub-Weibull tail behaviors. The following lemma comes as a corollary of Theorem 4, \citep{bong2023tight}. See \cref{app:pf_sum_phi2} for the definition of sub-Weibull random variables and the proof.
\begin{lemma} \label{lem:sum_phi2}
\begin{equation*}
    \Pr\left[
    \begin{aligned}
        \max_{i,j} \max \left\{
        \sum_{l=1}^m \abs*{\tilde\phi^{(\mathcal{S},l)}_{ii}\tilde\phi^{(\mathcal{S},l)}_{jj}},
        \sum_{l=1}^m \abs*{\tilde\phi^{(\mathcal{S},l)}_{ii}\tilde\phi^{(\mathcal{S},l)}_{ij}} \right\} & \\
        \geq C(\kappa_1,\kappa_3) ~\frac{m + \log(mn_0pq)}{n_0 p} &
    \end{aligned}
    \right] \leq (mn_0pq)^{-1/2}.
\end{equation*}
\end{lemma}
For the linear terms of $\hat\phi^{(\mathcal{S},l)}$, we use the following upperbound. See \cref{app:pf_sum_psi} for the proof.
\begin{lemma} \label{lem:sum_psi}
\begin{equation*}
    \Pr\left[
    \begin{aligned}
        \max_{i,j} \sum_{l=1}^m \abs*{ \hat\phi^{(\mathcal{S},l)}_{ij} }
        \geq C(\kappa_1,\kappa_3) ~d ~\frac{m + \log(q m n_0 p)}{n_0 p}
    \end{aligned}
    \right] \leq C(q m n_0 p)^{-1/2}.
\end{equation*}
\end{lemma}
The desired result is a straightforward result of the two above lemmas.

\subsection{Proof of Lemma~\ref{lem:temporal_beta_Phi_F}} \label{app:pf_temporal_beta_Phi_F}

Because 
\begin{equation*}
    \frac{1}{p}\norm*{\hat{\Phi}^{(\mathcal{T},l)} - \frac{\tr(\Sigma^{(\mathcal{S},l)})}{q} \Phi^{(\mathcal{T},l)}}_F^2
    = \frac{1}{p} \sum_{t=1}^p \abs*{\Phi^{(\mathcal{T},l)}_{tt} - \frac{\tr(\Sigma^{(\mathcal{T},l)})}{q} \Phi^{(\mathcal{T},l)}_{tt}}^2,
\end{equation*}
\cref{thm:temporal_regression} implies that
\begin{equation*}
    \frac{1}{p} \norm*{\hat{\Phi}^{(\mathcal{T},l)} - \frac{\tr(\Sigma^{(\mathcal{S},l)})}{q} \Phi^{(\mathcal{T},l)}}_F^2
    \leq C(\kappa_3,\kappa_5) \frac{\nu+1}{(n_lq)^{1-1/(2+2\alpha_l)}}
\end{equation*}
with probability at least $1 - C p h_l^2 e^{-\nu}$. Under the same event, 
%For $\hat{\Phi}^{(\mathcal{T},l)-1}$, 
\begin{equation*}
\begin{aligned}
    & \frac{1}{p} \norm*{\hat{\Phi}^{(\mathcal{T},l)-1} - \frac{q}{\tr(\Sigma^{(\mathcal{S},l)})} \Phi^{(\mathcal{T},l)-1}}_F^2 \\
    & \leq \frac{1}{p} \norm*{\hat{\Phi}^{(\mathcal{T},l)} - \frac{\tr(\Sigma^{(\mathcal{S},l)})}{q} \Phi^{(\mathcal{T},l)}}_F^2 \norm*{\frac{q}{\tr(\Sigma^{(\mathcal{S},l)})} \Phi^{(\mathcal{T},l)-1}}_\mathrm{op}^2 \norm*{\hat\Phi^{(\mathcal{T},l)-1}}_\mathrm{op}^2 \\
    & \leq C(\kappa_3) \frac{1}{p} \norm*{\hat{\Phi}^{(\mathcal{T},l)} - \frac{\tr(\Sigma^{(\mathcal{S},l)})}{q} \Phi^{(\mathcal{T},l)}}_F^2
\end{aligned}
\end{equation*}
%with probability at least $1-C(mn_0pq)^{-1/2}$ 
due to \cref{assmp:eigenvalues}.
%and the bounded operator norm of $\hat{\Phi}^{(\mathcal{T},l)-1}$ for sufficiently large $n_0$ (guaranteed by \cref{thm:temporal_regression}). 
The first conclusion follows the second result of \cref{thm:temporal_regression}.
For $\hat{\beta}$, \cref{thm:temporal_regression} implies that
\begin{equation*}
\begin{aligned}
    \frac{1}{p} \norm*{(I-\hat{\beta}^{(\mathcal{T},l)})- (I-\beta^{(\mathcal{T},l)})}_F^2
    & = \frac{1}{p} \sum_{t=1}^p \norm*{\hat{\beta}^{(\mathcal{T},l)}_{\cdot,t} - \beta^{(\mathcal{T},l)}_{\cdot,t}}_2^2 \\
    & \leq C(\kappa_3,\kappa_5) \frac{\nu+1}{(n_l q)^{1-1/(2+2\alpha_l)}}
\end{aligned}
\end{equation*}
with probability at least $1 - C p h_l^2 e^{-\nu}$. Because our estimator uses eigenvalue truncation, we use Lemma B.1 in \citet{liu2017} to guarantee the error bound for $P_\eta(I - \hat{\beta}^{(\mathcal{T},l)})$: given that $\eta \leq \lambda_1(I - \beta)$,
\begin{equation*}
\begin{aligned}
    \frac{1}{p} \norm*{P_\eta(I-\hat{\beta}^{(\mathcal{T},l)}) - (I - \beta^{(\mathcal{T},l)})}_F^2
    & \leq \frac{C}{p} \norm*{(I - \hat{\beta}^{(\mathcal{T},l)}) - (I - \beta^{(\mathcal{T},l)})}_F^2 \\
    & \leq C(\kappa_3,\kappa_5) \frac{\nu+1}{(n_l q)^{1-1/(2+2\alpha_l)}}
\end{aligned}
\end{equation*}
with probability at least $1 - C(mn_0pq)^{-1/2}$. Then,
\begin{equation*}
\begin{aligned}
    & \frac{1}{p} \norm*{P_\eta(I-\hat{\beta}^{(\mathcal{T},l)})^{-1} - (I - \beta^{(\mathcal{T},l)})^{-1}}_F^2 \\
    & \leq \frac{1}{p} \norm*{P_\eta(I-\hat{\beta}^{(\mathcal{T},l)}) - (I-\beta^{(\mathcal{T},l)})}_F^2 \norm*{P_\eta(I - \hat{\beta}^{(\mathcal{T},l)})^{-1}}_\mathrm{op}^2 \norm*{(I-\beta^{(\mathcal{T},l)})^{-1}}_\mathrm{op}^2 \\
    & \leq C(\kappa_3) \frac{1}{p} \norm*{P_\eta(I-\hat{\beta}^{(\mathcal{T},l)}) - (I - \beta^{(\mathcal{T},l)})}_F^2
\end{aligned}
\end{equation*}
due to Lemma B.2 in \citet{liu2017} and the bounded operator norm of $P_\eta(I - \hat{\beta}^{(\mathcal{T},l)})$. It verifies the second conclusion.

%%%%%%%%%%%%%%%%%%%%%%%%%%%%%%%%%%
%%%%%%%%%%%%%%%%%%%%%%%%%%%%%%%%%%
%%%%%%%%%%%%%%%%%%%%%%%%%%%%%%%%%%

\subsection{Proof of Lemma~\ref{lem:sum_phi2}} \label{app:pf_sum_phi2}

We first refer to the definitions of Orlicz norms and sub-Weibull random variables in \citet{kuchibhotla2018moving}. 
\begin{definition} \label{def:orlicz_norm}
    Let $g$ be a non-decreasing non-negative function on $[0, \infty)$ with $g(0) = 0$. For a random variable $X$, the $g$-Orlicz norm is defined as
    \begin{equation*}
        \norm{X}_g := \inf\{\nu > 0: \Exp[g(\abs{X}/\nu)] \leq 1\}.
    \end{equation*}
\end{definition}
\begin{definition} \label{def:sub_Weibull}
    A random variable $X$ is sub-Weibull of order $\alpha > 0$, denoted by $X \sim \text{sub-Weibull}(\alpha)$, if
    \begin{equation*}
        \norm{X}_{\psi_\alpha} < \infty, ~~\text{where} ~~\psi_\alpha(x) := \exp(x^{\alpha}) - 1 ~~\text{for} ~~x \geq 0.
    \end{equation*}
\end{definition}
Two important examples of sub-Weibull random variables are sub-Gaussian and sub-exponential random variables, which are $\text{sub-Weibull}(2)$ and $\text{sub-Weibull}(1)$, respectively. Theorem 4 in \citet{bong2023tight} induces, for any given $i, j \in [q]$, and $\nu > 0$,
\begin{equation*}
    \Pr\left[ \sum_{l=1}^m \abs{\tilde\phi^{(\mathcal{S},l)}_{ii} \tilde\phi^{(\mathcal{S},l)}_{jj}} 
    \geq C(\kappa_1,\kappa_3) \left(
        \frac{m + \sqrt{m\nu} + \nu}{n_0 p}
        + \frac{\nu^2}{n_0^2 p^2}
    \right) \right] \leq e^{-\nu},
\end{equation*}
and by setting $\nu = C \log (mn_0pq)$, due to \cref{assmp:spatial_sample},
\begin{equation*}
\begin{aligned}
    & \Pr\left[ \sum_{l=1}^m
    \abs{\tilde\phi^{(\mathcal{S},l)}_{ii} \tilde\phi^{(\mathcal{S},l)}_{jj}} 
    \geq \sum_{l=1}^m \Exp\abs{\tilde\phi^{(\mathcal{S},l)}_{ii} \tilde\phi^{(\mathcal{S},l)}_{jj}} + C(\kappa_1,\kappa_3) \frac{m + \log(mn_0pq)}{n_0 p}\right] \\
    & \leq \frac{2}{ q^2} (mn_0pq)^{-1/2},
\end{aligned}
\end{equation*}
which similarly applies to $\tilde\phi^{(\mathcal{S},l)}_{ii} \tilde\phi^{(\mathcal{S},l)}_{ij}$. Thus, by the maximal inequality,
\begin{equation*}
\begin{aligned}
    & \Pr\left[ 
    \max_{i,j} \max\left\{ \sum_{l=1}^m
    \abs{\tilde\phi^{(\mathcal{S},l)}_{ii} \tilde\phi^{(\mathcal{S},l)}_{jj}}, 
    \sum_{l=1}^m
    \abs{\tilde\phi^{(\mathcal{S},l)}_{ii} \tilde\phi^{(\mathcal{S},l)}_{ij}} \right\} \geq C(\kappa_1,\kappa_3) \left( \frac{m + \log(mn_0pq)}{n_0 p} \right)
    \right] \\
    & \leq (mn_0pq)^{-1/2}.
\end{aligned}
\end{equation*}

\subsection{Proof of Lemma~\ref{lem:sum_psi}} \label{app:pf_sum_psi}

Because $\hat\epsilon^{(\mathcal{S},l)}_{\cdot,i} = X^{(\mathcal{S},l)}_{\cdot,i} - X^{(\mathcal{S},l)} \hat\beta^{(\mathcal{S},l)}_{\cdot,i}$ and $\epsilon^{(\mathcal{S},l)}_{\cdot,i} = X^{(\mathcal{S},l)}_{\cdot,i} - X^{(\mathcal{S},l)} \beta^{(\mathcal{S},l)}_{\cdot,i}$,
\begin{equation} \label{eq:delta_hat_decomp}
\begin{aligned}
    & \sum_{l=1}^m \abs{\hat\phi^{(\mathcal{S},l)}_{ij}} \\
    & = \sum_{l=1}^m \frac{1}{n_l p} \abs*{
    \hat\epsilon^{(\mathcal{S},l)\top}_{\cdot,i} \hat\epsilon^{(\mathcal{S},l)}_{\cdot,j} - \epsilon^{(\mathcal{S},l)\top}_{\cdot,i} \epsilon^{(\mathcal{S},l)}_{\cdot,j} 
    + \left(
        \norm{\epsilon^{(\mathcal{S},l)}_{\cdot,i}}_2^2 \Delta^{(\mathcal{S},l)}_{ij}
        + \norm{\epsilon^{(\mathcal{S},l)}_{\cdot,j}}_2^2 \Delta^{(\mathcal{S},l)}_{ji}
    \right) \mathbb{I}(i \neq j)}\\
    % \sum_{l=1}^m \frac{1}{n_l p} (X^{(\mathcal{S},l)}_{\cdot,i} - X^{(\mathcal{S},l)} \hat\beta^{(\mathcal{S},l)}_{\cdot,i})^\top (X^{(\mathcal{S},l)}_{\cdot,j} - X^{(\mathcal{S},l)} \hat\beta^{(\mathcal{S},l)}_{\cdot,j})
    & = \sum_{l=1}^m \frac{1}{n_l p} \abs*{ \begin{aligned}
    \Delta^{(\mathcal{S},l)\top}_{\cdot,i} X^{(\mathcal{S},l)\top} X^{(\mathcal{S},l)} \Delta^{(\mathcal{S},l)}_{\cdot,j} 
    - \epsilon^{(\mathcal{S},l)\top}_{\cdot,j} X^{(\mathcal{S},l)}  \Delta^{(\mathcal{S},l)}_{\cdot,i} 
    - \epsilon^{(\mathcal{S},l)\top}_{\cdot,i} X^{(\mathcal{S},l)} \Delta^{(\mathcal{S},l)}_{\cdot,j} & \\
    + \left(
    \norm{\epsilon^{(\mathcal{S},l)}_{\cdot,i}}_2^2 \Delta^{(\mathcal{S},l)}_{ij}
    + \norm{\epsilon^{(\mathcal{S},l)}_{\cdot,j}}_2^2 \Delta^{(\mathcal{S},l)}_{ji}
    \right) \mathbb{I}(i \neq j) & \\
    \end{aligned} }
\end{aligned}
\end{equation}
For the first term, it follows \cref{thm:spatial_regression} that
\begin{equation} \label{eq:delta_hat_first}
    \sum_{l=1}^m \frac{1}{n_l p} \abs*{\Delta^{(\mathcal{S},l)\top}_{\cdot,i} X^{(\mathcal{S},l)\top} X^{(\mathcal{S},l)} \Delta^{(\mathcal{S},l)}_{\cdot,j}}
    \leq C(\kappa_1,\kappa_3) ~d ~\frac{m + \log(mn_0pq)}{n_0 p}
\end{equation}
uniformly over $i$ and $j$, with probability at least $1 - 2(mn_0pq)^{-1/2}$. For the rest terms,
\begin{equation*}
\begin{aligned}
    & \sum_{l=1}^m \frac{1}{n_l p} \abs*{
    \epsilon^{(\mathcal{S},l)\top}_{\cdot,j} X^{(\mathcal{S},l)}  \Delta^{(\mathcal{S},l)}_{\cdot,i}
    - \norm{\epsilon^{(\mathcal{S},l)}_{\cdot,j}}_2^2 \Delta^{(\mathcal{S},l)}_{ij} 
    \mathbb{I}(i \neq j)} \\
    & = \sum_{l=1}^m \frac{1}{n_l p} \abs*{ 
    \epsilon^{(\mathcal{S},l)\top}_{\cdot,j} X^{(\mathcal{S},l)}_{\cdot,j}  \Delta^{(\mathcal{S},l)}_{ji} \mathbb{I}(i \neq j)
    - \norm{\epsilon^{(\mathcal{S},l)}_{\cdot,j}}_2^2 \Delta^{(\mathcal{S},l)}_{ij} 
    \mathbb{I}(i \neq j)
    + \sum_{k:k \neq i, j} \epsilon^{(\mathcal{S},l)\top}_{\cdot,j}  X^{(\mathcal{S},l)}_{\cdot,k} \Delta^{(\mathcal{S},l)}_{ki} } \\
    & \leq \sum_{l=1}^m \frac{1}{n_l p} \left( \begin{aligned}
    \abs*{\epsilon^{(\mathcal{S},l)\top}_{\cdot,j} X^{(\mathcal{S},l)} \beta^{(\mathcal{S},l)}_{\cdot,j} \Delta^{(\mathcal{S},l)}_{ji} \mathbb{I}(i \neq j)}
    + \sum_{k:k \neq i, j} \abs*{\epsilon^{(\mathcal{S},l)\top}_{\cdot,j}  X^{(\mathcal{S},l)}_{\cdot,k} \Delta^{(\mathcal{S},l)}_{ki}} 
    \end{aligned} \right).
\end{aligned}
\end{equation*}
We notice that \cref{thm:spatial_regression,lem:l_2_EX} implies that
\begin{equation*}
\begin{aligned}
    & \sum_{l=1}^m \frac{1}{n_l p}  \sum_{k:k \neq i, j} \abs*{\epsilon^{(\mathcal{S},l)\top}_{\cdot,j}  X^{(\mathcal{S},l)}_{\cdot,k} \Delta^{(\mathcal{S},l)}_{ki}}  \\
    & \leq \sum_{k:k \neq i, j} \frac{1}{n_0 p}\left[\sum_{l=1}^m \left(\frac{\sqrt{n_l p}}{\norm{X^{(\mathcal{S},l)}_{\cdot,k}}_2} \epsilon^{(\mathcal{S},l)\top}_{\cdot,i} X^{(\mathcal{S},l)}_{\cdot,k}\right)^2\right]^{1/2} \norm{\underline{\Delta}^{(\mathcal{S},\cdot)}_{ki}}_2 \\
    & \leq \left(\max_{k:k \neq i, j} \frac{1}{n_0 p}\left[\sum_{l=1}^m \left(\frac{\sqrt{n_l p}}{\norm{X^{(\mathcal{S},l)}_{\cdot,k}}_2} \epsilon^{(\mathcal{S},l)\top}_{\cdot,i} X^{(\mathcal{S},l)}_{\cdot,k}\right)^2\right]^{1/2} \right)
    \left( \sum_{k:k \neq i, j} \norm{\underline{\Delta}^{(\mathcal{S},\cdot)}_{ki}}_2 \right) \\
    & \leq C(\kappa_1,\kappa_3) ~ d ~ \frac{m+\log(mn_0pq)}{n_0 p}
\end{aligned}
\end{equation*}
and
\begin{equation*}
\begin{aligned}
    & \sum_{l=1}^m \frac{1}{n_l p} \epsilon^{(\mathcal{S},l)\top}_{\cdot,j} X^{(\mathcal{S},l)} \beta^{(\mathcal{S},l)}_{\cdot,j} \Delta^{(\mathcal{S},l)}_{ji} \\
    & \leq \frac{1}{n_0 p}\left[\sum_{l=1}^m \left(\frac{\sqrt{n_l p}}{\norm{X^{(\mathcal{S},l)}_{\cdot,j}}_2} \epsilon^{(\mathcal{S},l)\top}_{\cdot,i} X^{(\mathcal{S},l)}\beta^{(\mathcal{S},l)}_{\cdot,j}\right)^2 \right]^{1/2}
    \norm{\underline{\Delta}^{(\mathcal{S},\cdot)}_{ji}}_2 \\
    & \leq C(\kappa_1,\kappa_3) ~ d ~ \frac{m+\log(mn_0pq)}{n_0 p}
\end{aligned}
\end{equation*}
satisfy simultaneously uniformly over $i$ and $j$ with probability at least $1 - C (mn_0pq)^{-1/2}$. We obtain the desired result by plugging the above probability bounds into \cref{eq:delta_hat_decomp}.

\section{Derivation of the asymptotic covariance between the edge-wise test statistics} \label{app:asymp_cov}

In this section, we derive the asymptotic covariance of $\hat{T}_{i_1j_1}$ and $\hat{T}_{i_2j_2}$ in \cref{eq:inf_S}:
\begin{equation*}
\begin{split}
    & S_{(i_1,j_1),(i_2,j_2)} \\
    & := \frac{1}{m} \sum_{l=1}^m \frac{\norm{\Sigma^{(\mathcal{T},l)}}_F^2}{p} 
    \left[ \begin{split} 
    & \rho^{(\mathcal{S},l)}_{i_1 i_2} \rho^{(\mathcal{S},l)}_{j_1 j_2}
    + \rho^{(\mathcal{S},l)}_{i_1 j_2} \rho^{(\mathcal{S},l)}_{i_2 j_1}
    + \frac{1}{2} \rho^{(\mathcal{S},l)}_{i_1 j_1} \rho^{(\mathcal{S},l)}_{i_2 j_2} 
    \Big( \rho^{(\mathcal{S},l)2}_{i_1 i_2} + \rho^{(\mathcal{S},l)2}_{j_1 j_2}
    + \rho^{(\mathcal{S},l)2}_{i_1 j_2} + \rho^{(\mathcal{S},l)2}_{i_2 j_1} \Big) \\
    & - \rho^{(\mathcal{S},l)}_{i_1 i_2} 
    \rho^{(\mathcal{S},l)}_{i_2 j_2} \rho^{(\mathcal{S},l)}_{i_2 j_1}
    - \rho^{(\mathcal{S},l)}_{i_1 i_2}
    \rho^{(\mathcal{S},l)}_{i_1 j_1} \rho^{(\mathcal{S},l)}_{i_1 j_2}
    - \rho^{(\mathcal{S},l)}_{j_1 j_2} 
    \rho^{(\mathcal{S},l)}_{i_2 j_2} \rho^{(\mathcal{S},l)}_{i_1 j_2}
    - \rho^{(\mathcal{S},l)}_{j_1 j_2} \rho^{(\mathcal{S},l)}_{i_2 j_1} \rho^{(\mathcal{S},l)}_{i_1 j_1}
    \end{split} \right].
\end{split}
\end{equation*}
By \cref{thm:spatial_est_rho}, $\hat{\rho}^{(\mathcal{S},l)}_{i j}-\rho^{(\mathcal{S},l)}_{i j}$ has the leading term $\Theta^{(\mathcal{S},l)}_{ij}$. Hence, $\hat{T}_{i_1 j_1}$ and $\hat{T}_{i_2 j_2}$ have asymptotic covariance,
\begin{equation} \label{eq:S_by_Theta_sq} 
\begin{aligned}
    S_{(i_1,j_1),(i_2,j_2)} 
    &= \Exp\left[
    \left( \frac{1}{\sqrt{m}} \sum_{l=1}^m \sqrt{n_lp} \Theta^{(\mathcal{S},l)}_{i_1 j_1} \right)
    \left( \frac{1}{\sqrt{m}} \sum_{l=1}^m \sqrt{n_lp} \Theta^{(\mathcal{S},l)}_{i_2 j_2} \right) \right] \\
     &=\frac{1}{m} \sum_{l=1}^m n_lp  
     \Exp\left[ \Theta^{(\mathcal{S},l)}_{i_1 j_1} \Theta^{(\mathcal{S},l)}_{i_2 j_2} \right]. 
\end{aligned}
\end{equation}
%     
% Therefore, we simply need to calculate each entry in $\mat{W}_t^\rho = \mathbb{E}\{\mat{\Theta}_{tS} (\mat{\Theta}_{tS})^{\T} \}$. 
% 
% To proceed for the calculation of $\mat{W}_t^\rho$, we omit $t$ here since the calculation is the same for each graph. Denote by $\tilde{r}_{tij} = \frac{1}{n_tp}\sum_{k=1}^{n_t} \sum_{l=1}^p {\varepsilon}^{(k)}_{tli}{\varepsilon}^{(k)}_{tlj}$, and
% \begin{equation}\label{eqn:tilde_delta}
%  \tilde{\delta}_{tij} = \tilde{r}_{tij} - r_{tij}=\frac{1}{n_tp}\sum_{k=1}^{n_t} \sum_{l=1}^p {\varepsilon}^{(k)}_{tli}{\varepsilon}^{(k)}_{tlj}- \frac{b_{tij}}{b_{tii}b_{tjj}}.   
% \end{equation}
Based on \cref{eq:spatial_Theta}, % and the following definition of $\tilde{\delta}_{ij}^{(\mathcal{S},l)}$,
for each $l=1,\dots,m$,
\begin{equation} \label{eq:Theta_sq_by_delta_sq}
\begin{aligned}
    & \Exp\left[ \Theta^{(\mathcal{S},l)}_{i_1 j_1} \Theta^{(\mathcal{S},l)}_{i_2 j_2} \right] \\
    & = \Exp\left[ \begin{aligned}
    & \left( \frac{\tilde{\delta}^{(\mathcal{S},l)}_{i_1j_1}}
    {\sqrt{\Phi^{(\mathcal{S},l)}_{i_1i_1} \Phi^{(\mathcal{S},l)}_{j_1j_1}}} 
    - \frac{\Phi^{(\mathcal{S},l)}_{i_1j_1} \tilde{\delta}^{(\mathcal{S},l)}_{j_1j_1}}
    {2 \Phi^{(\mathcal{S},l)}_{j_1j_1} 
    \sqrt{\Phi^{(\mathcal{S},l)}_{i_1i_1} \Phi^{(\mathcal{S},l)}_{j_1j_1}}}
    - \frac{\Phi^{(\mathcal{S},l)}_{i_1j_1} \tilde{\delta}^{(\mathcal{S},l)}_{i_1i_1}}
    {2 \Phi^{(\mathcal{S},l)}_{i_1i_1} 
    \sqrt{\Phi^{(\mathcal{S},l)}_{i_1i_1} \Phi^{(\mathcal{S},l)}_{j_1j_1}}} \right)\\
    & \times \left( \frac{\tilde{\delta}^{(\mathcal{S},l)}_{i_2j_2}}
    {\sqrt{\Phi^{(\mathcal{S},l)}_{i_2i_2} \Phi^{(\mathcal{S},l)}_{j_2j_2}}} 
    - \frac{\Phi^{(\mathcal{S},l)}_{i_2j_2} \tilde{\delta}^{(\mathcal{S},l)}_{j_2j_2}}
    {2 \Phi^{(\mathcal{S},l)}_{j_2j_2} 
    \sqrt{\Phi^{(\mathcal{S},l)}_{i_2i_2} \Phi^{(\mathcal{S},l)}_{j_2j_2}}}
    - \frac{\Phi^{(\mathcal{S},l)}_{i_2j_2} \tilde{\delta}^{(\mathcal{S},l)}_{i_2i_2}}
    {2 \Phi^{(\mathcal{S},l)}_{i_2i_2} 
    \sqrt{\Phi^{(\mathcal{S},l)}_{i_2i_2} \Phi^{(\mathcal{S},l)}_{j_2j_2}}}\right)
    \end{aligned} \right],
\end{aligned}
\end{equation}
which is a summation of the second moments of $\tilde{\delta}^{(\mathcal{S},l)}$. For each second moment,
\begin{equation} \label{eq:delta_sq_by_ksi_sq}
\begin{aligned}
    & \Exp[\tilde{\delta}^{(\mathcal{S},l)}_{i_1 j_1} \tilde{\delta}^{(\mathcal{S},l)}_{i_2 j_2}] \\
    & = \Exp\left[ \frac{1}{n_l^2 p^2} \sum_{k=1}^{n_l} \sum_{t=1}^{p} \left( 
        \epsilon^{(\mathcal{S},k,l)}_{ti_1} \epsilon^{(\mathcal{S},k,l)}_{tj_1}
        - \Sigma^{(\mathcal{T},l)}_{tt} \Phi^{(\mathcal{S},l)}_{i_1j_1} \right)
        \cdot \sum_{k=1}^{n_l} \sum_{t=1}^{p} \left( 
        \epsilon^{(\mathcal{S},k,l)}_{ti_2} \epsilon^{(\mathcal{S},k,l)}_{tj_2}
        - \Sigma^{(\mathcal{T},l)}_{tt} \Phi^{(\mathcal{S},l)}_{i_2j_2}\right) \right] \\
    & = \frac{1}{n_l^2 p^2} \sum_{t=1}^p \lambda^{(\mathcal{T},l)}_t  \cdot
        \Exp\left[ \sum_{k=1}^{n_l} \left(  \xi^{(k,l)}_{ti_1} \xi^{(k,l)}_{tj_1}
        - \Phi^{(\mathcal{S},l)}_{i_1j_1} \right)
        \cdot \sum_{k=1}^{n_l} \left( \xi^{(k,l)}_{ti_2} \xi^{(k,l)}_{tj_2}
        - \Phi^{(\mathcal{S},l)}_{i_2j_2} \right) \right] \\
    & = \frac{\norm{\Sigma^{(\mathcal{T},l)}}_F^2}{n_l p^2}
        \cdot \Exp\left[ \left(  \xi^{(k,l)}_{ti_1} \xi^{(k,l)}_{tj_1}
        - \Phi^{(\mathcal{S},l)}_{i_1j_1} \right)
        \cdot \left( \xi^{(k,l)}_{ti_2} \xi^{(k,l)}_{tj_2}
        - \Phi^{(\mathcal{S},l)}_{i_2j_2} \right) \right],
\end{aligned}
\end{equation}
where
\begin{equation} \label{eq:ksi_sq_by_rho}
\begin{aligned}
    \Exp\left[ \left(  \xi^{(k,l)}_{ti_1} \xi^{(k,l)}_{tj_1}
        - \Phi^{(\mathcal{S},l)}_{i_1j_1} \right)
        \cdot \left( \xi^{(k,l)}_{ti_2} \xi^{(k,l)}_{tj_2}
        - \Phi^{(\mathcal{S},l)}_{i_2j_2} \right) \right]
    = \Phi^{(\mathcal{S},l)}_{i_1 i_2} \Phi^{(\mathcal{S},l)}_{j_1 j_2}
    + \Phi^{(\mathcal{S},l)}_{i_1 j_1} \Phi^{(\mathcal{S},l)}_{i_2 j_2}.
\end{aligned}
\end{equation}
Plugging it into \cref{eq:Theta_sq_by_delta_sq},
\begin{equation*}
\begin{aligned}
    & \Exp\left[ \Theta^{(\mathcal{S},l)}_{i_1 j_1} \Theta^{(\mathcal{S},l)}_{i_2 j_2} \right] \\
    & = \frac{\norm{\Sigma^{(\mathcal{T},l)}}_F^2}{n_l p^2} 
    \frac{1}{\sqrt{\Phi^{(\mathcal{S},l)}_{i_1i_1} \Phi^{(\mathcal{S},l)}_{j_1j_1}
    \Phi^{(\mathcal{S},l)}_{i_2i_2} \Phi^{(\mathcal{S},l)}_{j_2j_2}}}\\
    & \hspace{0.65cm} \times \left[ \begin{aligned}
        & \Phi^{(\mathcal{S},l)}_{i_1i_2} \Phi^{(\mathcal{S},l)}_{j_1j_2} + \Phi^{(\mathcal{S},l)}_{i_1j_2} \Phi^{(\mathcal{S},l)}_{j_2i_1}
        + \frac{\Phi^{(\mathcal{S},l)}_{i_1j_1} \Phi^{(\mathcal{S},l)}_{i_2j_2}}
        {2 \Phi^{(\mathcal{S},l)}_{i_1i_1} \Phi^{(\mathcal{S},l)}_{i_2i_2}} 
        ( \Phi^{(\mathcal{S},l)2}_{i_1i_2} + \Phi^{(\mathcal{S},l)2}_{i_1j_2} + \Phi^{(\mathcal{S},l)2}_{j_1i_2} + \Phi^{(\mathcal{S},l)2}_{j_1j_2} ) \\
        & - \frac{\Phi^{(\mathcal{S},l)}_{i_2j_2}}{\Phi^{(\mathcal{S},l)}_{i_2i_2}} \Phi^{(\mathcal{S},l)}_{i_1i_2} \Phi^{(\mathcal{S},l)}_{j_1i_2}
        - \frac{\Phi^{(\mathcal{S},l)}_{i_2j_2}}{\Phi^{(\mathcal{S},l)}_{j_2j_2}} \Phi^{(\mathcal{S},l)}_{i_1j_2} \Phi^{(\mathcal{S},l)}_{j_1j_2}
        - \frac{\Phi^{(\mathcal{S},l)}_{i_1j_1}}{\Phi^{(\mathcal{S},l)}_{i_1i_1}} \Phi^{(\mathcal{S},l)}_{i_2i_1} \Phi^{(\mathcal{S},l)}_{j_2i_1}
        - \frac{\Phi^{(\mathcal{S},l)}_{i_1j_1}}{\Phi^{(\mathcal{S},l)}_{j_1j_1}} \Phi^{(\mathcal{S},l)}_{i_2j_1} \Phi^{(\mathcal{S},l)}_{j_2j_1}
    \end{aligned} \right]\\
    & = \frac{\norm{\Sigma^{(\mathcal{T},l)}}_F^2}{n_l p^2} 
    \left[ \begin{aligned} 
        & \rho^{(\mathcal{S},l)}_{i_1 i_2} \rho^{(\mathcal{S},l)}_{j_1 j_2}
        + \rho^{(\mathcal{S},l)}_{i_1 j_2} \rho^{(\mathcal{S},l)}_{i_2 j_1}
        + \frac{1}{2} \rho^{(\mathcal{S},l)}_{i_1 j_1} \rho^{(\mathcal{S},l)}_{i_2 j_2} 
        \Big( \rho^{(\mathcal{S},l)2}_{i_1 i_2} + \rho^{(\mathcal{S},l)2}_{j_1 j_2}
        + \rho^{(\mathcal{S},l)2}_{i_1 j_2} + \rho^{(\mathcal{S},l)2}_{i_2 j_1} \Big) \\
        & - \rho^{(\mathcal{S},l)}_{i_1 i_2} 
        \rho^{(\mathcal{S},l)}_{i_2 j_2} \rho^{(\mathcal{S},l)}_{i_2 j_1}
        - \rho^{(\mathcal{S},l)}_{i_1 i_2}
        \rho^{(\mathcal{S},l)}_{i_1 j_1} \rho^{(\mathcal{S},l)}_{i_1 j_2}
        - \rho^{(\mathcal{S},l)}_{j_1 j_2} 
        \rho^{(\mathcal{S},l)}_{i_2 j_2} \rho^{(\mathcal{S},l)}_{i_1 j_2}
        - \rho^{(\mathcal{S},l)}_{j_1 j_2} \rho^{(\mathcal{S},l)}_{i_2 j_1} \rho^{(\mathcal{S},l)}_{i_1 j_1}
    \end{aligned} \right].
\end{aligned}
\end{equation*}
Plugging it into \cref{eq:S_by_Theta_sq} retrieves the desired result.

\end{document}